\documentclass[fleqn,usenatbib]{mnras}

\usepackage{newtxtext,newtxmath}
\usepackage[T1]{fontenc}

\DeclareRobustCommand{\VAN}[3]{#2}
\let\VANthebibliography\thebibliography
\def\thebibliography{\DeclareRobustCommand{\VAN}[3]{##3}\VANthebibliography}

\usepackage{epsfig}
\usepackage{amsmath}
\usepackage{color}
\usepackage{booktabs}
\usepackage{pdflscape}
\usepackage{textcomp}

\usepackage{soul}
\usepackage{aas_macros}  % required to handle the ADS bibtex journal macros
\usepackage{chngcntr}
\usepackage{array}
\newcolumntype{P}[1]{>{\raggedright\arraybackslash}p{#1}}
\graphicspath{{figures/}}

  \newcommand{\Teff}{\mbox{\,\em T$_{\rm eff}$}}         % effective temperature
 \newcommand{\teff}{\mbox{\,$T_{\rm eff}$}}      % Teff 
\newcommand{\lgcs}{\mbox{\,$\log g / {\rm cm\,s^{-2}}$}}        % log g
\newcommand{\nH}{\mbox{\,$n_{\rm H}$}}                 % hydrogen abundance
  \newcommand{\nHe}{\mbox{\,$n_{\rm He}$}}               % helium abundance
  \newcommand{\vsini}{\mbox{\,$v\,\sin i$}}              % V sin i
  \newcommand{\kmsec}{\,\mbox{$\mbox{km}\,\mbox{s}^{-1}$}}    % kilometres/second
  \def\simge{\mathrel{\raise1.16pt\hbox{$>$}\kern-7.0pt
    \lower3.06pt\hbox{{$\scriptstyle \sim$}}}}           % approx ge
  \def\simle{\mathrel{\raise1.16pt\hbox{$<$}\kern-7.0pt
    \lower3.06pt\hbox{{$\scriptstyle \sim$}}}}           % approx le
\title[SALT subdwarfs I]{The SALT survey of helium-rich hot subdwarfs: methods, classification, and coarse analysis.}
\author[Jeffery et al.]{C. S. Jeffery$^{1}$, B. Miszalski$^{2}$ and E. Snowdon$^{1}$\\
$^1$Armagh Observatory and Planetarium, College Hill, Armagh BT61 9DG, United Kingdom\\
$^3$Australian Astronomical Optics - Macquarie, Faculty of Science and Engineering, Macquarie University, North Ryde, NSW 2113, Australia
}

\begin{document}

\date{Accepted \ldots. Received \ldots; in original form \ldots}

\pagerange{\pageref{firstpage}--\pageref{lastpage}} \pubyear{2020}

\maketitle

\label{firstpage}

\begin{abstract}
A medium- and high-resolution spectroscopic survey of helium-rich hot subdwarfs  is being carried out using the Southern African Large Telescope (SALT).
Objectives include the discovery of exotic hot subdwarfs and of sequences connecting chemically-peculiar subdwarfs of different types. 
The first phase consists of medium-resolution spectroscopy of over 100 stars selected from low-resolution surveys.
This paper describes the selection criteria, and the observing, classification and analysis methods. 
It presents 107 spectral classifications on the MK-like Drilling system and 106 coarse analyses ($\Teff, \log g, \log y$) based on a hybrid grid of zero-metal non-LTE and line-blanketed LTE model atmospheres.  
For 75 stars, atmospheric parameters have been derived for the first time. 
The sample may be divided into 6 distinct groups including the classical `helium-rich' sdO stars with spectral types (Sp) sdO6.5 - sdB1 (74) comprising carbon-rich (35) and carbon-weak (39) stars,  very hot He-sdO's with Sp $\lesssim$ sdO6 (13), extreme helium stars with luminosity class $\lesssim 5$ (5), intermediate helium-rich subdwarfs with helium class 25 -- 35 (8), and intermediate helium-rich subdwarfs with helium class $10 - 25$ (6). 
The last covers a narrow spectral range (sdB0 -- sdB1) including two known and four candidate heavy-metal subdwarfs. 
Within other groups are several stars of individual interest, including an extremely metal-poor helium star, candidate double-helium subdwarf binaries, and a candidate low-gravity He-sdO star.  
\end{abstract}


\section{Introduction}
Hot subluminous stars can be divided into three major groups. 
These include i) the hydrogen-rich subdwarf B (sdB) stars, often characterized as extreme horizontal-branch stars, ii) the sdOB and sdO stars lying on or around the helium-main sequence, and iii) the more luminous sdO stars on post-AGB evolution tracks \citep{heber16}.       
Apart from the sdB stars which have hydrogen-rich surfaces, a substantial fraction of hot subdwarfs have hydrogen-deficient or hydrogen-weak surfaces. 
Amongst these, there is evidence for sequences extending either away from or towards the helium main-sequence, connecting with cooler extreme helium stars, or with the white dwarf cooling sequence; many have atmospheres enriched in carbon or nitrogen or both.      
Amongst the subdwarfs with hydrogen-weak surfaces, several show extraordinary overabundances of heavy metals (trans-iron elements) including zirconium and lead \citep{naslim11,naslim13}.
This diversity is apparent in the helium subclasses identified by \citet{drilling13} (D13 hereafter), who noted that certain classes of helium-rich hot subdwarf and extreme helium stars are difficult to distinguish at low resolution. 
In order to trace  these sequences of hydrogen-deficient  and hydrogen-weak subdwarfs with greater clarity, to discover how they relate to other categories of hydrogen-deficient star, and to study the physics that transforms their surface chemistries, we commenced a survey of chemically-peculiar hot subdwarfs. 
The object of the survey would be to obtain spectra of sufficient quality to measure effective temperature, surface gravities, and surface hydrogen, helium, carbon and nitrogen abundances, as well as to identify any  exotic elements that might be present.  
This paper reports the initial part of the survey including selection criteria, observing procedures, data products and primary classifications. 

\begin{figure}
\begin{center}
\includegraphics[width=1.0\linewidth]{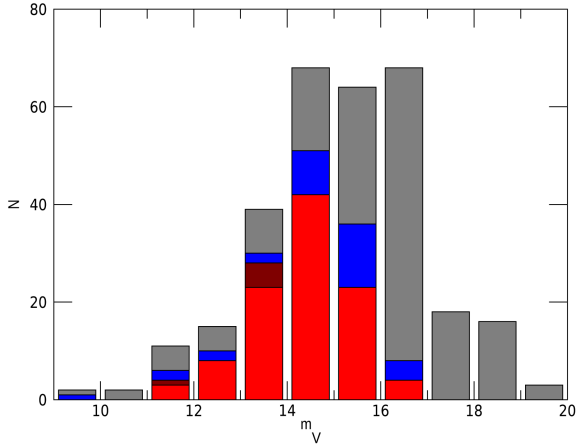}
\caption{Relative numbers of known or suspected He-sdO, He-sdOB and He-sdB stars visible from SALT ordered by brightness. The coloured segments represent the numbers observed in the current sample with SALT/RSS only or with SALT/RSS {\it and} SALT/HRS (103 stars: red), with SALT/HRS only (6: dark red), or with another high-resolution spectrograph and {\it not} SALT (33: blue). As of 2020 September 30, the total known to us is 306. } 
\label{f:bright}
\end{center}
\end{figure}

\section{Observations}
\subsection{Target selection}
The primary motivation for this survey was the classification of several stars in the Edinburgh-Cape (EC) survey of faint blue stars as `He-sdB' \citep{stobie97a,Cat.EC2}, a classification similar to sdOD in the Palomar-Green (PG) survey \citep{green86}  and indistinguishable at the survey resolutions from that of `extreme helium stars' (D13). 
Efforts to explore this category by \citet{ahmad03a} and by \citet{naslim10} were limited by telescope aperture and observing time. 
The construction of the Southern African Large Telescope (SALT) offered the perfect opportunity to extend previous studies. 

Initial target selection was made on the basis of low-resolution classifications of He-sdB, He-sdOB and He-sdO in the EC survey. 
These classifications are described by \citet{moehler90b,geier17} and \citet{lei20a}.
To these were added similar stars classified He-sdB, sdOD, or similar in one or more  of the compilations by \citet{carnochan83,green86,kilkenny82,kilkenny88,beers92,stobie97a,Cat.EC2,sdss4,ostensen06,nemeth12,Cat.EC3,Cat.EC4,kepler15.sdss10,Cat.EC5} or \citet{geier17}. 
Stars which had been observed at high resolution with \'echelle spectrographs at either VLT/UVES \citep{stroeer07}, AAT/UCLES \citep{ahmad07c,naslim11,naslim12}, ESO/FEROS
\citep{naslim13}, or Subaru/HDS \citep{jeffery17a,naslim20} were  not included at first, but the benefits of having medium resolution spectra available for class prototypes meant that some were included later. 

Since the boundaries between He-sdB, He-sdOB and He-sdO are  spectroscopic and therefore artificial in terms of exploring connections between stars in closely related classes, stars from all three categories were included as the survey progressed. 
Moreover, since some chemically-peculiar subdwarfs simply have solar or slightly super-solar abundances of helium, we included a number of sdOB and sdO stars.
The principal exclusions were stars classified sdB, since these usually have weak or absent He{\sc i} and no He{\sc ii} lines.  

Our helium-rich list currently (2020 September) contains over 600 subdwarfs, of which $306$ lie between the declinations of $-75^{\circ}$ and  $+8^{\circ}$, the effective limits of SALT. 
Some 33 of the latter have been observed at high resolution in  campaigns cited above and are not included here. 
Approximately 30 are common to previous campaigns and to the SALT observations presented here.     
Figure \ref{f:bright} shows the brightness distribution of known or suspected southern helium-rich subdwarfs accessible to SALT. 

\begin{table}
    \caption{Observation dates in the form {\tt yyyymmdd}.  The full table is given in the Supplementary Online Data.  }
    \label{t:salt_obs}
    \setlength{\tabcolsep}{2pt}
    \centering
    \begin{tabular}{P{25mm}P{27mm}P{27mm}}
    \hline
Name & RSS Dates & HRS Dates \\
\hline
Ton S 144 &  20181101 &  20180611\\
Ton S 148 &  20191101 &  20180616 20180705 20180722 20190619 20190715\\
\ldots & \ldots & \ldots \\
    \hline
    \end{tabular}
\end{table}

\subsection{SALT/HRS}
From 2016 to the present, observations have been obtained with the SALT High Resolution Spectrograph  \citep[HRS: $R\approx43\,000$, $\lambda\lambda = 4100 - 5200$\AA,][]{bramall10}.  
Observation dates are given in Table\,\ref{t:salt_obs}.
HRS spectra obtained prior to 2019 were reduced to order-by-order wavelength calibrated rectified form using the SALT pipeline pyHRS \citep{crawford16}; orders were stitched into a single spectrum using our own software. 
In general, spectra were obtained in pairs (or a higher multiple) which were coadded to provide a single observation for each date.
The pyHRS pipeline ceased to be supported after the beginning of 2019.
HRS spectra obtained after that date will be described in a subsequent paper. 

\subsection{SALT/RSS}
To reduce errors arising from poor blaze correction, which is difficult for broad-lined spectra, and also to extend the sample to stars too faint for HRS, observations were also obtained with the SALT Robert Stobie Spectrograph (RSS: resolution $R\approx 3\,600$, \cite{burgh03,kobulnicky03}). 
Observation dates are given in Table\,\ref{t:salt_obs}.
Since the RSS detector consists of three charge-coupled devices separated by two gaps,  double exposures were taken at two different grating angles. 
This provides a continuous spectrum in the wavelength range 3850 -- 5150 \AA\, and assists in the removal of cosmic-ray contamination. 
Basic data processing used the \textsc{pysalt}\footnote{http://pysalt.salt.ac.za} package \citep{crawford10}.
Reduction used standard \textsc{iraf} tasks and the \textsc{lacosmic} package \citep{dokkum01} as described by \citet{koen17}. 
The one-dimensional wavelength-calibrated and sky-subtracted spectra were extracted using the \textsc{apall} task.  
These were rectified using low-order polynomials fitted to regions of continuum identified automatically. 
The three segments from both observations at both grating angles were merged  using weights based on the number of photons detected in each segment. 
The wavelengths of each spectrum were adjusted to correct for Earth motion. 

\subsection{The Drilling sample}
The complete sample of normalised spectra used by D13 (the `Drilling sample') has been used to validate the classification procedure.   

\subsection{Nomenclature}
The SALT sample is described in Table\,\ref{t:saltclass}, which gives positions (J2000.0), Gaia magnitudes, names, and classifications. By convention, we adopt the catalogue name at which the star was first identified as a helium-rich subdwarf. Other catalogues which include the star are indicated by abbreviation; a full list is available for each star from {\sc simbad} \citep{wenger00}. For brevity, we have contracted BPS\,CS to BPS and, except in Table\,\ref{t:saltclass}, the full GALEX identifier to GLX Jhhmmm+ddmm, with positions rounded down to tenths of a minute in right ascension and arcminutes in declination.

\begin{landscape}
\begin{figure}
\includegraphics[width=0.50\linewidth]{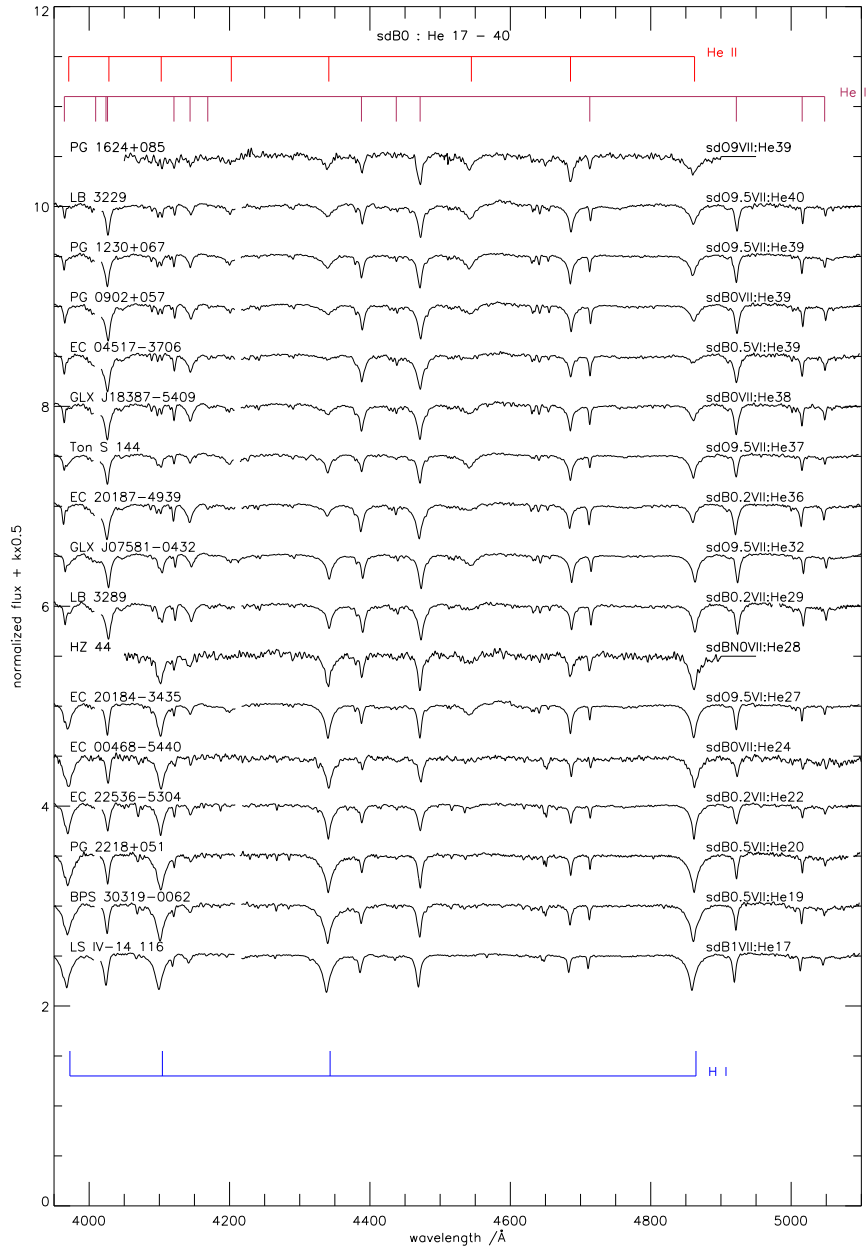}
\includegraphics[width=0.50\linewidth]{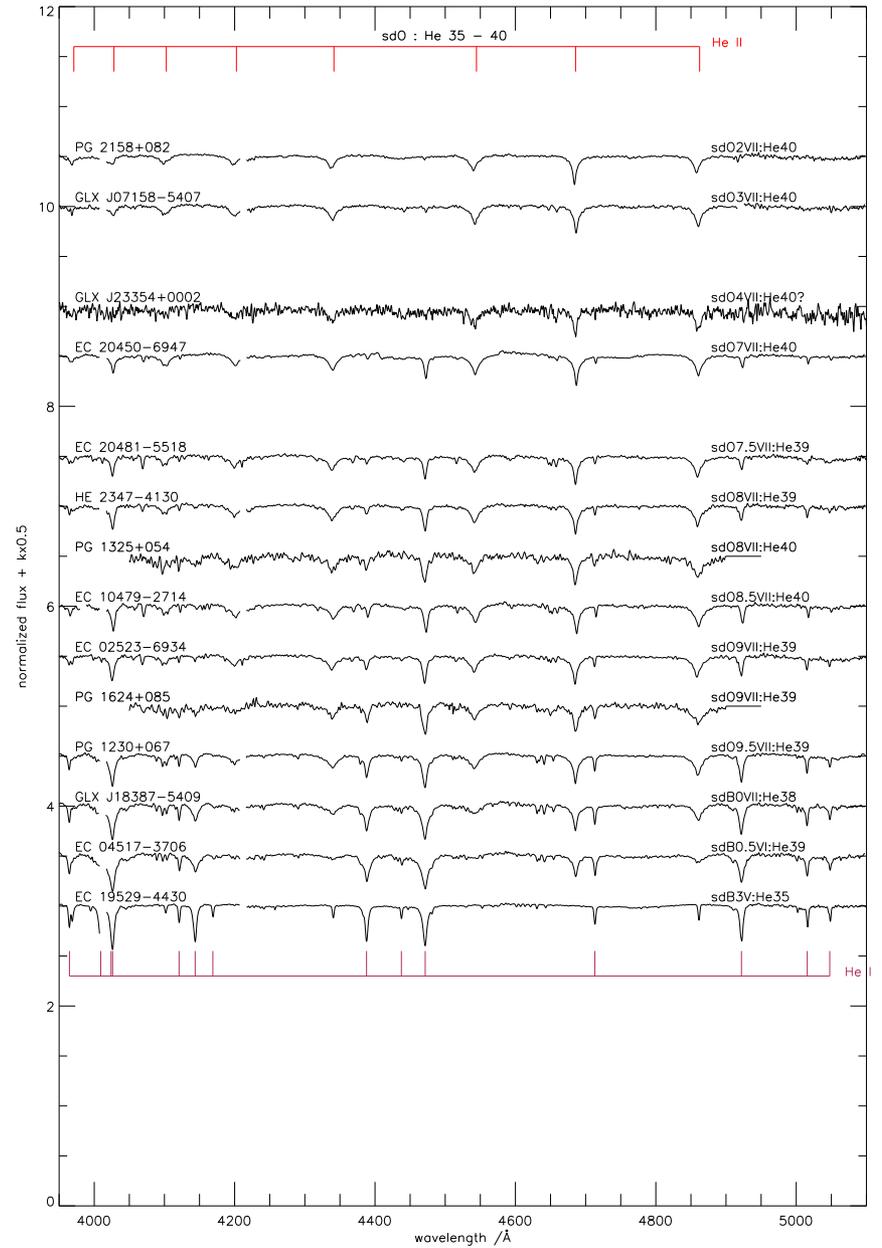}
\caption{SALT atlas of early-type helium-rich hot subdwarfs showing (left)  helium classes from He17 to He40 at spectral type around sdB0
and (right) spectral types from sdO2 to sdB3 with helium class 38 -- 40 that do not show strong C or N lines.
Principal lines are indicated in colour.
In some cases star names are obvious contractions of names given in Table\,2.
Bold labels and spectra represent standards from D13.
}
\label{f:HE}
\end{figure}
\end{landscape}
\begin{landscape}
\begin{figure}
\includegraphics[width=0.50\linewidth]{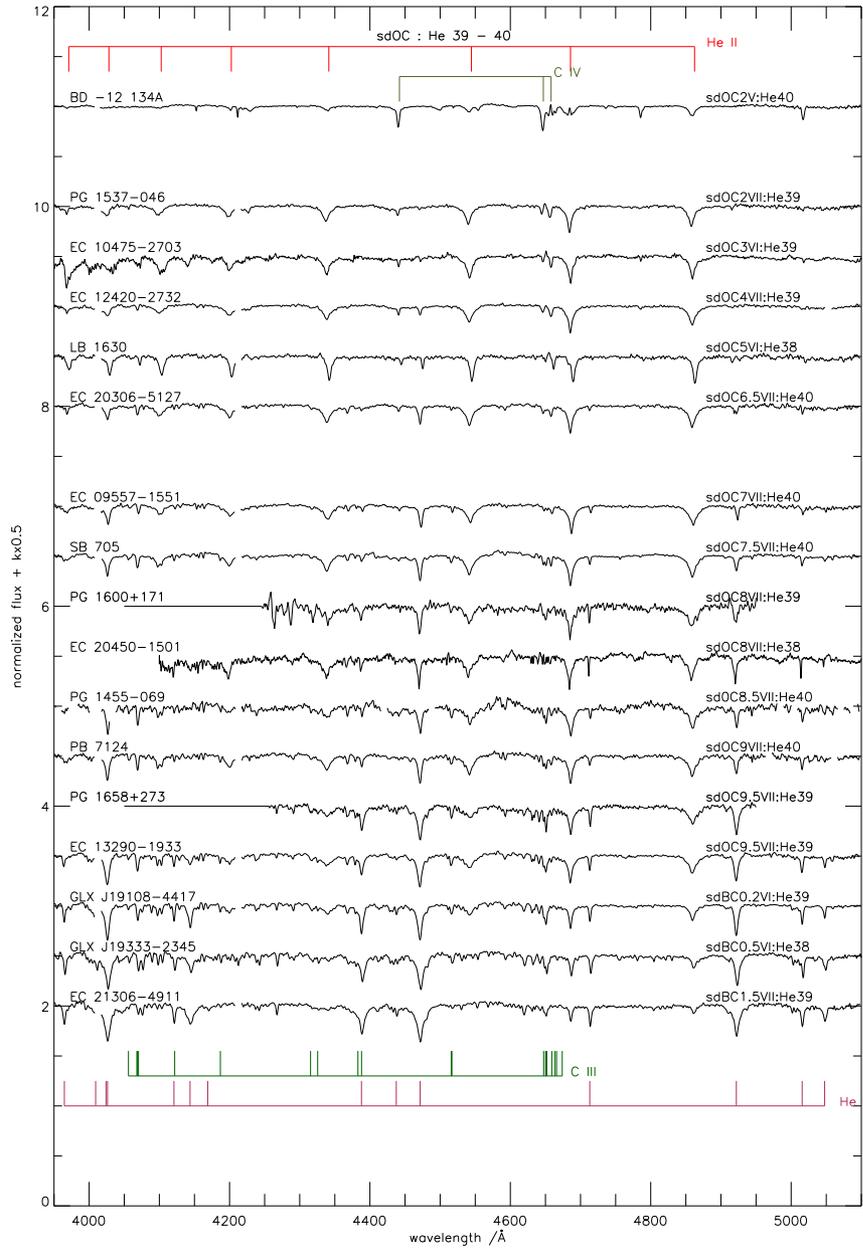}
\includegraphics[width=0.50\linewidth]{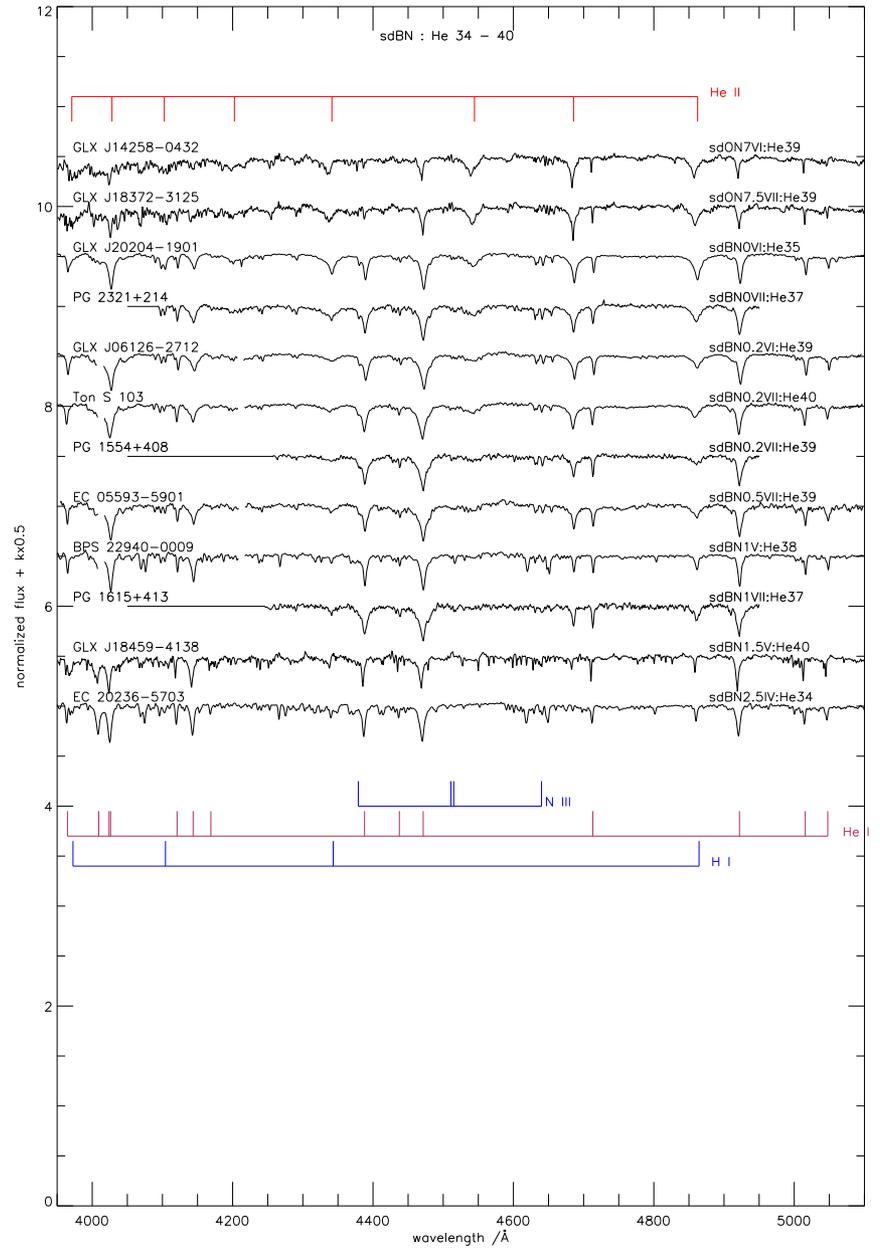}
\caption{As Fig. \ref{f:HE} for (left) C-rich subdwarfs with spectral types from sdOC2 to sdBC1.5 with helium class 38 -- 40 and 
(right) N-rich subdwarfs with spectral types from sdON7 to sdBN2.5 with helium class 34 -- 40. }
\label{f:OC}
\end{figure}
\end{landscape}

\begin{figure}
\begin{center}
\includegraphics[clip, width=1.0\linewidth]{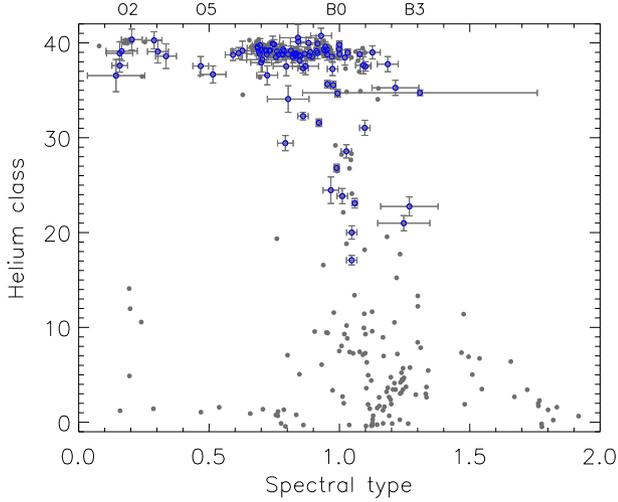}
\caption{The Sp -- He classification diagram for helium-rich subdwarfs (blue circles). The distribution from D13 is shown by grey dots; a uniform jitter covering $\pm$ half a division has been applied to both datasets in both axes. } 
\label{f:saltclass}
\end{center}
\end{figure}

\section{Classification}

\subsection{Method}
Classification using the D13 system gives proxies for effective temperature $\teff$ (spectral type), surface gravity $g$ (luminosity class), and helium / hydrogen ratio $y$ (helium class). 
The criteria for spectral type and helium class are based on relative line strengths and depths assuming a spectral resolution $R\approx 2\,000$.   
Both HRS and RSS spectra are therefore degraded to this resolution for classification. 

Spectral type, luminosity class and helium class are evaluated numerically from the digital spectra. 
For helium class (He) formulae based on fractional line depths $d$ are given by D13:
\[{\rm He} < 20: \\
{\rm He} = 20 * (d_{4471} + d_{4541}) / (d_{\gamma} - 0.83 d_{4541})\]
\[{\rm He} \geq 20: \\
{\rm He} = 40 - 20 * (d_{\gamma} - 0.83 d_{4541}) / (d_{4471} + d_{4541})\]

Spectral types (Sp) for helium-rich classes are based on He{\sc i}/He{\sc ii} line ratios as follows:
\[ d_{4686} \geq d_{4471}: {\rm Sp} =  0.1 + 0.8 d_{4471}/d_{4686}, \]
\[ d_{4686} < d_{4471}: {\rm Sp} =  0.9 + 0.4 ( 1 - d_{4686}/d_{4471}). \]
Sp corresponds to a numerical scale on which spectral type O2 = 0.2, O5 = 0.5, B0 = 1.0, etc.
D13 derived spectral types for hydrogen-rich classes using the depths of  H$\beta$ ($d_{\beta}$) and H$\gamma$ ($d_{\gamma}$). 
For helium classes He\,$<12$ and spectral types later than O8, we define:
\[{\rm Sp} = ((3.8 d_{\gamma} - 0.5) + (4.8 d_{\beta} - 0.8)) / 2. \]

\begin{table}
    \caption{Line depth criteria for He-strong stars to be sub-classified C or N. }
    \label{t:CNcrit}
    \centering
    \begin{tabular}{ccl}
    \hline
    C &  C{\sc iv} 4658 $|$ 4442  & $d > 0.08$  \\
    C &  C{\sc iii} 4647 $|$ 4650  & $d > 0.10$  \\
    C &  C{\sc ii} 4267 $|$ 4619  & $d > 0.13$  \\
    N &  N{\sc iii} 4379 $|$ 4513 $|$ 4640 & $d > 0.09$  \\
    N &  N{\sc ii} 4530 $|$ 4447  & $d > 0.20 \,\,\&\,\, {\rm Sp} > 1.0$ \\
    \hline
    \end{tabular}
\end{table}

Luminosity classes are harder to quantify from simple line criteria. 
Line widths for  gravity-sensitive lines were calibrated against spectral type and helium class using the sample of spectra from D13, and used with  partial success. 

Criteria for identifying carbon- and nitrogen-strong spectra were established by measuring depths of carbon and nitrogen lines for stars identified as sdOC and sdBN by D13. Criteria valid for He-strong spectra (He$>25$) are summarized in Table~\ref{t:CNcrit} 

Errors are based on the signal-to-noise in each spectrum estimated from a region of continuum and then propagated formally through the line depth formulae. 

Figures comparing automatic classification of the Drilling sample with the D13 manual classifications are shown in Appendix A. 

\begin{figure}
\begin{center}
\includegraphics[width=1.0\linewidth]{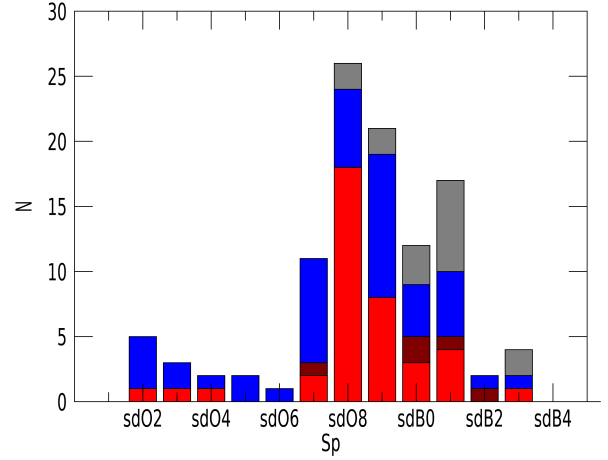}
\caption{The distribution of SALT helium-rich subdwarf stars by spectral type. Coloured segments represent stars in helium class He$\geq35$, subdivided into carbon-rich (C: blue), nitrogen-rich (N: maroon) and no qualification (red). Stars with He $<35$ are enumerated in grey. The total number of stars in the sample is 106. }
\label{f:spstats}
\end{center}
\end{figure}

\subsection{Results}

To achieve sufficient signal-to-noise for subsequent analysis, spectra for several stars were obtained over more than one observing block.
Originally, the reduced spectrum from each block was classified as a separate spectrum, which gave a good indication of the errors associated with noise.  
The final classification was obtained from a single spectrum constructed from all SALT/RSS observations combined.
Where an RSS spectrum was not available, the weighted average HRS spectrum was degraded by convolution with a Gaussian FWHM = 1.2\AA. 
This is essential because the relative depths of broad and sharp lines change with spectral resolution. 
Final classifications are shown in  Table \ref{t:saltclass}. 
The prefix `sd' implies a D13 classification as distinct from an MK classification; it does not of itself imply that the object is a subdwarf.  
Fig.\,\ref{f:saltclass} shows the spectral-type helium-class distribution obtained from automatic SALT classifications. 
The distribution of the sample by spectral type is shown in Fig.,\,\ref{f:spstats}. 
Both Figs.\,\ref{f:saltclass} and \ref{f:spstats} suggest a break in the distribution at 
spectral type sdO6. 
One may identify a minimum of three groups containing stars with: 
a) spectral type earlier than sdO6 (all have helium class $\geq 35$), 
b) helium class $<35$ (and spectral types between sdO9 and sdB1), and
c) those having helium class $\geq 35$ and spectral type between sdO6 and sdB3.
These groups include all but two or three outliers.
By further considering the luminosity class, it may be shown that other groups exist. 
However, errors currently associated with the luminosity criteria determine that 
other physical characteristics associated with the spectra should be examined first. 

Although there are 58 stars with He$\geq15$ in D13, only 7 are common to the SALT sample.
The majority of the D13 sample are northern hemisphere stars. 
The latter range from spectral type sdO2 to sdB1,  from luminosity class VI to VIII, and from helium class He18 to He40. 
In six, the  differences are less than one subclass in spectral type, luminosity class and helium class. 
The seventh is sdO2 in D13 and sdO4 here.

\begin{table*}
    \caption{Fundamental data and classifications for helium-rich hot subdwarfs observed with SALT. }
    \label{t:saltclass}
    \setlength{\tabcolsep}{2pt}
    \centering
    \begin{tabular}{llc lP{28mm} P{14mm}P{10mm} cc c l}
    \hline
\multicolumn{3}{l}{FUNDAMENTAL DATA} & 
\multicolumn{2}{l}{NAMES} & 
\multicolumn{2}{c}{LORES} & 
\multicolumn{2}{c}{SALT} 
& CLASS & Notes \\
$\alpha_{2000}$ & $\delta_{2000}$ & $m_G$  & Adopted & Other & Class & Ref  &  &  &  & \\
\hline
00:10:07 & $-$26:12:56 & 12.8 & Ton S 144 & PHL, SB, FB, MCT, BPS, EC & HesdB sdO6He4 & EC5 lam00 & RSS & HRS & sdO9.5VII:He37 & \\
00:18:53 & $-$31:56:02 & 14.4 & Ton S 148 & PHL, HE, MCT, GLX, EC & HesdB sdO7He3 & EC5 lam00 & RSS & HRS & sdBC0.2VI:He37 & \\ 
00:49:05 & $-$54:24:39 & 16.2 & EC 00468$-$5440 &  & HesdB & EC5 & RSS &  & sdBC0VII:He24 & \\ 
01:16:53 & $-$22:12:09 & 14.8 & BPS 22946$-$0005 & MCT & B, pAGB & BPS & RSS &  & sdB2.5II:He24 & \\ 
01:43:08 & $-$38:33:16 & 13.0 & SB 705 & GLX, EC & HesdO & kil89 & RSS & HRS & sdOC7.5VII:He40 & \\ % ucles 
01:47:17 & $-$51:33:39 & 13.5 & LB  3229 & JL, GLX & HesdO & kil89 & RSS & HRS & sdO9.5VII:He39 & \\ % ucles
02:10:54 & $+$01:47:47 & 13.7 & Feige 19 & PB, PG, GLX & HesdO & moe90 & RSS & HRS & sdO9VII:He37 & D13 \\ % PG 0208+016 
02:33:26 & $-$59:12:31 & 15.1 & LB 1630 & EC & HesdO & EC5 & RSS &  & sdOC5VI:He38 & \\  
02:43:23 & $+$04:50:36 & 14.1 & PG 0240+046 & GLX & sdOB & PG & RSS & HRS & sdBC0.5VII:He25 & D13 \\ 
02:51:21 & $-$72:34:33 & 14.6 & LB 3289 & EC, GLX & HesdO & EC3 & RSS & HRS & sdB0.2VII:He30 & \\ 
02:52:51 & $-$69:22:34 & 16.1 & EC 02523$-$6934 &  & HesdO & EC4 & RSS &  & sdO9VII:He39 & \\ 
03:06:08 & $-$14:31:52 & 15.6 & PHL 1466 & PB, EC & HesdO & EC5 & RSS &  & sdOC4V:He40 & noisy\\
03:50:38 & $-$69:20:57 & 14.7 & EC 03505$-$6929 &  & HesdO & EC4 & RSS &  & sdO9VII:He40 & \\ 
04:03:05 & $-$40:09:41 & 14.4 & EC 04013$-$4017 &  & HesdB & EC5 & RSS & HRS & sdBC1VII:He32 & \\ 
04:11:10 & $-$00:48:48 & 14.1 & \multicolumn{2}{l}{GLX J041110.1$-$004848}  & HesdO & {\O}G & RSS & HRS & sdO8VII:He40 & \\ 
04:13:19 & $-$13:41:03 & 12.5 & EC 04110$-$1348 &  & HesdO & EC3 & RSS &  & sdOC7.5VII:He39 & \\ 
04:15:30 & $-$54:21:59 & 14.9 & HE 0414$-$5429 & EC, GLX & HesdO & {\O}G & RSS &  & sdO8VII:He39 & \\ % uves, str07
04:20:35 & $+$01:20:41 & 12.3 & \multicolumn{2}{l}{GLX J042034.8+012041}  & HesdO & ven11 & RSS & HRS & sdOC8.5VII:He40 & \\
04:22:37 & $-$54:08:50 & 14.0 & LB 1721 & EC, GLX & HesdO & EC4 & RSS &  & sdOC9VII:He38 & \\ % HE 0421$-$5415
04:29:11 & $-$29:02:48 & 14.1 & EC 04271$-$2909 & BPS, GLX & HesdO & BPS & RSS &  & sdO8.5VI:He39 & \\
04:29:33 & $-$47:31:44 & 15.8 & EC 04281$-$4738 & GLX & sdB? & EC4 & RSS &  & sdOC6.5VII:He39 & \\ 
04:36:15 & $-$53:43:34 & 12.5 & LB 1741 & EC & HesdO & kil92 & RSS &  & sdO9VII:He39 & \\ 
04:37:34 & $-$61:57:43 & 14.6 & BPS 29520$-$0048 & EC, GLX & HesdO & rod07 & RSS &  & sdOC9VII:He39 & \\ 
04:42:26 & $-$32:06:01 & 14.8 & EC 04405$-$3211 & GLX & HesdO & EC4 & RSS & & sdO7.5VII:He39 & \\
04:53:32 & $-$37:01:43 & 15.7 & EC 04517$-$3706 &  & HesdB & EC3 & RSS &  & sdB0.5VI:He40 & \\
%% 05:06:56 & $-$25:12:46 & 13.8 & EC 05048$-$2516 &  & Hdef? & EC2 & RSS &  &  & \\ 
05:13:48 & $-$19:44:18 & 15.1 &  \multicolumn{2}{l}{GLX J051348.2$-$194417} & HesdOB & {\O}G & RSS &  & sdO7.5VII:He39 & \\ 
05:17:57 & $-$30:47:50 & 13.3 & Ton S 415 & EC, GLX & HesdO & EC3 & RSS & HRS & sdO8VII:He30 & \\
05:26:12 & $-$28:58:25 & 15.6 &  EC 05242-2900 & GLX  & HesdB  & EC3  & RSS &   & sdOC7VII:He39 & \\
05:58:05 & $-$29:27:09 & 15.2 & \multicolumn{2}{l}{GLX J055804.5$-$292708} & HesdOB & {\O}G & RSS &  & sdOC7VII:He39 & \\ 
06:00:01 & $-$59:01:03 & 16.1 & EC 05593$-$5901 &  & HesdB & EC3 & RSS &  & sdBN0.5VII:He39 & \\ 
06:12:37 & $-$27:12:55 & 13.4 & \multicolumn{2}{l}{GLX J061237.5$-$271254} & HesdOB & {\O}G & RSS & HRS & sdBN0.2VI:He40 & \\ 
07:07:39 & $-$62:22:41 & 14.5 & \multicolumn{2}{l}{GLX J070738.9$-$622241} & HesdOB & {\O}G & RSS & HRS & sdOC6.5VII:He40 & \\ 
07:15:50 & $-$54:07:57 & 14.4 & \multicolumn{2}{l}{GLX J071549.6$-$540755} & HesdO & {\O}G & RSS & HRS & sdO3VII:He40 & \\ 
07:58:08 & $-$04:32:05 & 13.1 & \multicolumn{2}{l}{GLX J075807.5$-$043203} & HesdO & nem12 & RSS & HRS & sdO9.5VII:He33 & \\ 
08:35:24 & $-$01:55:53 & 11.4 & [CW83] 0832$-$01 &  & sdOp & CW83 & RSS & & sdO8VII:He40 & \\ 
08:45:29 & $-$12:14:10 & 14.0 & \multicolumn{2}{l}{GLX J084528.7$-$121410}   & HesdOB & {\O}G & RSS & HRS & sdOC9.5VI:He39 & \\ 
09:05:05 & $+$05:33:01 & 14.1 & PG 0902+057 & GLX & sdOD & PG & RSS & HRS & sdB0VII:He39 & D13 \\ 
09:07:08 & $-$03:06:14 & 11.9 & [CW83] 0904$-$02 &  & sdOp(He) & ber80 & RSS &  & sdO7.5VI:He39 & \\ 
09:18:56 & $-$57:04:25 & 12.9 & LSS 1274 &  & HesdO & {\O}G & RSS &  & sdOC8VI:He39 & \\
09:58:11 & $-$16:05:52 & 14.3 & EC 09557$-$1551 & BPS, GLX & HesdO & EC2 & RSS &  & sdO7VII:He40 & \\
10:00:43 & $-$12:05:59 & 14.0 & PG 0958$-$119 & HE, EC, GLX & HesdO & EC2 & RSS &  & sdO8VI:He39 & \\
10:49:55 & $-$27:19:09 & 13.4 & EC 10475$-$2703 & GLX & HesdO & EC2 &  & HRS & sdOC3VII:He39 & \\
10:50:18 & $-$27:30:37 & 13.9 & EC 10479$-$2714 & GLX & HesdO & EC2 & RSS &  & sdO8.5VII:He40 & \\ 
11:26:11 & $-$20:01:39 & 14.4 & EC 11236$-$1945 & GLX & HesdO & EC2 & RSS &  & sdOC2VII:He40 & \\
11:30:04 & $+$01:37:37 & 13.8 & PG 1127+019 & GLX & sdOD & PG & RSS & HRS & sdOC9.5VII:He39 &  D13\\
12:22:59 & $-$05:53:05 & 14.7 & PG 1220$-$056 & GLX & sdOC & PG & RSS &  & sdO4VII:He39 & D13\\
12:33:23 & $-$06:25:18 & 13.1 & PG 1230+067 & GLX &  & PG & RSS &  & sdON9.5VII:He39 & D13 \\
12:37:35 & $-$28:41:01 & 14.8 & EC 12349$-$2824 & GLX & HesdO & EC2 & RSS &  & sdO8VII:He40 & \\
12:44:42 & $-$27:48:58 & 14.7 & EC 12420$-$2732 & GLX & HesdO & EC2 & RSS &  & sdOC4VII:He40 & \\
13:20:44 & $+$05:59:01 & 14.7 & PG 1318+062 &  & sdOC & PG & RSS & HRS & sdOC9VI:He39 & \\
13:31:46 & $-$19:48:26 & 14.4 & EC 13290$-$1933 & GLX & HesdB & EC2 & RSS &  & sdOC9.5VII:He39 & sen15\\
14:25:50 & $-$04:32:33 & 14.0 & \multicolumn{2}{l}{GLX J142549.8$-$043231}  & HesdOB & {\O}G &  & HRS & sdO9VII:He39 & \\
14:57:57 & $-$07:05:05 & 16.4 & PG 1455$-$069 &  & sdOB & PG & RSS &  & sdOC8.5VII:He40 & \\
15:23:32 & $-$18:17:26 & 13.9 & \multicolumn{2}{l}{GLX J152332.2$-$181726}  & HesdOB & {\O}G & RSS & HRS & sdCO9VII:He39 & \\
15:30:56 & $+$02:42:23 & 15.4 & PG 1528+029 & GLX & sdOC & PG & RSS &  & sdO8VII:He40 & \\
15:37:40 & $-$17:02:15 & 15.1 & EC 15348$-$1652 & GLX & HesdO & EC2 & RSS &  & sdO8VII:He39 & \\
15:40:33 & $-$04:48:12 & 15.0 & PG 1537$-$046 & BPS, GLX & HesdO & PG & RSS &  & sdOC2VII:He40 & D13\\
16:28:36 & $-$03:32:38 & 15.5 & PG 1625$-$034 & BPS, GLX  & HesdO & BPS & RSS &  & sdO8VII:He39 & noisy \\
16:54:38 & $+$03:18:47 & 15.1 & \multicolumn{2}{l}{GLX J165438.5+031847}   & HesdO & {\O}G & RSS & & sdOC3VII:He40 & \\
17:05:06 & $-$71:56:09 & 13.8 & \multicolumn{2}{l}{GLX J170506.0$-$715609} & HesdO & {\O}G & RSS & HRS & sdO7.5VII:He39 & \\
18:32:32 & $-$47:44:38 & 13.5 & \multicolumn{2}{l}{GLX J183231.7$-$474435} & HesdOB & {\O}G & & HRS & sdOC9VII:He38 & \\
18:37:17 & $-$31:25:16 & 13.9 & \multicolumn{2}{l}{GLX J183716.7$-$312514} & HesdOB &  {\O}G & & HRS & sdO7.5VII:He39 & \\

    \hline
    \end{tabular}
\end{table*}

\begin{table*}
    \contcaption{}
    \setlength{\tabcolsep}{2pt}
    \centering
    \begin{tabular}{llc lP{28mm} P{12mm}P{8mm} cc c l}
    \hline
    \multicolumn{3}{l}{FUNDAMENTAL DATA} & 
    \multicolumn{2}{l}{NAMES} & 
    \multicolumn{2}{c}{LORES} & 
    \multicolumn{2}{c}{SALT} 
    & CLASS & Notes \\
    $\alpha_{2000}$ & $\delta_{2000}$ & $m_G$  & Adopted & Other & Class & Ref  &  &  &  & \\
    \hline
    18:38:46 & $-$54:09:34 & 13.6 & \multicolumn{2}{l}{GLX J183845.6$-$540935} & HesdOB & {\O}G & RSS & HRS & sdB0VII:He39 & \\
18:46:00 & $-$41:38:28 & 14.6 & \multicolumn{2}{l}{GLX J184559.8$-$413827} & HesdB & ven11 & RSS & HRS & sdBN2V:He38 & jef17\\
19:05:56 & $-$44:38:40 & 13.6 & \multicolumn{2}{l}{GLX J190555.7$-$443838} & HesdOB & {\O}G &  & HRS & sdO8.59VI:He39 & \\
19:10:50 & $-$44:17:14 & 12.9 & \multicolumn{2}{l}{GLX J191049.5$-$441713} & HesdOB & {\O}G & RSS & HRS & sdBC0.2VI:He39 & \\
19:11:09 & $-$14:06:53 & 11.9 & \multicolumn{2}{l}{GLX J191109.3$-$140654} & HesdO & ven11 & & HRS & sdOC6.5VII:He39 & \\
19:15:04 & $-$42:35:04 & 14.0 & \multicolumn{2}{l}{GLX J191504.3$-$423502} & HesdOB & {\O}G & RSS & HRS & sdO8.5VII:He40 & \\
%% 19:18:50 & $-$31:04:41 & 13.7 & \multicolumn{2}{l}{GLX J191849.6$-$310441} & HesdOB & {\O}G &  & HRS & B:He0 & \\   %% sdB 
%% 19:30:46 & $-$30:49:59 & 14.4 & \multicolumn{2}{l}{GLX J193046.0$-$305000} & HesdO & {\O}G & RSS & HRS & G high PM: V=11 & csj\\  :: wrong star
19:33:24 & $-$23:45:53 & 14.8 & \multicolumn{2}{l}{GLX J193323.6$-$234553} & HesdOB & {\O}G & RSS &  & sdBC0.5VI:He38 & \\
19:37:40 & $-$43:03:56 & 13.4 & \multicolumn{2}{l}{GLX J193740.3$-$430356} & HesdB & {\O}G & RSS & HRS & sdB2.5V:He21 & \\
19:41:04 & $-$52:46:57 & 15.7 & BPS 22896$-$0128 & EC & HesdO & BPS & RSS &  & sdOC7VII:He39 & \\
19:56:31 & $-$44:22:19 & 11.8 & EC 19529$-$4430 &  & B  & EC3 & RSS & HRS & sdB3IV:He35 & metal poor\\ 
20:13:19 & $-$12:01:18 & 13.8 & \multicolumn{2}{l}{GLX J201318.8$-$120119}  & HesdO & {\O}G & RSS & HRS & sdOC2VII:He37 & \\
20:14:23 & $-$37:15:42 & 13.4 & EC 20111$-$3724 & GLX & HesdO & EC3 & RSS & & sdO9VII:He33 & \\
20:16:09 & $-$68:53:33 & 15.9 & EC 20111$-$6902 & GLX & HesdB & EC3 & RSS & & sdBC1.5VII:He38 & \\
20:20:26 & $-$19:01:50 & 14.8 & \multicolumn{2}{l}{GLX J202026.0$-$190150} & HesdOB & {\O}G & RSS &  & sdBN0VI:He35 & \\
20:21:39 & $-$34:25:46 & 14.4 & EC 20184$-$3435 & GLX & HesdO & EC3 & RSS &     & sdO9.5VI:He28 & \\
20:22:22 & $-$49:29:40 & 13.4 & EC 20187$-$4939 & GLX & HesdB & EC3 & RSS & HRS & sdB0.2VII:He36 & \\
20:25:06 & $-$08:04:18 & 13.9 & \multicolumn{2}{l}{GLX J202506.0$-$080419} & HesdO & {\O}G & RSS & HRS & sdOC2VII:He39 & \\
20:26:30 & $-$62:40:07 & 14.1 & EC 20221$-$6249 & GLX & HesdO & EC3 & RSS & HRS & sdOC9.5VII:He39 & \\
20:27:37$^1$ & $-$56:53:56$^1$ & 14.8$^1$ & EC 20236$-$5703 &  & H.def & EC3 & RSS & HRS & sdBC2.5IV:He35 & \\
20:30:20 & $-$59:50:39 & 14.0 & BPS 22940$-$0009 & EC, GLX & HesdB & BPS & RSS & HRS & sdBN1VI:He37 & \\
20:34:21 & $-$51:17:16 & 14.3 & EC 20306$-$5127 & GLX & HesdO & EC3 & RSS & HRS & sdOC6.5VII:He39 & \\
20:47:48 & $-$14:50:27 & 13.9 & EC 20450$-$1501 & GLX & HesdB & EC3 & RSS & HRS & sdO8.5VII:He38 & \\
20:49:54 & $-$69:36:31 & 14.7 & EC 20450$-$6947 &     & HesdO & EC3 & RSS &     & sdO7VII:He40 & \\
20:51:54 & $-$55:07:34 & 15.6 & EC 20481$-$5518 & GLX & HesdO & EC3 & RSS &  & sdO7.5VII:He40 & \\
20:57:38 & $-$14:25:44 & 13.0 & LS IV$-$14 116 & GLX, EC & HesdO & vit91 & RSS &  & sdB1VII:He18 & D13, nas11\\
21:01:30 & $-$56:29:43 & 16.2 & EC 20577$-$5641 & GLX & HesdB/O & EC3 & RSS &  &  & noisy \\
21:04:18 & $-$27:11:43 & 15.2 & Ton S 14 & EC & HesdO & EC3 & RSS &  & sdOC9VII:He40 & \\
21:11:11 & $-$48:02:57 & 15.3 & EC 21077$-$4815 &  & HesdO & EC4 & RSS &  & sdOC7.5VII:He39 & \\
21:11:21 & $-$23:48:14 & 14.4 & BPS 30319$-$0062 & GLX, EC & HesdOB & BPS & RSS & HRS & sdB0.5VII:He20 & \\
21:17:09 & $-$70:01:04 & 14.1 & EC 21125$-$7013 & GLX & HesdO & EC3 & RSS &  & sdOC6.5VII:He40 & \\
21:33:58 & $-$48:58:03 & 15.1 & EC 21306$-$4911 & GLX & HesdB & EC4 & RSS &  & sdBC1VII:He40 & \\
21:44:38 & $-$36:31:47 & 15.3 & EC 21416$-$3645 &  & HesdO & EC4 & RSS &  & sdO8.5VII:He34 & \\
21:47:52 & $-$12:35:44 & 14.5 & PHL 149 & BPS, GLX, EC & HesdO & BPS & RSS &  & sdO7.5VII:He40 & \\
21:51:13 & $-$21:07:04 & 13.0 & PHL 178 & EC & HesdO & kil89 & RSS &  & sdO7.5VII:He40 & \\
22:01:02 & $+$08:30:48 & 13.1 & PG 2158+082 & GLX & HesdO & PG & RSS &  & sdO2VII:He40 & D13 \\
22:14:58 & $-$63:41:45 & 14.5 & BPS 22956$-$0090 & GLX & HesdB & BPS & RSS & HRS & sdO9VII:He40 & \\
22:16:04 & $-$17:19:47 & 14.6 & BPS 22892$-$0051 & GLX, EC & HesdO & BPS & RSS &  & sdOC7VII:He40 & \\
22:17:22 & $-$05:27:50 & 14.3 & PB 7124 & GLX & HesdOB & {\O}G & RSS &  & sdOC9VII:He40 & \\
22:19:02 & $-$41:23:32 & 13.9 & BPS 22875$-$0002 & GLX, EC & HesdO & BPS & RSS & HRS & sdOC9VII:He40 & \\
22:21:23 & $+$05:24:58 & 15.3 & PG 2218+052 & & HesdB & PG & RSS &   & sdB0.5VII:He21 & \\
22:36:50 & $-$68:22:20 & 16.1 & EC 22332$-$6837 & GLX & HesdO & EC4 & RSS &  & sdO7.5VII:He37 & \\
22:52:20 & $-$63:15:55 & 15.3 & BPS 22938$-$0044 & GLX, EC & HesdB & BPS & RSS &     & sdO7.5VII:He40 & \\
22:56:36 & $-$52:48:36 & 13.3 & EC 22536$-$5304 & GLX & sdB & EC5 & RSS & HRS & sdB0.2VI:He23 & \\
23:10:54 & $-$63:03:25 & 14.3 & BPS 22938$-$0073 & GLX & HesdO & BPS & RSS & HRS & sdO7.5VII:He39 & \\
23:29:10 & $-$10:06:06 & 13.3 & PHL 540 & GLX & sdO & kil88 & RSS &  & sdO7.5VII:He40 & \\
23:34:02 & $-$28:51:38 & 14.7 & Ton S 103 & FB, PHL, BPS, GLX & HesdB & BPS & RSS & HRS & sdBN0.2VII:He40 & \\
23:35:41 & $+$00:02:19 & 15.9 & PB 5462 & PG, BPS, GLX & HesdO & BPS & RSS & & & noisy \\
23:50:20 & $-$41:14:02 & 15.3 & HE 2347$-$4130 & GLX, EC & HesdO & str07 & RSS &  & sdO8VII:He39 & \\[2mm]
02:53:08 & $-$70:58:56 & 16.1 & EC 02527$-$7111 &  & HesdB & EC4 & RSS &  & DB & \\ 
04:32:14 & $-$16:45:09 & 15.4 & EC 04299$-$1651 & HE & HesdB & EC2 & RSS & HRS & DB+dM & vos07 \\ %% Hem should not have been included : DB+dM from ESO SPY survey 
19:31:57 & $-$58:22:45 & 16.6 & EC 19277$-$5829 &  & HesdB & EC2 & RSS &  &  DB &  \\    %% DB white dwarf
20:36:46 & $-$25:14:41 & 15.1 & EC 20337$-$2525 &  & HesdB & EC3 & RSS &  & DB & \\
22:23:58 & $-$25:10:44 & 16.4 & EC 22211$-$2525 & GLX  & HesdB & EC5 & RSS &  & DB & \\

    \hline
    \end{tabular}

    \centering
    \begin{tabular}{P{175mm}}
    \parbox{175mm}{Fundamental Data: $\alpha_{2000}$, $\delta_{2000}$, $m_G$: \citet{gaia18.dr2}, 1: EC3 ($m_V$).  }
    \parbox{175mm}{Selected catalogues: 
    BPS=\citet{beers92}, 
    [CW83]=\citet{carnochan83}, 
    ${\rm EC}={\rm EC}n:n=1,5=$\citet{Cat.EC1,Cat.EC2,Cat.EC3,Cat.EC4,Cat.EC5}, % EC is the superset containin all on ECn. 
    Feige=\citet{Cat.Feige}, 
    FB=\citet{greenstein74}, 
    GLX=\citet{Cat.Galex17}, 
    HE=\citet{Cat.HE}, 
    JL=\citet{jaidee69}, 
    KUV=\citet{Cat.KUV2}, 
    LB=\cite[][et seq.]{Cat.LB}, 
    LS\,IV=\citet{Cat.LS4}, 
    LSS=\citet{Cat.LSS}, 
    MCT=\citet{Cat.MCT}, 
    PB=\citet{Cat.PB2}, 
    PG=\citet{green86}, 
    PHL=\citet{Cat.PHL}, 
    SB=\citet{Cat.SB}, 
    Ton\,S=\citet{Cat.TonS}, 
    UVO=\citet{carnochan83} }
    \parbox{175mm}{LORES references: as above plus 
    ber80=\citet{berger80b},
    {\O}G=\citet{ostensen06,geier17}, 
    kil89=\citet{kilkenny89},
    kil92=\citet{kilkenny92},
    lam00=\citet{lamontagne00}, 
    moe90=\citet{moehler90a}, 
    rod07=\citet{rodriguez07},
    str07=\citet{stroeer07},
    ven11=\citet{vennes11},
    vit91=\citet{viton91} }
    \parbox{175mm}{Notes. a CLASS has also given by:
    D13=\citet{drilling13}, 
    sen15=\citet{sener15.thesis}, 
    jef17=\citet{jeffery17b}, 
    nas11=\citet{naslim11}
    vos07=\citet{voss07}; 
    metal poor = very weak metal lines;  
    noisy = too noisy to classify.  }\\
    \hline
\end{tabular}
\end{table*}

\begin{figure}
\begin{center}
\includegraphics[clip, width=0.98\linewidth]{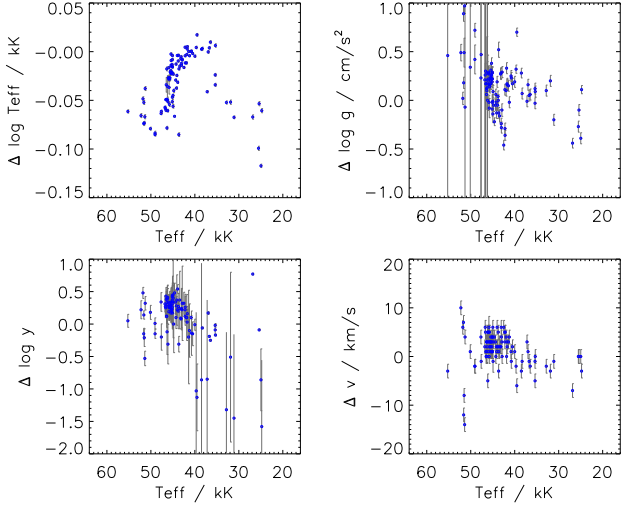}
\caption{Residuals in the sense {\it salt\_p00} -- {\it XTgrid} for $\Teff$, $\log g$, $\log y$ and $v$ as functions of $\Teff$.  
Offscale error bars are due to fit failures with  {\it salt\_p00} at high \Teff\ and  with  {\it XTgrid}  at low \Teff\ and $y$.  } 
\label{f:system}
\end{center}
\end{figure}

\begin{figure*}
\begin{center}
\includegraphics[clip, width=0.98\linewidth]{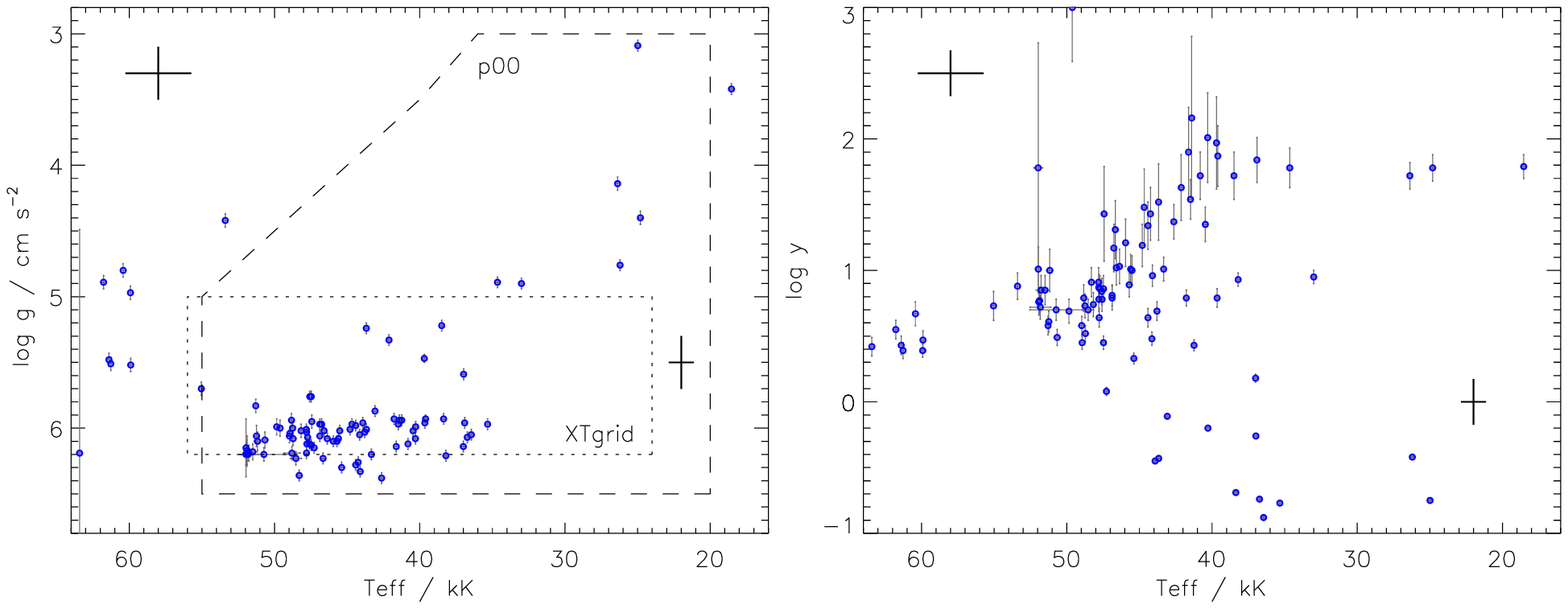}
\caption{$\Teff-\log g$ (left) and $\Teff-\log y$ (right) diagrams for helium-rich subdwarfs from
the SALT sample obtained from the {\it salt\_p00} model grid. 
The long-dashed line shows the boundary of the LTE solar-mix grid.
The short-dashed line shows the boundary of the non-LTE zero-metal grid \citep{nemeth12}.  
Gray error bars associated with each datum are formal errors as given in Table\,\ref{t:saltpars}. 
Stars with $\delta\log g>2$ are omitted.
Dark error bars upper-left and lower-right represent estimated mean observational errors (see text). 
 } 
\label{f:saltpars}
\end{center}
\end{figure*}

\section{Physical properties}

\subsection{Model atmospheres}
\label{s:models}
An alternative to classification is to match observed spectra within a grid of theoretical spectra (models) in order to estimate physical properties pertaining to the atmosphere of each star. 
Here, these comprise the star's effective temperature \Teff,  surface gravity $g$ and $y\equiv\nHe/\nH$, the ratio of hydrogen to helium atoms by number.  

Ideally, such models should also match the abundances of other significant chemical species, carbon, nitrogen and iron all having significant effects on the stellar spectrum.
To minimize computational costs, the present study considers only the following grids:

\noindent {\it salt\_p00:} models were computed with the Armagh LTE radiative transfer package {\sc lte-codes} \citep{jeffery01b,behara06} on a grid\footnote{http://193.63.77.2:2805/$\sim$SJeffery/m45.models/index.html}:  
\[
\begin{split} 
\teff/{\rm kK} & = [ 08(01)16(02)40(02.5)55 ], \\
\lgcs & =  [1.50(0.25)6.50],  \\
\nHe & = [0.01,0.05,0.10,0.3,0.5,0.7,0.8,0.9,0.99,1.0].
\end{split}
\]
Emergent spectra were computed on a self-adapting wavelength grid which optimally samples the local opacity structure and yields between 50\,000 and 200\,000 wavelength points in the range 3500 -- 6800 \AA . 
The abundance distribution for elements heavier than helium based was assumed to be solar; hence the label p00 which is shorthand for +0.0 dex. 
In the course of this investigation, it was realised that the abundance normalisation within {\sc sterne}, the model atmosphere component of {\sc lte\_codes}, assumed conservation of relative number fractions for metals when replacing hydrogen by helium. 
It is more natural that relative {\it mass fractions} should be conserved following, say, the fusion of 4 protons to a $^{4}{\rm He}$ nucleus. 
{\sc sterne} was consequently modified and the entire model grid, currently comprising $\sim 4000$ models for a single metallicity and microturbulent velocity ($v_{\rm turb}$), was recomputed.
For the latter, $v_{\rm turb}= 0\kmsec$ was assumed for both the calculation of line opacities in the model atmosphere (which affects the temperature stratification of the models) and for the formal solution, which affects relative line strengths and widths.
For comparison with the SALT RSS spectra, a subset of these models having
\[
\begin{split} 
\teff/{\rm kK} & = [20,24,28,32,36,40,42.5,45,47.5,50,55], \\
\lgcs & =  [3.0,3.5,4.0,4.5,5.0,5.5,6.0,6.5],  \\
\nHe & = [0.3,0.5,0.7,0.9,0.99,1.0]
\end{split}
\]
was sampled over the wavelength interval  $3600 - 5200$\AA\ on an interval 0.2\AA . 

\noindent {\it XTgrid:} models computed with the non-LTE radiative transfer codes {\sc tlusty} and {\sc synspec} \citep{hubeny94} for the analysis of hot subdwarfs observed in the Sloan Digital Sky Survey \citep{vennes11,nemeth12} were made available by  \citet{nemeth14}\footnote{http://stelweb.asu.cas.cz/$\sim$nemeth/work/sd\_grid/}. The original grid\footnote{ described by three parameters $[p_1$, $p_2$, $p_3]$ and  
three triplets $p_{\rm min} (\delta p) p_{\rm max}$ implying 
 $p \in p_{\rm min}, p_{\rm min}+\delta p, p_{\rm min}+2 \delta p, \ldots, p_{\rm max}$.} was defined as 
\[
\begin{split} 
&[\teff/{\rm kK},\lgcs, \log y ]  \\
& = [ 20(1)56 , 5.0(0.1)6.2, -5.0(0.3)+2.0 ],
\end{split}
\]
with no contribution from elements heavier than helium, and a microturbulent velocity $v_{\rm turb}= 0\kmsec$ throughout. This grid contains some 1390 model spectra each computed on a wavelength range 3130 - 7530 \AA\ with typically 38\, 000 wavelength points. To conserve memory a subset having
\[
\begin{split} 
\teff/{\rm kK} & = [24,28,32,36,40,44,48,52,56], \\
\lgcs & =  [5.0,5.3,5.6,5.9,6.2],  \\
\log y & = [ -1.0, -0.4, +0.2, +0.8, +1.4, +2.0 ] \\
\end{split}
\]
was sampled over the wavelength interval  $3600 - 5200$\AA\ on an interval 0.2\AA (as above). 

Ideally, the model atmosphere and emergent spectrum would be adapted and iterated to match the heavy-element distribution, and microturbulent velocity  measured on a first iteration, since these strongly influence the atmosphere structure at low hydrogen abundances. 
At present, a full fine analysis is only practical for limited numbers of stars. 
For this paper, only approximate values are required in order to identify overall trends and stars of  interest for further analysis. 
The LTE approximation is known to break down increasingly for stars with $\Teff \gtrsim 30\,000$\,K, but is less important than the contribution of metal opacities otherwise  \citep{anderson91,loebling20}.
These systematics are also discussed for restricted cases by \citet{napiwotzki97,latour11,pereira11,latour14b} and \citet{schindewolf18a}.  
Chemical stratification due to radiative levitation provides an additional vector of free parameters not considered in the current models \citep{behara08}. 

\subsection{Method}
\label{s:method}
{\sc lte\_codes} include the optimization code {\sc sfit} \citep{jeffery01b}. Here we use the Levenburg-Marquardt option to minimize the square residual between each observed normalized spectrum and the grid of models described above. 

Before optimization, the radial velocity $v_{\rm rad}$ of the observed spectrum relative to the laboratory rest-frame is established by cross-correlation with a representative theoretical spectrum; the wavelengths of the observed spectrum are then corrected by this amount so that the radial velocity is not a free parameter of the fit.

Inputs to the optimization include the normalized spectrum shifted to the local rest-frame velocity, 
a definition of regions of spectrum representative of continuum, masks to exclude non-stellar features (e.g. interestellar calcium H and K), masks to give additional weight to key lines, full-width half-maximum for the instrumental broadening profile (1.25\AA), and a threshold for excluding cosmic-ray features ($1.5\times$continuum). 
Regions of spectrum weighted 10 times other regions included H$\beta$, H$\gamma$, He{\sc ii} 4686 and 4540 \AA, He{\sc i} 4471, 4381, 4121, 4144 and 4169 \AA.     

The initial optimisation commences with starting values for
$\Teff=33.0$kK, $\log g/{\rm cm\,s^{-2}}=4.9$, $n_{\rm He}=0.9$ and with the projected rotation velocity fixed at $\vsini = 0$. 
Two renormalization steps are carried out using the high-pass procedure outlined by \citet{jeffery98a}, with filter widths set at 200\AA\ and 50\AA\ respectively.  
An optimization step is carried out after each renormalization; for the second of these $\vsini$ is free.  

Outputs from the optimization include \Teff, $\log g$, $n_{\rm He}$ (or equivalently, $\log y = n_{\rm He}/n_{\rm H}$), and  $\vsini$, as well as the renormalized  observed and best-fit model spectra.
$v_{\rm rad}$ determined prior to optimization is included in the overall set of outputs. 

Various sets of starting values for \Teff, $\log g$, and $n_{\rm He}$ were investigated.  
The adopted values were chosen because a converged solution was obtained in all cases. 
Choosing $\Teff$ too high or too low yielded consistent solutions in some fraction of cases,  but also led to divergence for a fraction of late or early-type  spectra, respectively.

Care is required in the choice of template used to determine $v_{\rm rad}$, especially for very hot stars. 
Because of the offset between hydrogen Balmer lines and \ion{He}{ii} lines, a template helium-hydrogen ratio which does not match the observed spectrum produces a systematic velocity shift and hence degrades the model optimisation. 
A second iteration was therefore introduced in which the first best-fit model was used as the velocity template, the second $v_{\rm rad}$ measurement being retained.   

From experience, the zero-points for both RSS and HRS wavelength calibrations must be treated with caution. Undocumented evidence for seasonal drifts might be associated with thermal drift in RSS, which sits on the tracker some 15\,m above the primary mirror assembly. 
%% No comparable drifts have been seen in the HRS calibration.
Early implementations of the HRS calibration pipeline suffered zero-point errors (Crawford, private communication). %% email on 2018 January 17.  
Consequently individual measurements of $v_{\rm rad}$ should be treated with caution.  Excessively high values may indicate an object of interest. 

Since $\vsini$ is an output from {\sc sfit}, and required to ensure the solution is self-consistent, it has a lower limit represented by the instrumental resolution.
For RSS spectra, $c/R \approx 83\kmsec$. 
For stars with $\Teff \gtrsim 50$\,kK, there are no sharp lines in the zero-metal models  with which to constrain the rotational broadening; hence $\vsini$ is degenerate with $\log g$.
For the sample of 71 RSS spectra with $\Teff < 50$\,kK,  $\langle \vsini \rangle = 83 \pm 22\kmsec$, all except EC\,20111--6902 have $\vsini < \langle \vsini \rangle + 2\sigma$. 

For the HRS spectra, the nominal $c/R \approx 7\kmsec$ was degraded by resampling so that the mean  $\langle \vsini \rangle = 12 \pm 2\kmsec$ obtained for four (excluding EC\,10475--2703) using the line-blanketed LTE models is satisfactory. 
Again the absence of sharp lines in the high-\Teff\ zero-metal models required that $\vsini = 12\kmsec$ be fixed for the final fits. 

\begin{table*}
    \caption{Atmospheric parameters for helium-rich hot subdwarfs observed and classified with SALT/RSS. 
    Stars observed only with SALT/HRS are marked '$\ast$',
    $\sigma$ represents standard deviation of fluxes about the mean in a continuum region $\lambda 4810 - 4845 $\AA\ ({i.e.} $\sigma=0.01 \Rightarrow {\rm S/N} = 100$.)
    All errors are formal; rather, mean errors $\delta\Teff/\Teff \approx \pm 0.028$, $\delta \log g \approx \pm 0.27$,  $\delta \log y \approx \pm 0.29$, and $\delta v_{\rm rad} \approx \pm 3.0 \kmsec$ should be adopted (\S\,4.3).
    For cases where {\sc sfit} finds $\nHe \geq 1$, we set $\nHe = 1$ and $\log y = 3$. 
    Tests suggest $\langle \delta v_{\rm rad}\rangle\approx\pm 3\kmsec$ (see tex). 
    $v_{\rm wid}$ is a nominal measure of the line broadening, dominated by the instrumental width $c/R \approx 83 \kmsec $ (RSS) (see text);  a smaller value $\sim 12 \kmsec$ is indicated where only an HRS spectrum is available (marked !).  } 
    \label{t:saltpars}
    \setlength{\tabcolsep}{5pt}
    \centering
    \begin{tabular}{llc r@{$\pm$}l r@{$\pm$}l r@{$\pm$}l r@{$\pm$}l r r c }
    \hline
Star            & Class        & $\sigma$ & \multicolumn{2}{c}{\Teff} & \multicolumn{2}{c}{$\log g$} & \multicolumn{2}{c}{$\log y$} & \multicolumn{2}{c}{\nHe} & {$v_{\rm rad}$} & {$v_{\rm wid}$} & grid  \\
  &  &  & \multicolumn{2}{c}{kK} & 
  \multicolumn{2}{c}{cm s$^{-2}$} & 
  \multicolumn{2}{c}{ } & 
  \multicolumn{2}{c}{ } & 
  {\kmsec} & {\kmsec} & \\
    \hline
    %# Star, Drilling_Class, sigma, Teff, TeffErr, Logg, LoggErr, Logy, LogyErr, NHe, NHeErr, RV, vsini, grid
Ton S 144       & sdO9.5VII:He37     & 0.010 & 43.79 &  0.07 &  6.03 &  0.04 &  0.69 &  0.07 &  0.83 &  0.03 &  --17 &   87 & xt \\
Ton S 148       & sdBC0.2VI:He37     & 0.012 & 38.49 &  0.09 &  5.22 &  0.04 &  1.72 &  0.18 &  0.98 &  0.01 &  159 &  109 & p00 \\
EC 00468--5440   & sdBC0VII:He25      & 0.020 & 38.36 &  0.10 &  5.93 &  0.04 & --0.69 &  0.01 &  0.17 &  0.02 &   57 &  117 & p00 \\
BPS 22946--0005  & sdB2.5II:He24      & 0.010 & 24.99 &  0.11 &  3.09 &  0.04 & --0.75 &  0.01 &  0.15 &  0.03 &  --64 &  102 & p00 \\
SB 705          & sdOC7.5VII:He40    & 0.013 & 48.72 &  0.20 &  6.08 &  0.05 &  0.52 &  0.06 &  0.77 &  0.03 &  --14 &   65 & xt \\
LB 3229         & sdO9.5VII:He39     & 0.012 & 44.10 &  0.08 & $\dagger$ 6.33 &  0.04 &  0.96 &  0.08 &  0.90 &  0.02 &   48 &   79 & xt \\
Feige 19        & sdO9VII:He37       & 0.017 & 44.14 &  0.08 &  6.05 &  0.04 &  0.48 &  0.05 &  0.75 &  0.03 &   19 &   61 & xt \\
LB 1630         & sdOC5VI:He38       & 0.020 & 53.39 &  0.15 & $\dagger$ 4.42 &  0.05 &  0.88 &  0.10 &  0.88 &  0.03 &  233 &  105 & xt \\
PG 0240+046     & sdBC0.5VII:He25    & 0.011 & 37.00 &  0.09 &  6.14 &  0.04 &  0.18 &  0.03 &  0.60 &  0.02 &   52 &  105 & p00 \\
LB 3289         & sdBN0.2VII:He29    & 0.013 & 39.65 &  0.10 &  5.96 &  0.04 &  0.79 &  0.07 &  0.86 &  0.02 &   98 &   92 & p00 \\
EC 02523--6934   & sdO9VII:He39       & 0.013 & 46.37 &  0.10 &  6.08 &  0.04 &  1.03 &  0.13 &  0.91 &  0.02 &  --44 &   80 & xt \\
EC 03505--6929   & sdO9VII:He40       & 0.018 & 45.51 &  0.10 &  6.02 &  0.04 &  1.00 &  0.11 &  0.91 &  0.02 &   --2 &   71 & xt \\
EC 04013--4017   & sdBC1VII:He32      & 0.012 & 38.20 &  0.10 & $\dagger$ 6.21 &  0.04 &  0.93 &  0.05 &  0.89 &  0.01 &    9 &   94 & p00 \\
GLX J04111--0048 & sdO8VII:He40       & 0.017 & 46.88 &  0.10 &  5.97 &  0.04 &  0.79 &  0.09 &  0.86 &  0.03 &   41 &   74 & xt \\
EC 04110--1348   & sdOC7.5VII:He39    & 0.013 & 48.94 &  0.18 &  6.04 &  0.05 &  0.45 &  0.05 &  0.74 &  0.03 &   36 &   70 & xt \\
HE 0414--5429    & sdO8VII:He39       & 0.016 & 47.79 &  0.10 &  6.03 &  0.04 &  0.78 &  0.09 &  0.86 &  0.03 &   15 &   78 & xt \\
GLX J04205+0120 & sdOC8.5VII:He39    & 0.009 & 47.80 &  0.10 &  6.19 &  0.04 &  0.87 &  0.10 &  0.88 &  0.03 &   49 &   83 & xt \\
LB 1721         & sdOC9VII:He38      & 0.024 & 45.60 &  0.10 &  6.08 &  0.04 &  1.01 &  0.11 &  0.91 &  0.02 &   41 &   86 & xt \\
EC 04271--2909   & sdO8.5VI:He39      & 0.011 & 47.47 &  0.10 &  6.13 &  0.04 &  0.86 &  0.10 &  0.88 &  0.03 &   15 &   60 & xt \\
EC 04281--4738   & sdOC6.5VII:He39    & 0.027 & 52.97 &  0.13 &  6.07 &  4.32 &  0.68 &  0.09 &  0.83 &  0.04 &   64 &   78 & xt \\
LB 1741         & sdO9VII:He39       & 0.009 & 44.41 &  0.10 &  5.98 &  0.04 &  0.64 &  0.07 &  0.81 &  0.03 &   18 &   75 & xt \\
BPS 29520--0048  & sdOC9VII:He39      & 0.015 & 45.95 &  0.10 &  6.10 &  0.04 &  1.21 &  0.18 &  0.94 &  0.02 &   71 &   76 & xt \\
EC 04405--3211   & sdO7.5VII:He39     & 0.013 & 51.50 &  0.30 &  6.18 &  0.06 &  0.85 &  0.11 &  0.88 &  0.03 &    3 &   73 & xt \\
EC 04517--3706   & sdB0.5VI:He40      & 0.013 & 40.30 &  0.08 &  6.08 &  0.04 &  2.01 &  0.34 &  0.99 &  0.01 &   14 &  118 & p00 \\
GLX J05138--1944 & sdO7.5VII:He39     & 0.036 & 48.53 &  0.38 & $\dagger$ 6.23 &  0.05 &  0.70 &  0.08 &  0.83 &  0.03 &   --7 &   48 & xt \\
Ton S 415       & sdO8VII:He30       & 0.007 & 43.92 &  0.10 &  5.96 &  0.04 & --0.45 &  0.01 &  0.26 &  0.02 &  233 &   85 & xt \\
EC 05242--2900   & sdO8VII:He28       & 0.008 & 43.67 &  0.10 &  6.01 &  0.04 & --0.43 &  0.01 &  0.27 &  0.02 &  148 &  105 & xt \\
GLX J05580--2927 & sdOC7VII:He39      & 0.014 & 53.60 &  0.13 &  6.05 &  4.34 &  0.58 &  0.08 &  0.79 &  0.04 &   25 &   85 & xt \\
EC 05593--5901   & sdB0.5VII:He39     & 0.019 & 40.81 &  0.08 &  6.12 &  0.04 &  1.72 &  0.18 &  0.98 &  0.01 &   22 &  101 & p00 \\
GLX J06126--2712 & sdB0.2VI:He40      & 0.010 & 41.62 &  0.06 &  6.14 &  0.04 &  1.90 &  0.34 &  0.99 &  0.01 &   85 &   76 & xt \\
GLX J07076--6222 & sdOC6.5VII:He40    & 0.009 & 53.62 &  0.13 &  6.02 &  4.53 &  0.58 &  0.08 &  0.79 &  0.04 &    3 &   85 & xt \\
GLX J07158--5407 & sdO3VII:He40       & 0.018 & $\dagger$ 61.02 &  0.15 &  6.02 &  4.58 &  0.30 &  0.05 &  0.67 &  0.04 &   37 &  133 & xt \\
GLX J07581--0432 & sdO9.5VII:He33     & 0.009 & 41.24 &  0.07 &  5.94 &  0.04 &  0.43 &  0.04 &  0.73 &  0.02 &  107 &   78 & xt \\
GLX J08454--1214 & sdOC9.5VI:He39     & 0.008 & 43.68 &  0.08 &  5.24 &  0.04 &  1.52 &  0.29 &  0.97 &  0.02 &  113 &   77 & xt \\
PG 0902+057     & sdB0VII:He39       & 0.010 & 42.63 &  0.07 & $\dagger$ 6.38 &  0.04 &  1.37 &  0.13 &  0.96 &  0.01 &   55 &   60 & xt \\
UVO 0904--02     & sdO7.5VI:He39      & 0.010 & 51.88 &  0.21 &  6.17 &  0.11 &  0.77 &  0.10 &  0.85 &  0.04 &   12 &   80 & xt \\
LSS 1274        & sdO8VI:He39        & 0.011 & 46.88 &  0.10 &  6.06 &  0.04 &  0.81 &  0.08 &  0.86 &  0.03 &   22 &   73 & xt \\
EC 09557--1551   & sdO7VII:He40       & 0.014 & 51.92 &  0.22 &  6.19 &  0.10 &  0.76 &  0.10 &  0.85 &  0.04 &   75 &   94 & xt \\
PG 0958--119     & sdO8VII:He39       & 0.013 & 47.57 &  0.11 &  5.76 &  0.04 &  0.78 &  0.09 &  0.86 &  0.03 &   40 &   61 & xt \\
EC 10475--2703   & sdOC3VI:He39     * & 0.010 & 59.90 &  0.17 &  5.52 &  0.05 &  0.47 &  0.07 &  0.75 &  0.04 &   --8 & ! 12 & xt \\
EC 10479--2714   & sdO8.5VII:He40     & 0.014 & 47.59 &  0.10 &  6.12 &  0.04 &  0.84 &  0.10 &  0.87 &  0.03 &   89 &   73 & xt \\
EC 11236--1945   & sdOC2VII:He40      & 0.012 & $\dagger$ 61.77 &  0.18 &  4.89 &  0.05 &  0.55 &  0.07 &  0.78 &  0.04 &   32 &  136 & xt \\
PG 1127+019     & sdOC9.5VII:He39    & 0.010 & 44.41 &  0.10 & $\dagger$ 6.28 &  0.04 &  1.34 &  0.18 &  0.96 &  0.02 &   35 &   71 & xt \\

    \hline
    \multicolumn{13}{l}{$\ast$: HRS spectrum, $\dagger$: extrapolated}\\
    \end{tabular}
\end{table*}

\begin{table*}
    \contcaption{}
    \label{t:saltpars2}
    \setlength{\tabcolsep}{5pt}
    \centering
    \begin{tabular}{llc r@{$\pm$}l r@{$\pm$}l r@{$\pm$}l r@{$\pm$}l r r c  }
    \hline
Star            & Class        & $\sigma$ & \multicolumn{2}{c}{\Teff} & \multicolumn{2}{c}{$\log g$} & \multicolumn{2}{c}{$\log y$} & \multicolumn{2}{c}{\nHe} & {$v_{\rm rad}$} & {$v_{\rm wid}$} & grid \\
  &  &  & \multicolumn{2}{c}{kK} & 
  \multicolumn{2}{c}{cm s$^{-2}$} & 
  \multicolumn{2}{c}{ } & 
  \multicolumn{2}{c}{ } & 
  {\kmsec} & {\kmsec} &  \\
    \hline
    %# Star, Drilling_Class, sigma, Teff, TeffErr, Logg, LoggErr, Logy, LogyErr, NHe, NHeErr, RV, vsini, grid
PG 1220--056     & sdO4VII:He39       & 0.021 & $\dagger$ 59.04 &  0.14 &  6.04 &  4.46 &  0.51 &  0.07 &  0.76 &  0.04 &   --9 &   90 & xt \\
PG 1230+067     & sdON9.5VII:He39    & 0.011 & 43.33 &  0.07 &  6.20 &  0.04 &  1.01 &  0.09 &  0.91 &  0.02 &  --18 &   57 & xt \\
EC 12349--2824   & sdO8VII:He40       & 0.014 & 47.72 &  0.10 &  6.07 &  0.04 &  0.86 &  0.10 &  0.88 &  0.03 &   --9 &   72 & xt \\
EC 12420--2732   & sdOC4VII:He40      & 0.013 & $\dagger$ 60.37 &  0.14 &  6.03 &  4.52 &  0.59 &  0.09 &  0.80 &  0.04 &  --28 &  110 & xt \\
PG 1318+062     & sdOC9VII:He39      & 0.023 & 46.76 &  0.10 &  5.97 &  0.04 &  1.17 &  0.18 &  0.94 &  0.03 &   20 &   66 & xt \\
EC 13290--1933   & sdOC9.5VII:He39    & 0.012 & 44.24 &  0.08 &  $\dagger$ 6.26 &  0.04 &  1.43 &  0.20 &  0.96 &  0.02 &  --26 &   72 & xt \\
GLX J14258--0432 & sdON7VI:He39     * & 0.015 & 51.97 &  0.29 &  6.20 &  0.08 &  1.78 &  0.95 &  0.98 &  0.04 & --137 & ! 12 & xt \\
PG 1455--069     & sdOC8.5VII:He40    & 0.033 & 48.30 &  0.19 &  $\dagger$ 6.36 &  0.04 &  0.91 &  0.11 &  0.89 &  0.03 &   18 &   83 & xt \\
GLX J15235--1817 & sdOC9VII:He39      & 0.016 & 44.80 &  0.10 &  6.01 &  0.04 &  1.19 &  0.16 &  0.94 &  0.02 &   18 &   68 & xt \\
PG 1528+029     & sdO8VII:He40       & 0.013 & 48.83 &  0.14 &  5.94 &  0.05 &  0.79 &  0.10 &  0.86 &  0.03 &  --48 &   78 & xt \\
EC 15348--1652   & sdO8VII:He39       & 0.015 & 48.97 &  0.19 &  6.06 &  0.05 &  0.58 &  0.07 &  0.79 &  0.03 &   28 &   76 & xt \\
PG 1537--046     & sdOC2VII:He40      & 0.015 & $\dagger$ 61.27 &  0.16 &  5.51 &  0.05 &  0.39 &  0.06 &  0.71 &  0.04 & --109 &  118 & xt \\
PG 1625--034     & sdO8VII:He39       & 0.043 & 45.38 &  0.10 & $\dagger$ 6.30 &  0.04 &  0.33 &  0.04 &  0.68 &  0.03 &  --92 &   50 & xt \\
GLX J16546+0318 & sdOC3VII:He40      & 0.021 & $\dagger$ 61.30 &  0.15 &  6.07 &  4.24 &  0.68 &  0.10 &  0.83 &  0.04 &   36 &  106 & xt \\
GLX J17051--7156 & sdOC6VII:He40      & 0.009 & 54.93 &  0.13 &  6.14 &  3.77 &  0.56 &  0.08 &  0.79 &  0.04 &  --48 &   97 & xt \\
GLX J18325--4744 & sdOC9VII:He38    * & 0.010 & 44.67 &  0.11 &  5.97 &  0.04 &  1.48 &  0.29 &  0.97 &  0.02 & --104 & ! 12 & xt \\
GLX J18372--3125 & sdOC8VII:He38    * & 0.017 & 49.62 &  0.18 &  6.00 &  0.06 &  3.00 &  0.41 &  1.04 &  0.04 &  --37 & ! 12 & xt \\
GLX J18387--5409 & sdB0VII:He39       & 0.012 & 41.48 &  0.06 &  5.97 &  0.04 &  1.54 &  0.15 &  0.97 &  0.01 &   --9 &   78 & xt \\
GLX J18459--4138 & sdBN2V:He38        & 0.017 & 24.81 &  0.12 &  4.40 &  0.05 &  1.78 &  0.10 &  0.98 &  0.00 &  --69 &   98 & p00 \\
GLX J19059--4438 & sdOC8.5VI:He39   * & 0.013 & 47.43 &  0.11 &  5.95 &  0.05 &  1.43 &  0.36 &  0.96 &  0.03 &  --62 & ! 12 & xt \\
GLX J19108--4417 & sdBC0.2VI:He39     & 0.008 & 39.69 &  0.09 &  5.47 &  0.03 &  1.97 &  0.35 &  0.99 &  0.01 &    1 &   81 & p00 \\
GLX J19111--1406 & sdOC6.5VII:He39  * & 0.012 & $\dagger$ 55.04 &  0.16 &  5.70 &  0.05 &  0.73 &  0.11 &  0.84 &  0.04 & --240 & ! 12 & xt \\
GLX J19150--4235 & sdO8.5VII:He40     & 0.013 & 48.17 &  0.12 &  6.02 &  0.05 &  0.74 &  0.09 &  0.85 &  0.03 &  --15 &   71 & xt \\
GLX J19333--2345 & sdBC0.5VI:He38     & 0.016 & 39.60 &  0.10 &  5.93 &  0.03 &  1.87 &  0.23 &  0.99 &  0.01 &   84 &   86 & p00 \\
GLX J19376--4303 & sdB2.5V:He21       & 0.013 & 26.21 &  0.12 &  4.76 &  0.04 & --0.42 &  0.01 &  0.28 &  0.02 &  --50 &   89 & p00 \\
BPS 22896--0128  & sdOC7VII:He39      & 0.018 & 51.29 &  0.13 &  5.83 &  0.05 &  0.58 &  0.07 &  0.79 &  0.04 & --102 &   93 & xt \\
EC 19529--4430   & sdB3IV:He35        & 0.005 & $\dagger$ 18.54 &  0.09 &  3.42 &  0.04 &  1.79 &  0.09 &  0.98 &  0.00 &    5 &  113 & p00 \\
GLX J20133--1201 & sdOC2VII:He37      & 0.014 & $\dagger$ 60.42 &  0.18 &  4.80 &  0.05 &  0.67 &  0.09 &  0.82 &  0.03 &  --70 &  135 & xt \\
EC 20111--3724   & sdO9VII:He33       & 0.010 & 43.08 &  0.09 &  5.87 &  0.04 & --0.11 &  0.02 &  0.44 &  0.03 &   65 &   66 & xt \\
EC 20111--6902   & sdBC1.5VII:He38    & 0.019 & 34.10 &  0.11 &  5.68 &  0.04 &  1.99 &  0.22 &  0.99 &  0.01 & --81 &  153 & p00 \\
GLX J20204--1901 & sdBN0VI:He35       & 0.008 & 41.77 &  0.07 &  5.93 &  0.04 &  0.79 &  0.06 &  0.86 &  0.02 &   65 &   77 & xt \\
EC 20184--3435   & sdO9.5VI:He28      & 0.011 & 40.28 &  0.10 &  5.99 &  0.04 & --0.20 &  0.02 &  0.39 &  0.02 &  --11 &   72 & p00 \\
EC 20187--4939   & sdB0.2VII:He36     & 0.012 & 40.46 &  0.08 &  6.02 &  0.04 &  1.35 &  0.13 &  0.96 &  0.01 &  --69 &   94 & p00 \\
GLX J20251--0804 & sdOC2VII:He39      & 0.014 & $\dagger$ 59.92 &  0.17 &  4.97 &  0.05 &  0.39 &  0.05 &  0.71 &  0.04 &   64 &  118 & xt \\
EC 20221--6249   & sdOC9.5VII:He39    & 0.016 & 42.12 &  0.08 &  5.33 &  0.04 &  1.63 &  0.25 &  0.98 &  0.01 &   72 &   79 & xt \\
EC 20236--5703   & sdBC2.5IV:He35     & 0.011 & 26.38 &  0.13 &  4.14 &  0.05 &  1.72 &  0.10 &  0.98 &  0.00 &  --74 &   89 & p00 \\
BPS 22940--0009  & sdBC1V:He38        & 0.010 & 34.65 &  0.11 &  4.89 &  0.04 &  1.78 &  0.15 &  0.98 &  0.01 &   28 &   93 & p00 \\
EC 20306--5127   & sdOC6.5VII:He39    & 0.017 & 53.66 &  0.13 &  5.97 &  4.73 &  0.55 &  0.07 &  0.78 &  0.04 &  --17 &   81 & xt \\
EC 20450--1501   & sdOC8VII:He38      & 0.040 & 47.47 &  0.09 &  5.76 &  0.04 &  0.45 &  0.05 &  0.74 &  0.03 & --100 &   13 & xt \\
EC 20450--6947   & sdO7VII:He40       & 0.009 & 51.24 &  0.19 &  6.06 &  0.08 &  0.61 &  0.08 &  0.80 &  0.03 &   69 &   76 & xt \\
EC 20481--5518   & sdO7.5VII:He39     & 0.011 & 51.96 &  0.19 &  6.15 &  0.22 &  1.01 &  0.17 &  0.91 &  0.03 &  --10 &   73 & xt \\
LS IV--14 116    & sdB1VII:He18       & 0.019 & 35.33 &  0.11 &  5.97 &  0.04 & --0.77 &  0.01 &  0.15 &  0.02 & --163 &  105 & p00 \\
Ton S 14        & sdOC9VII:He40      & 0.022 & 46.59 &  0.10 &  6.02 &  0.04 &  1.02 &  0.13 &  0.91 &  0.03 &  --35 &   59 & xt \\
EC 21077--4815   & sdOC7.5VII:He39    & 0.025 & 47.77 &  0.10 &  6.12 &  0.04 &  0.64 &  0.07 &  0.81 &  0.03 &   41 &   55 & xt \\
BPS 30319--0062  & sdB0.5VII:He20     & 0.020 & 36.73 &  0.08 &  6.07 &  0.04 & --0.74 &  0.01 &  0.15 &  0.02 &  --42 &   86 & p00 \\
EC 21125--7013   & sdOC6.5VII:He40    & 0.019 & 54.11 &  0.13 &  6.16 &  3.10 &  0.69 &  0.09 &  0.83 &  0.04 &   --8 &   92 & xt \\
EC 21306--4911   & sdBC1VII:He40      & 0.019 & 36.91 &  0.09 &  5.96 &  0.04 &  1.84 &  0.17 &  0.99 &  0.01 &   29 &  105 & p00 \\
EC 21416--3645   & sdO8.5VII:He34     & 0.019 & 51.17 &  0.21 &  6.10 &  0.07 &  1.00 &  0.16 &  0.91 &  0.03 &   43 &  130 & xt \\
PHL 149         & sdO7.5VII:He40     & 0.017 & 51.77 &  0.32 &  6.19 &  0.06 &  0.85 &  0.11 &  0.88 &  0.03 &   --5 &   69 & xt \\
PHL 178         & sdO7.5VII:He40     & 0.012 & 50.66 &  0.21 &  6.09 &  0.06 &  0.49 &  0.06 &  0.76 &  0.03 &   20 &   66 & xt \\
PG 2158+082     & sdO2VII:He40       & 0.012 & $\dagger$ 63.42 &  0.15 &  6.19 &  1.70 &  0.42 &  0.07 &  0.72 &  0.04 & --102 &  105 & xt \\
BPS 22956--0090  & sdO9VII:He40       & 0.012 & 45.70 &  0.10 &  6.10 &  0.04 &  0.89 &  0.09 &  0.89 &  0.02 &  --83 &   80 & xt \\
BPS 22892--0051  & sdOC7VII:He40      & 0.018 & 51.82 &  0.71 &  6.20 &  0.05 &  0.72 &  0.09 &  0.84 &  0.03 & --100 &   56 & xt \\
PB 7124         & sdOC9VII:He40      & 0.021 & 47.81 &  0.10 &  6.01 &  0.04 &  0.91 &  0.11 &  0.89 &  0.03 &   --7 &   75 & xt \\
BPS 22875--0002  & sdOC9VII:He40      & 0.013 & 46.66 &  0.10 & $\dagger$ 6.23 &  0.04 &  1.31 &  0.22 &  0.95 &  0.02 &  --47 &   63 & xt \\
PG 2218+051     & sdB0.5VII:He20     & 0.014 & 36.45 &  0.08 &  6.05 &  0.04 & --0.88 &  0.01 &  0.12 &  0.02 &   14 &   94 & p00 \\
EC 22332--6837   & sdO7.5VII:He37     & 0.030 & 47.27 &  0.11 &  6.15 &  0.04 &  0.08 &  0.03 &  0.54 &  0.03 & --102 &   58 & xt \\
BPS 22938--0044  & sdOC7.5VII:He40    & 0.012 & 50.73 &  1.80 &  6.20 &  0.05 &  0.70 &  0.08 &  0.83 &  0.03 &   16 &   69 & xt \\
EC 22536--5304   & sdB0.2VII:He23     & 0.012 & 36.98 &  0.09 &  5.59 &  0.04 & --0.26 &  0.02 &  0.36 &  0.02 &   21 &   97 & p00 \\
BPS 22938--0073  & sdO7.5VII:He39     & 0.016 & 48.75 &  0.16 &  6.00 &  0.05 &  0.73 &  0.09 &  0.84 &  0.03 &   47 &   83 & xt \\
PHL 540         & sdO7.5VII:He40     & 0.013 & 49.85 &  0.17 &  5.99 &  0.06 &  0.69 &  0.09 &  0.83 &  0.03 &    1 &   72 & xt \\
Ton S 103       & sdBN0.2VII:He40    & 0.013 & 41.40 &  0.06 &  5.94 &  0.04 &  2.16 &  0.62 &  0.99 &  0.01 &  --55 &   83 & xt \\
PB 5462         & sdOC5VII:He40      & 0.045 & $\dagger$ 61.40 &  0.17 &  5.48 &  0.05 &  0.43 &  0.07 &  0.73 &  0.04 &  --24 &  174 & xt \\
HE 2347--4130    & sdO8VII:He39       & 0.016 & 48.80 &  0.56 &  6.19 &  0.05 &  0.74 &  0.09 &  0.85 &  0.03 &   --8 &   83 & xt \\

    \hline
    \multicolumn{13}{l}{$\ast$: HRS spectrum, $\dagger$: extrapolated}\\
    \end{tabular}
\end{table*}

\subsection{Results}
\label{s:params}
Values for \Teff, $\log g$ and $\log y$ obtained for the full sample are shown in Table~\ref{t:saltpars}. 
Results lying outside respective grid boundaries are included for completeness. 
It is emphasized that these analyses have been carried out for the purpose of data exploration and discovery; their used in detailed investigations of individual stars may be ill advised. 

Large errors in $\log y$ occur for stars with very low hydrogen abundances.
The hydrogen abundance is difficult to measure precisely in hot helium-rich stars since the Balmer lines are completely dominated by the corresponding lines in the He{\sc ii} Pickering series and, since $y\equiv \nHe/\nH$, the increasing error in the smaller denominator dominates the error budget. 

Table~\ref{t:saltpars} also provides an estimate of the noise $\sigma$ in the spectrum used for the analysis, the radial velocity $v_{\rm rad}$ of said spectrum, and the parameter obtained as $v \sin i$ in the model atmosphere fit, but more precisely labelled as a line width $v_{\rm wid}$ in velocity units.   

Fig. E.1 in the Supplementary Material shows correlations between physical parameters and  spectral class indicators. 
Trends illustrate the systematics, and scatter provides an estimate of the random errors.

For $\Teff \lesssim 35$\,kK, (Sp $\gtrsim$ sdB1)  stars have surface gravities outside the {\it XTgrid} boundary; the {\it salt\_p00} grid results shown in Table\,\ref{t:saltpars} are to be preferred. 
For $42 \gtrsim \Teff/{\rm kK} \gtrsim 35$  (sdO9 $\lesssim$ Sp $\lesssim$ sdB1), both grids give comparable values for \Teff, with {\it salt\_p00} giving slightly higher $g$ (by 0 -- 0.3 dex) for  $\Teff \lesssim 42$\,kK and lower $g$ for  $\Teff \gtrsim 42$\,kK.
For $42 \gtrsim \Teff/{\rm kK} \gtrsim 35$ (sdO9.5 $\lesssim$ Sp $\lesssim$ sdB1),
{\it salt\_p00} gives $g$ higher than {\it XTgrid} by $\approx 0.1$ dex, 
but for  $\Teff/{\rm kK} \gtrsim 42$ (Sp $\lesssim$ sdO9.5), systematic trends appear in the residual. 
For $\Teff \gtrsim 42$\,kK  (Sp $\lesssim$ sdO9.5), {\it salt\_p00} increasingly underestimates \Teff\ compared with {\it XTgrid}. 
The latter provides a roughly linear correlation between \Teff\ and Sp between sdO7 and sdB1, although the gradient is markedly steeper than the equivalent relation reported by D13. 
The differences between results obtained from the two model grids are shown in Fig.\,\ref{f:system}. 
These are indicative of the  systematic errors introduced by assuming the overall metallicity and/or local thermodynamic equilibrium.

Experiments suggested that model grid spacings, metallicity and microturbulent velocity have a significant influence on the outcomes, but we conclude that the systematic errors introduced by the LTE assumption are unacceptable for $\Teff \geq 42$\,kK. 
To avoid this critical boundary, we therefore use the nLTE zero metal grid ({\it XTgrid}) for $\Teff \geq 41$\,kK. 
Below this value, metal-line blanketing and a model grid which extends to $\log g = 3$ are both required to obtain satisfactory fits, so the {\it salt\_p00} grid is used for $\Teff < 41$\,kK. 
Excluding outliers, there is a mean offset of $\approx 1.3 \pm 3.0 \kmsec$ between radial velocities obtained using models from the two grids. 

{\sc sfit} provides formal errors on the parameters governing the fit used based on a value of  $\chi^2$ which is not realistic because of the method used to increae the weight of specified  spectral lines as described above. 
From the scatter of points in Fig.\,\ref{f:system}, we estimate measurement errors to be $\delta \log \Teff \approx \pm 0.012$, $\delta \log g \approx \pm 0.27$,  $\delta \log y \approx \pm 0.29$, and $\delta v_{\rm rad} \approx \pm 3.0 \kmsec$. 
For convenience, the first three translate to fractional errors: $\delta\Teff/\Teff \approx \pm 0.028$, $\delta g/g \approx \pm 0.62$, and $\delta y/y \approx \pm 0.67$.
These should be used in preference to the formal errors cited in Table\,\ref{t:saltpars}.

A second approach to estimating the measurement errors was to use the best-fit models obtained with the {\it salt\_p00} grid as an independent low-noise sample with otherwise similar spectral properties to the observed sample. 
This sample was processed with the {\it XTgrid} models in exactly the same way as before, except that no velocity correction was necessary. 
Then the differences between the parameters obtained from the observed sample and the theoretical sample were formed, giving the following mean differences and standard deviations: $\langle \Delta \log \Teff \rangle = 0.000 \pm 0.017$, 
$\langle \Delta \log g \rangle = 0.253 \pm 0.203$, and  
$\langle \Delta \log y \rangle = -0.112\pm 0.262$. 
Restricting the test sample to $\Teff > 42$\,kK, these numbers change to 
$\langle \Delta \log \Teff \rangle = -0.005 \pm 0.015$, 
$\langle \Delta \log g \rangle = 0.236 \pm 0.201$, and  
$\langle \Delta \log y \rangle = -0.069\pm 0.174$.
Again, these translate to mean fractional errors 
$\delta\Teff/\Teff \approx \pm 0.039$, $\delta g/g \approx \pm 0.46$, and $\delta y/y \approx \pm 0.40$.

\begin{figure*}
\includegraphics[width=0.85\linewidth]{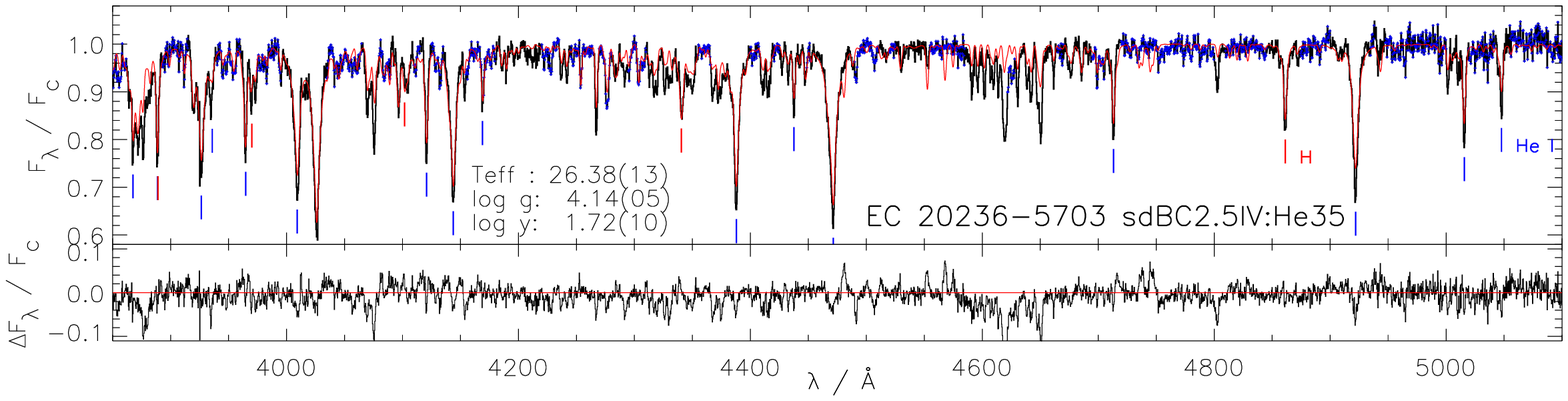}\\
\includegraphics[width=0.85\linewidth]{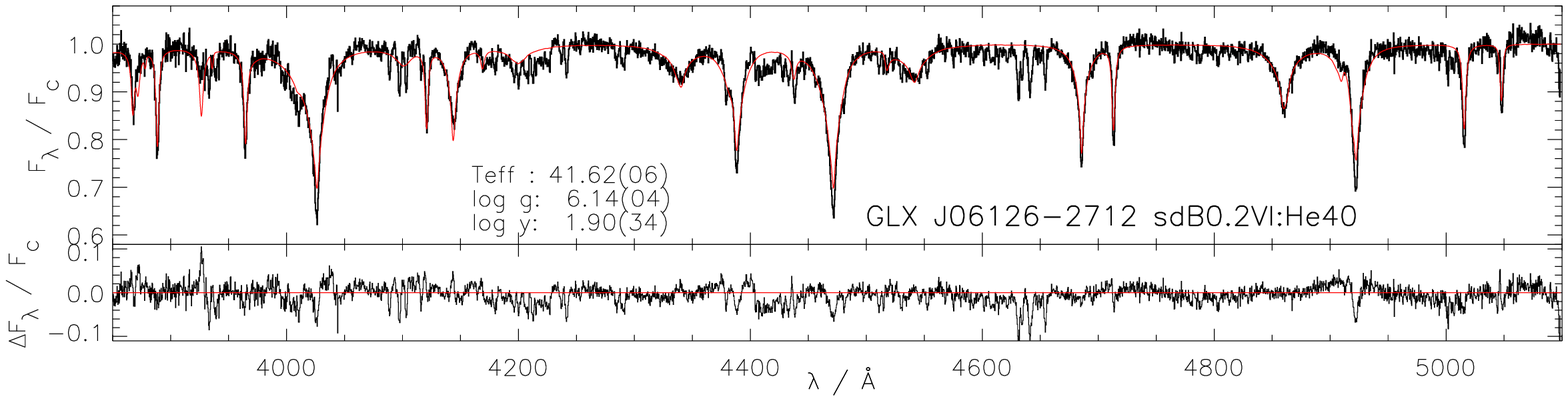}\\
\includegraphics[width=0.85\linewidth]{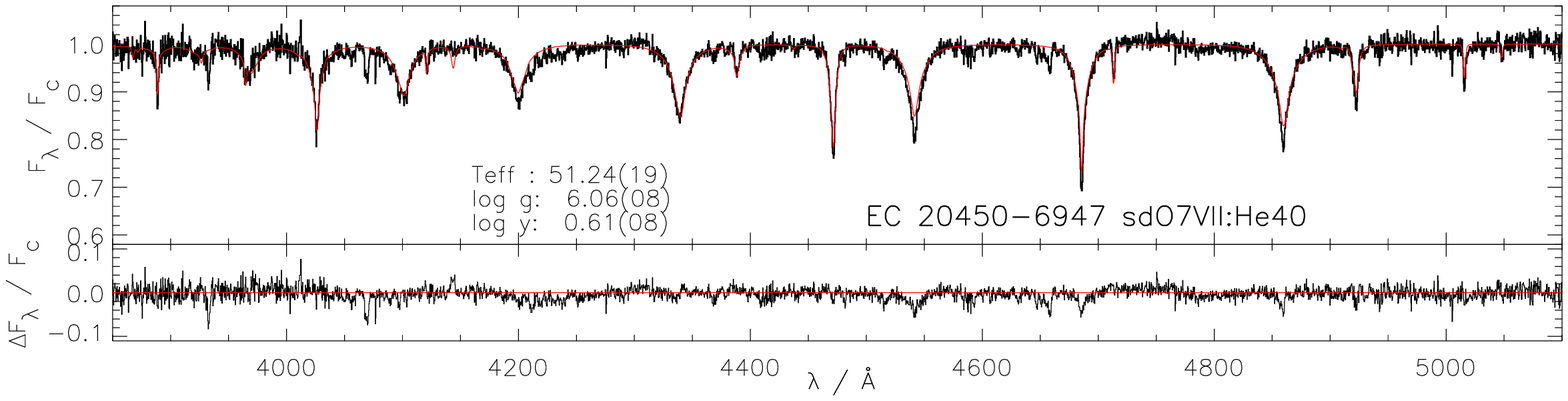}\\
\includegraphics[width=0.85\linewidth]{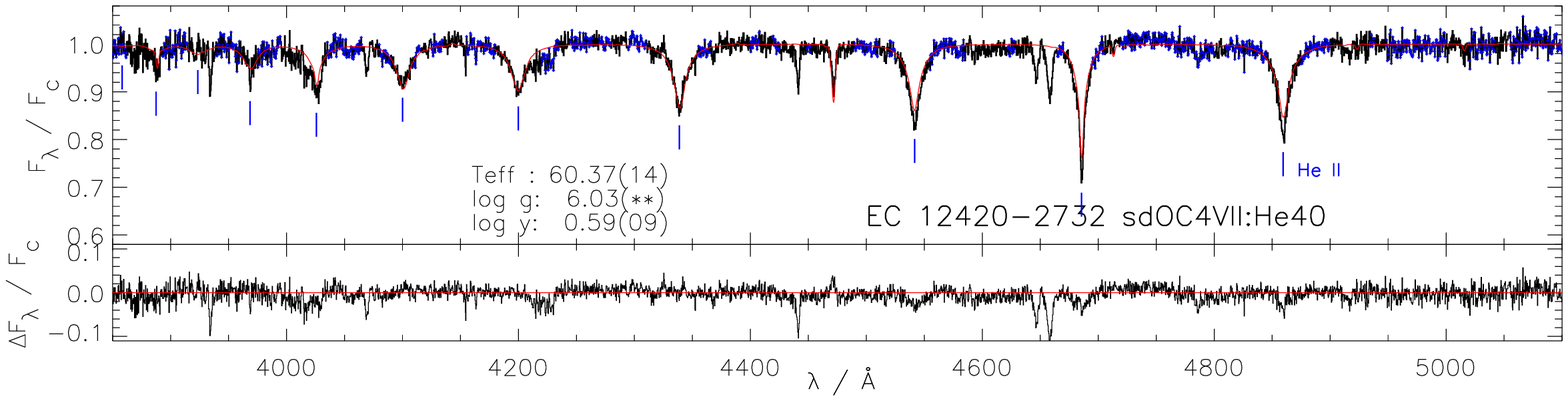}\\
\caption{RSS spectra and best fit solutions for a selection of representative stars.
Each panel shows the merged spectrum (black histogram) and the best-fit solution (red polyline). The residual (observed -- calculated) is plotted beneath. 
Each star's identifier, assigned spectral type and physical parameters ($\Teff$/kK, $\log g/{\rm cm\,s^{-2}}$, and $\log y$) are indicated, with formal errors in the last 2 digits in parentheses.  
Positions of hydrogen Balmer lines (red) and neutral helium lines (blue) are identified for EC\,20236-5703.
Ionized helium lines are identified for EC\,1240-2732. 
`Continuum' regions used to rectify all observed spectra are identified by blue crosses for the latter two stars. 
Equivalent plots for the entire sample of Table 3 are provided in the supplementary material. 
} 
\label{f:fits}
\end{figure*}

 Solutions obtained with both grids show a correlation between the upper limit of $\log y$ and \Teff\ for $\Teff\simge 40$kK (Fig.\,\ref{f:saltpars}).
As \Teff\ increases and the number of neutral hydrogen atoms becomes critically small, RSS spectra containing helium become increasingly degenerate in $y$ at high \Teff. 
The situation ameliorates at high resolution when the displacement between Balmer and ionized helium lines allows the former to be resolved.  

Representative spectra and best-fit solutions are illustrated in Fig.\,\ref{f:fits}. Equivalent plots for the entire sample of Table 3 are provided in Figs. E.1--E.7 of the Supplementary material.

\begin{figure*}
\includegraphics[width=0.98\linewidth]{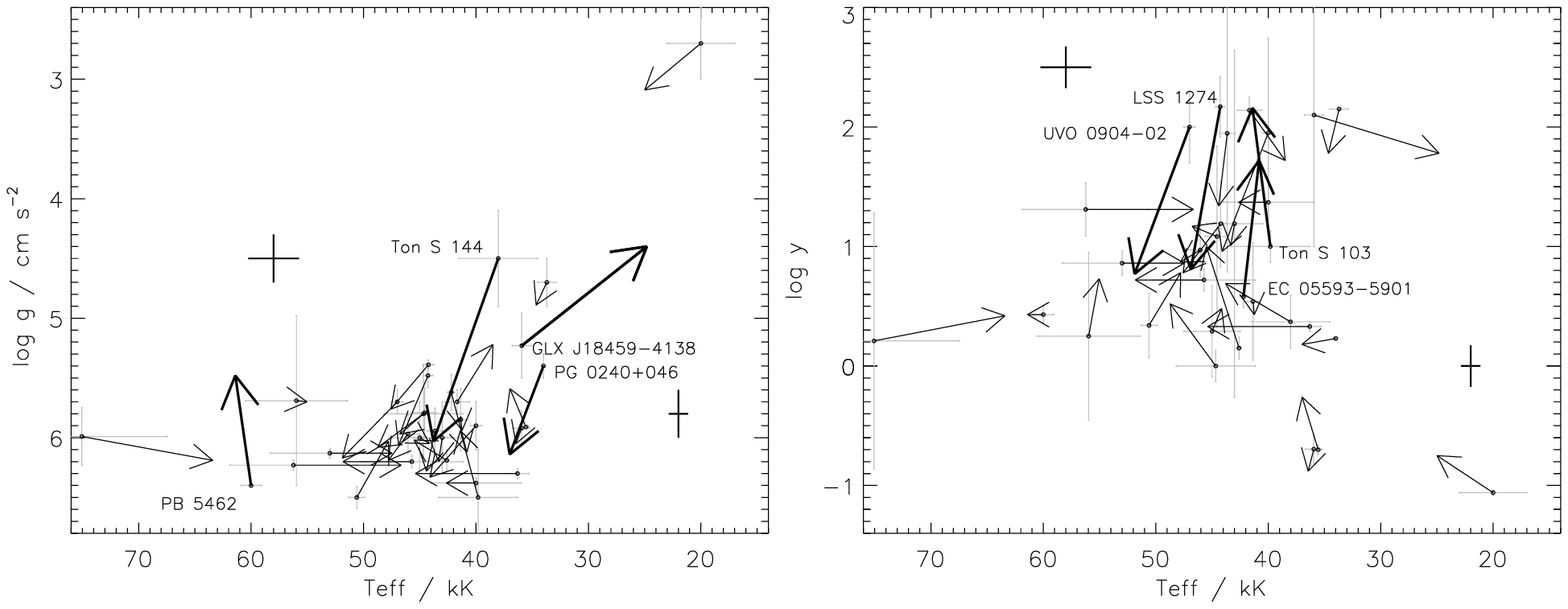}
\caption{As Fig.\,\ref{f:saltpars}, showing previous results for SALT sample members (grey error bars) linked to the current result (Table\,\ref{t:saltpars}) by an arrow. 
The largest differences are indicated by thicker lines and labelled (i.e. for $\delta \log \Teff > 0.1, \delta \log g > 0.6,$ or  $\delta \log y > 1.0$).
The mean SALT error bars are shown as in Fig.\,\ref{f:saltpars}. 
Data values and sources for the previous results are given in the Supplementary Material (Appendix F). Where $g$ or $y$ were not given in the earlier study, we have substituted values from Table\,\ref{t:saltpars}.  }
\label{f:prev}
\end{figure*}

\begin{table}
    \caption{Published atmospheric parameters for SALT sample members. Errors $<\pm0.1$ have been rounded up. }
    \label{t:prev}
    \setlength{\tabcolsep}{2pt}
    \centering
    \begin{tabular}{l r@{$\pm$}l r@{$\pm$}l r@{$\pm$}l l  }
    \hline
    Star  & \multicolumn{2}{c}{\Teff} &  \multicolumn{2}{c}{$\log g$} &  \multicolumn{2}{c}{$\log y$} & Reference \\
          &       \multicolumn{2}{c}{kK}   &  \multicolumn{2}{c}{cm\,s$^{-2}$} &  \multicolumn{2}{c}{ }  &    \\
    \hline
Ton S 144         & 38.0 & 3.5 & 4.5 & 0.4 & 0.4 & 0.2 & \citet{hunger81}  \\
 --- " ---        & 41.7 & 1.1 & 5.7 & 0.1 & 2.1 & 0.1 & \citet{stroeer07}  \\
BPS 22946–0005    & 20.0 & 3.0 & 2.7 & 0.3 & --1.1& 0.1 & \citet{kendall97}  \\
SB 705            & 45.1 & 8.1 & 5.6 & 0.6 & 1.0 & 0.4 & \citet{nemeth12}  \\
 --- " ---        & 44.7 & 3.5 & 5.8 & 0.4 & 0.0 & 0.1 & \citet{hunger81}  \\
LB 3229           & 40.0 & 0.5 & 5.2 & 0.2 & 1.9 & 0.8 & \citet{naslim10}  \\
Feige 19          & 40.0 & 2.5 & 5.0 & 0.3 & 1.0 & 1.5 & \citet{dreizler90}  \\
 --- " ---        & 45.0 & 2.5 & 6.0 & 0.3 & 0.3 & 0.4 & \citet{thejll94}  \\
PG 0240+046       & 37.0 & 2.5 & 5.3 & 0.3 & 0.1 & 0.3 & \citet{thejll94}  \\
 --- " ---        & 34.0 & 0.2 & 5.4 & 0.1 & 0.2 & 0.1 & \citet{ahmad03a}  \\
EC 03505--6929    & 42.6 & 0.2 & 6.2 & 0.1 & 0.2 & 0.1 & \citet{monibidin17}  \\
HE 0414--5429     & 50.6 & 0.7 & 6.5 & 0.1 & 0.3 & 0.3 & \citet{monibidin17}  \\
GLX J04205+0120   & 45.0 & 0.8 & 5.7 & 0.2 & \multicolumn{2}{c}{$>1.2$} & \citet{vennes11}  \\
 --- " ---        & 46.1 & 0.9 & 6.0 & 0.2 & 1.0 & 0.2 & \citet{nemeth12}  \\
EC 04271--2909    & \multicolumn{2}{c}{53.0} & \multicolumn{2}{c}{--}  & \multicolumn{2}{c}{--} & \citet{drilling95b}  \\
EC 05593--5901    & 42.2 & 0.3 & 5.6 & 0.2 & 0.6 & 0.2 & \citet{monibidin17}  \\
GLX J07581--0432  & 41.4 & 0.5 & 5.9 & 0.3 & 0.5 & 0.5 & \citet{nemeth12}  \\
PG 0902+057       & 43.0 & 2.5 & 6.0 & 0.3 & 1.5 & 0.4 & \citet{thejll94}  \\
UVO 0904--02      & 47.0 & 0.5 & 5.7 & 0.1 & 2.0 & 0.3 & \citet{schindewolf18}  \\
LSS 1274          & 44.3 & 0.4 & 5.5 & 0.1 & 2.2 & 0.3 & \citet{schindewolf18}  \\
PG 0958--119      & 44.2 & 0.5 & 5.4 & 0.1 & 1.2 & 0.4 & \citet{hirsch09}  \\
PG 1127+019       & 43.7 & 0.7 & 5.9 & 0.2 & 1.9 & 1.2 & \citet{luo16}  \\
PG 1230+067       & 43.0 & 2.5 & 5.5 & 0.3 & 1.2 & 1.5 & \citet{thejll94}  \\
PG 1318+062       & 44.6 & 1.0 & 5.8 & 0.2 & 1.1 & 0.8 & \citet{luo16}  \\
GLX J18459--4138  & 35.9 & 4.8 & 5.2 & 0.3 & 2.1 & 1.1 & \citet{nemeth12}  \\
 --- " ---        & 26.2 & 0.8 & 4.2 & 0.1 & 2.0 & 0.4 & \citet{jeffery17b}  \\
GLX J19111--1406  & 56.0 & 4.5 & 5.7 & 0.7 & 0.3 & 0.9 & \citet{nemeth12}  \\
BPS 22940--0009   & 33.7 & 0.8 & 4.7 & 0.2 & 2.2 & 0.1 & \citet{naslim10}  \\
LS IV--14 116     & 34.0 & 0.5 & 5.6 & 0.1 & --0.7 & 0.1 & \citet{naslim11}  \\
 --- " ---        & 35.0 & 0.3 & 5.9 & 0.1 & --0.6 & 0.1 & \citet{green11}  \\
 --- " ---        & 35.2 & 0.1 & 5.9 & 0.1 & --0.6 & 0.1 & \citet{randall15} \\
 --- " ---        & 35.5 & 1.0 & 5.9 & 0.9 & --0.6 & 0.1 & \citet{dorsch20} \\
PG 2158+082       & 75.1 & 7.6 & 6.0 & 0.2 & 0.2 & 1.1 & \citet{nemeth12}  \\
BPS 22892--0051   & \multicolumn{2}{c}{45.7}  & \multicolumn{2}{c}{--} & \multicolumn{2}{c}{--} & \citet{beers92}  \\
BPS 22875--0002   & \multicolumn{2}{c}{56.2}  & \multicolumn{2}{c}{--} & \multicolumn{2}{c}{--} & \citet{beers92}  \\
PG 2218+051       & 36.5 & 1.0 & 6.2 & 0.2 & --0.8 & 0.1 & \citet{saffer94}  \\
 --- " ---        & 36.0 & 0.7 & 5.9 & 0.1 & --0.7& 0.1 & \citet{luo16}  \\
EC 22536--5304    & 36.9 & 0.1 & 6.1 & 0.1 & --0.5& 0.1 & \citet{jeffery19b}  \\
Ton S 103         & 39.8 & 3.5 & 6.5 & 0.4 & 2 & 1 & \citet{hunger81}  \\
PB 5462           & \multicolumn{2}{c}{47.5} & \multicolumn{2}{c}{8.2} & \multicolumn{2}{c}{--} & \citet{kepler15.sdss10}  \\
 --- " ---        & 60.0 & 0.9 & 6.4 & 0.1 &    \multicolumn{2}{c}{--}     & \citet{hugelmeyer06}  \\
HE 2347--4130     & 44.9 & 1.2 & 5.8 & 1.5 & 1.4 & 0.4 & \citet{stroeer07}  \\
    \hline
    \end{tabular}
\end{table}

\subsection{Previous results}

Spectroscopic measurements of one or more of $\Teff, \log g, \log y$ have been published for some 30 members of the overall sample (Table\,\ref{t:prev}).
Fig.\,\ref{f:prev} compares those data with values in Table\,\ref{t:saltpars}. 
More than two thirds of the differences are within either the errors of the original observations or of the new measurements. 
Of the remainder:
PB\,5462 \citep{hugelmeyer06} lies outside the current model grid,
Ton\,S\,144 and Ton\,S,103 were measured using a restricted grid of nLTE models  \citep{hunger81}, 
the weakness of \ion{He}{ii}\,4686 in GLX\,J18459--4138 was overlooked by \citet{nemeth12} \citep[cf.][]{jeffery17b}, 
the hydrogen abundances measured from high-resolution spectra of LSS\,1274 and UVO\,0904--02 by \citet{schindewolf18a} are to be preferred, 
the published helium abundance of EC\,05593--5901 was 1.5 dex above the boundary of the model grid used by \citet{monibidin17}, 
and the spectrum of PG\,0240+046 used by \citet{ahmad03a} was limited to H$\gamma$, \ion{He}{i}4388 and 4471.

\begin{figure*}
\begin{center}
\includegraphics[clip, width=0.98\linewidth]{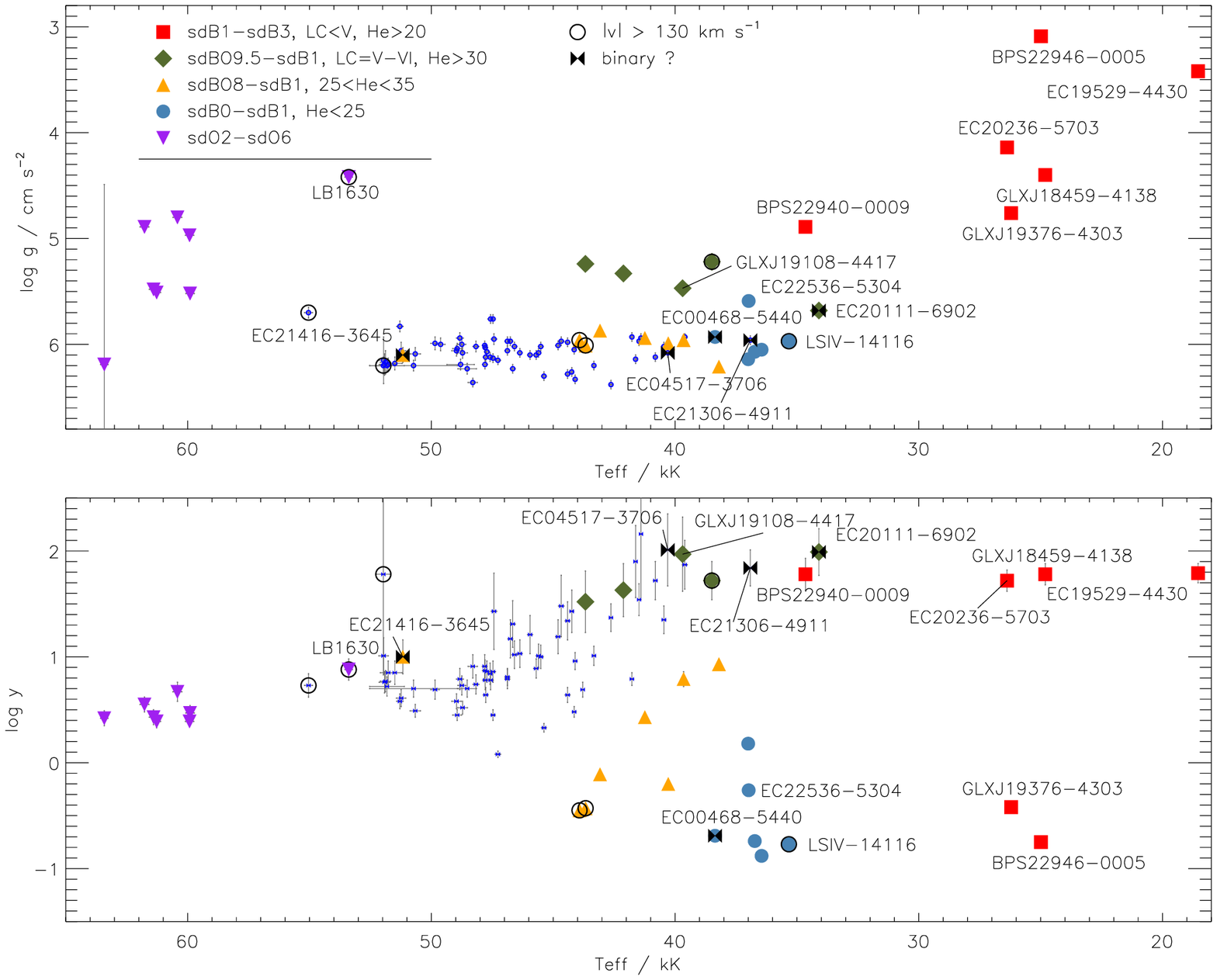}
\caption{ As Fig.\,\ref{f:saltpars} extended to identify subgroups and individual stars discussed in \S\,\ref{s:lights}. The groups identified in the key and by different colours and symbols correspond to \S\S\,\ref{s:cool} -- \ref{s:hot}. The small blue circles refer to the remaining stars  (cf. \S\,\ref{s:rest}). } 
\label{f:lights}
\end{center}
\end{figure*}

\section{Highlights}
\label{s:lights}

The primary objective of this part of the survey was to identify stars of particular interest for further investigation. 
For us this means:\\
a) stars at late spectral types (sdO9--sdB3) (or low \Teff\ and $g$) which might indicate links to other classes of helium-rich stars, \\
b) stars with intermediate helium classes (He10--He35) (or helium - to - hydrogen ratios) which might include heavy-metal stars and confronts the question of why hot subdwarfs are predominantly extremely helium-poor or helium-rich,  \\
c) stars with anomalous radial or rotational velocities which might indicate subdwarfs in close binary systems or otherwise high-velocity stars.\\
 The stars have been gathered into subgroups described in the following subsections. 
Within each group, stars are introduced  by  name, spectral class and model-dependent parameters as in Table \ref{t:saltpars}. 
The last are  expressed as ($\Teff/{\rm kK}, \log g/{\rm cm\,s^{-2}}, \log y$).
Groups and individual stars are identified in Fig.\,\ref{f:lights}. 

\subsection{Sp = sdB1 -- sdB3, LC $\lesssim$V, He $>20$}
\label{s:cool}

The first group includes stars which are classified sdB1 or later and have luminosity class V or less. 
These are indicated by filled red squares in Fig.\,\ref{f:lights}.
As such, they are not true subdwarfs since their surface gravity is similar to or lower than that of the main-sequence. 
Some or all might be shown to be subluminous on account of their mass and luminosity. 

\paragraph*{GLX\,J18459--4138 (sdB2V:He38)} (24.8, 4.4, 1.8) was identified as being similar to the pulsating helium star V652\,Her.  
The survey parameters are consistent with those given by \citet{jeffery17b}.  
It is the only nitrogen-rich member of this group. 

\paragraph*{GLX\,J19376--4303 (sdB2.5V:He21)} (26.2, 4.8, --0.4) appears similar to GLX\,J18459--4138 but has stronger Balmer lines, indicating  hydrogen and helium abundances of 63\% and 27\% respectively \citep{jeffery17c}.

\paragraph*{EC\,19529--4430 (sdB3IV:He35)} (18.5, 3.4, 1.8) shows no ionized helium lines (including He{\sc ii} 4686\AA) and Balmer lines much weaker than the neutral helium lines.
A defining feature is the weakness of all metal lines \citep[cf. HD144941:][]{harrison97} and the narrow wings of the H and He{\sc i} lines.

\paragraph*{EC\,20236--5703 (sdBC2.5IV:He35)} (26.4, 4.1, 1.7) has similar properties to EC\,19529-4430, but with slightly narrower \ion{He}{i} lines and a carbon rich metal-lined spectrum.
It is likely to have similarities to the carbon-rich pulsating helium star BX\,Cir \citep{woolf00}. 

\paragraph*{BPS\,22940--0009 (sdBC1V:He38)} (34.7, 4.9, 1.8) makes the fifth and hottest member of this group, all of which could be called extreme helium stars. 
Its closest well-studied counterparts are the hot extreme helium star LS\,IV$+6^{\circ}2$ \citep{jeffery98}, and PG\,1415+492 \citep[sdBC1VI:He39][]{ahmad03a} and and PG\,0135+243 \citep{moehler90b}.
A high-resolution spectral analysis was carried out by \citet{naslim10} who showed it to be the lowest gravity member of their sample of helium-rich subdwarfs. 
Our coarse analysis is in general agreement. 

\paragraph*{BPS\,22946--0005 (sdB2.5II:He24)} (25.0, 3.1, --0.8) has He $<30$ but otherwise fits this group. It is a post-AGB star analyzed by \citet{kendall97}. It was mistakenly included in our sample but provides  a useful control. In comparison with Kendall et al., our temperature is high and our gravity low, but still consistent with a post-AGB star.

\subsection{Sp = sdO9.5 -- sdB1, LC $\approx$ V -- VI, He $> 30$}
\label{s:b0}

If signal-to-noise ratios were higher, and luminosity classification was a more precise science, this section would isolate other sample members with LC $\leq$ VI, and hence identify the remaining high luminosity stars. 
Given the large errors associated with assigning luminosity class, 
$\log g \lesssim 5.7$ has been used as a proxy. 

\paragraph*{GLX\,J19108-4417 (sdBC0.2VI:He39)} (39.7, 5.5, 2.0) is just slightly hotter and less luminous than BPS\,22940--0009 (see above). 
Whilst it also resembles the extreme helium dwarf LS\,IV$+6^{\circ}2$ \citep{jeffery98}, it has a higher hydrogen abundance \citep{beliere18}. 
Again, connections with  PG\,1415+492 \citep{ahmad03a} and PG\,0135+243 \citep{moehler90b} should also be explored. 

\paragraph*{Other stars in this group} include: 
EC\,20111--6902 (sdBC1.5VII:He38 --- 34.1, 5.7, 2.0),
Ton\,S\,148 (sdBC0.2VI:He37 -- 38.5, 5.2, 1.7),
GLX\,J08454--1214 (sdOC9.5VI:He39 --- 43.4, 5.2, 1.5), 
and
EC20221-6249 (sdOC9.5VII:He39 --- 42.1, 5.3, 1.6). 
They are indicated by green diamonds in Fig.\,\ref{f:lights} and all await detailed analysis from high-resolution spectroscopy. 
These stars will be crucial in establishing any link between the low-luminosity helium stars  identified in \S\,\ref{s:cool} and helium-rich subdwarfs stars on the helium-main-sequence, such as the post-double white dwarf merger connection proposed by \cite{zhang12a}.

\begin{figure*}
\includegraphics[width=0.85\linewidth]{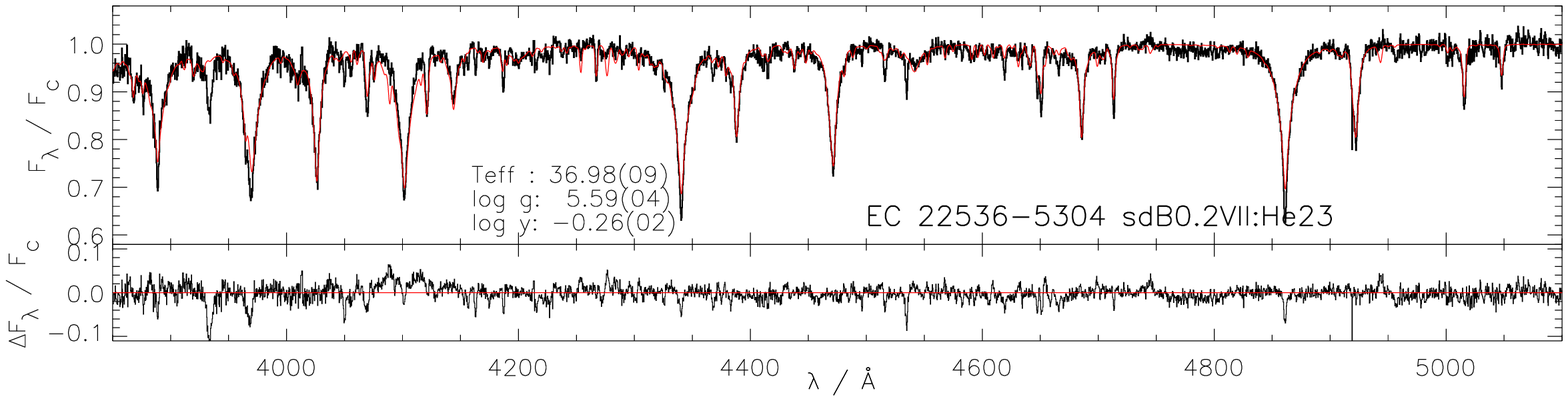}\\
\includegraphics[width=0.85\linewidth]{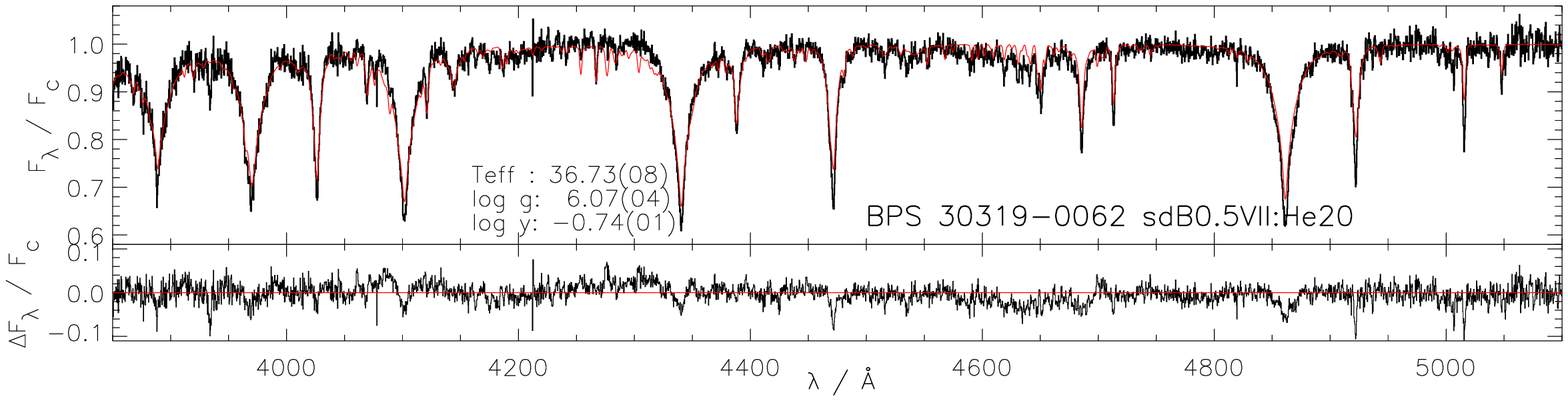}\\
\includegraphics[width=0.85\linewidth]{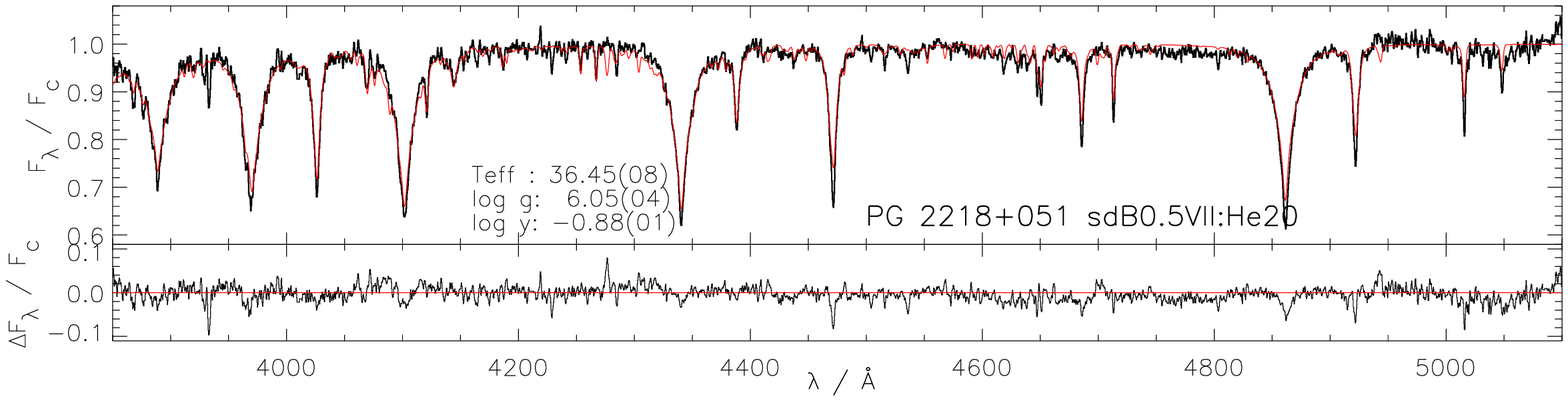}\\
\includegraphics[width=0.85\linewidth]{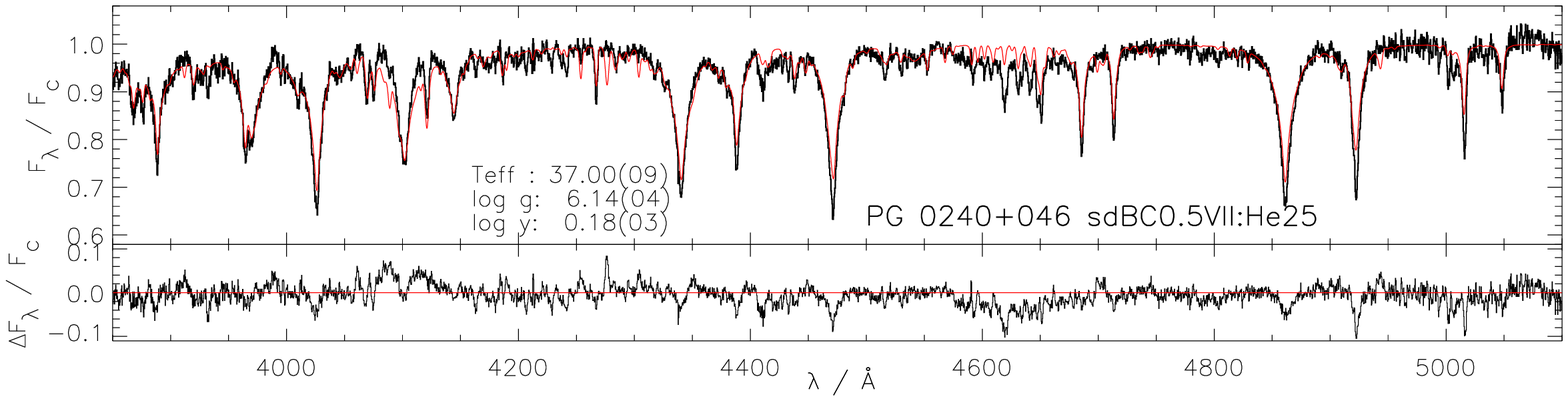}\\
\includegraphics[width=0.85\linewidth]{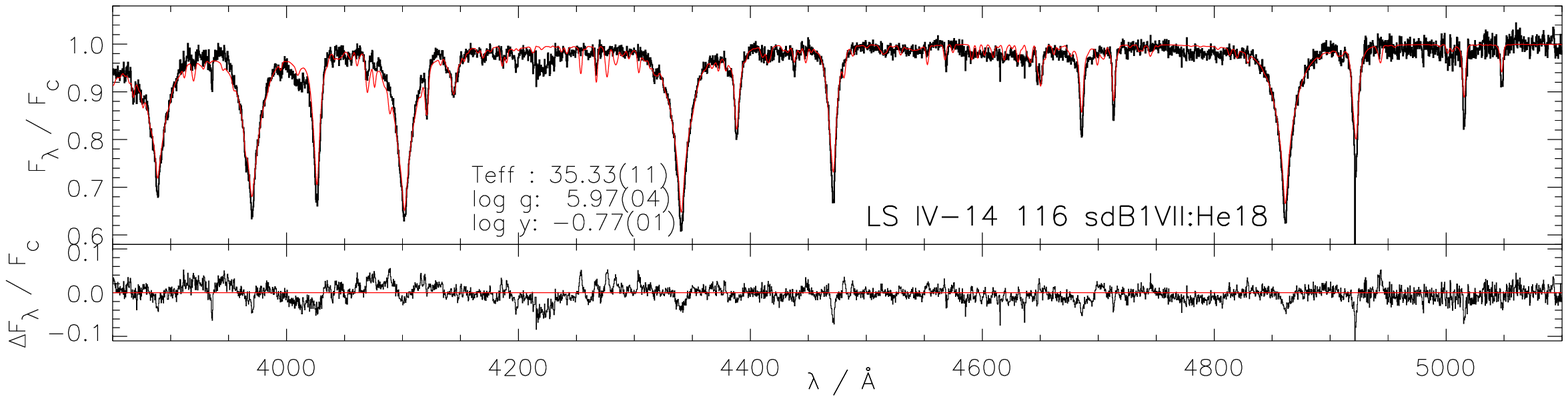} \\
\includegraphics[width=0.85\linewidth]{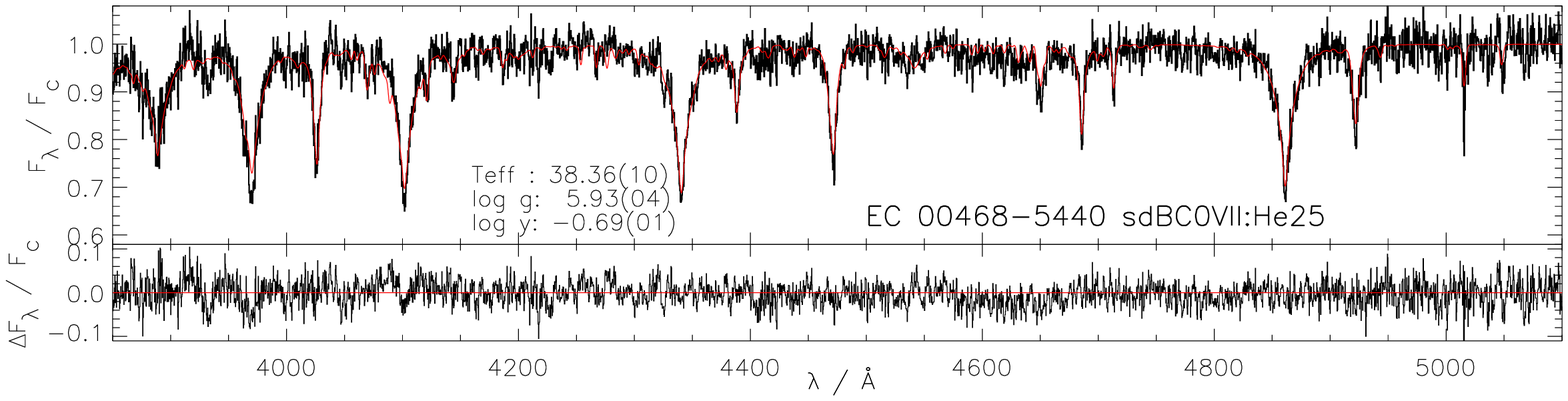}\\
\caption{As Fig.\,\ref{f:fits} for intermediate helium subdwarfs. LS\,IV$-14^{\circ}116$ and EC\,22536-5304 are known heavy-metal subdwarfs \citep{naslim10,jeffery19b};  PG\,0240+046, PG\,2218+051, BPS\,30319--0062, or EC\,00468--5440 have similar \Teff\, $g$ and $y$, but no heavy-metal detections yet. } 
\label{f:hiz}
\end{figure*}

\subsection{Sp = sdB0 -- sdB1, He $<25$}
\label{s:hiz}

As a primary indicator, helium class is a useful proxy for surface helium abundance, but is increasingly imperfect at spectral types earlier than sdO8 (Fig.\,\ref{f:saltclass}). 
There are 14 stars in the sample with sdO8 $\lesssim$ Sp $\lesssim$ sdB1 and He$<35$.
These are often referred to as intermediate helium-rich subdwarfs. 
Spectral characteristics vary enormously across the group, which covers transitions from  helium to hydrogen dominated  and \ion{He}{i} to \ion{He}{ii} dominated spectra. Identifying smaller subgroups is useful. 

The most distinctive and most hydrogen-rich group covers a narrow spectral range sdB0 $\lesssim$ Sp $\lesssim$ sdB1, He $<25$ and includes the heavy-metal subdwarfs.
These are indicated by large filled blue circles in Fig.\,\ref{f:lights}.

\paragraph*{LS\,IV\,--14\,116 (sdB1VII:He18)} (35.3, 6.0, --0.8) is a well-studied pulsating intermediate helium subdwarf with a remarkable surface chemistry \citep{ahmad05a,naslim11}.
It was included in the RSS sample as a control. The survey parameters are consistent with other recent measurements \citep{randall15,dorsch20}.

\paragraph*{EC\,22536--5304 (sdB0.2VII:He23)} (37.0, 5.6, --0.3) was identified from the 4495\AA\ line of triply-ionized lead in the HRS spectrum, and confirmed by the detection of both Pb{\sc iv} 4495 and  4049\AA\  in the RSS spectrum. It is the most lead-rich heavy-metal subdwarf so far, with a lead abundance 4.5 dex above solar \citep{jeffery19b}. 

\paragraph*{Four additional stars} have similar spectral type: 
EC\,00468--5440	(sdBC0VII:He25 --- 38.4, 5.9, --0.7),
PG\,2218+051	(sdB0.5VII:He20 --- 36.5, 6.1, --0.9),
BPS\,30319--0062	(sdB0.5VII:He20 --- 36.7, 6.1, --0.7),
and
PG\,0240+046	(sdBC0.5VII:He25 --- 37.0, 6.1, 0.2).
Their spectra are illustrated in Fig.\,\ref{f:hiz}.
For the known examples of this group, in which radiative levitation is regarded as the crucial driver of exotic chemistry, the sharp heavy-metal absorption lines are distinctive in high-resolution spectra because of their very low rotation velocity. 
The lines are much harder to recognise at the resolution of classification spectra.
Coarse analyses have been carried out previously for PG\,2218+051 \citep{saffer94,luo16} and  PG\,0240+046 \citep{aznar01,ahmad03a}, with similar results to those presented here.
Whilst all six stars show a clear signature from \ion{C}{iii} 4647,4650\AA, it is only  strong enough in two cases, PG\,0240+046 and EC\,00468--5440, to trigger a carbon-rich `C' classification.   
The four heavy-metal candidates  should be investigated at higher resolution and signal-to-noise for evidence of lead or zirconium absorption lines, and to determine whether the carbon abundance  is correlated with hydrogen-to-helium ratio.

\subsection{Sp = sdO8 -- sdB1, $25 <$ He $<35$}
\label{s:ihe}

Of the remaining intermediate helium stars,   
EC\,04013--4017    (sdBC1VII:He32 --- 38.2, 6.2, 0.9) is the coolest and could arguably have been included amongst the group in \S\,\ref{s:b0} since the helium class and $\log y$ appear contradictory. 

Six stars in the sample have similar spectra with some spread in H/He and \ion{He}{i/ii} ratios:
Ton\,S\,148 (sdBC0.2VI:He32 --- 38.5, 5.2, 1.7), 
LB\,3289	        (sdBN0.2VII:He29 --- 39.7, 6.0, 0.8),
GLX\,J07581--0432  (sdO9.5VII:He33 --- 41.2, 5.9, 0.4),
EC\,20184--3435    (sdO9.5VI:He28 --- 40.3, 6.0 --0.2),
EC\,20111--3724    (sdO9VII:He33 --- 43.1, 5.9, --0.1),
Ton\,S\,415     (sdO8VII:He30 --- 43.9, 6.0, --0.5),
and
EC\,05242--2900    (sdO8VII:He28 --- 43.7, 6.0, --0.4). 
These represent quintessentially typical intermediate helium-rich stars, concerning which  little is known. 
They are indicated by yellow upward triangles in Fig.\,\ref{f:lights}.

For GLX\,J07581-0432, \citet{nemeth12} give  $\Teff,\log g,\log y \approx 41.4, 5.9, 0.07$ in good agreement with our analysis.  
 Other stars in this group include BPS\,22956--0094 \citep{naslim10} and possibly HS\,1000+471 (sdBC0.2VII:He28)  and Ton\,107 (sdBC0.5VII:He28) \citep{ahmad03a}.

\paragraph*{EC\,21416--3645 (sdO8.5VII:He34)} (51.2, 6.1, 1.0) stands out. 
Most of the principal H and He lines are weaker than in the stars described above.
The spectrum is unique in our sample, showing strong broad features at calcium H and K. 
In hot stars, these normally correspond to either H$\epsilon$ or \ion{He}{ii} 3968\AA (or both), and \ion{He}{i} 3935\AA, and are rarely seen at similar strength.
The DSS2 image is elliptical, whilst the 2MASS image is circular and offset 2.7\arcsec\ to the west. 
It appears that the hot subdwarf spectrum is contaminated by that of a faint red star, 
Gaia DR2 6586406672826522112 ($\langle g \rangle=15.6, b_p-r_p=0.85$).
Having twice the parallax of EC\,21416--3645, the two stars are unlikely to be associated. 
The two stars would be unresolved under normal SALT seeing conditions.
The cool star  may also account for apparent noise in the combined spectrum.

\subsection{High radial velocity}
\label{s:hivrad}
As a consequence of the optical layout and from undocumented experience, we do not have full confidence in the SALT/RSS radial velocities and so the precision of $v_{\rm rad}$ in Table\,\ref{t:saltpars} may be worse than the statistical errors suggest. 
However, as a counter argument, LS\,IV\,-14\,116 has a well-established radial velocity of $-149\pm2$\kmsec \citep{randall15}. 
Table\,\ref{t:saltpars} gives $-163\pm1$\kmsec.  
Table\,\ref{t:saltpars} is therefore useful for identifying high velocity stars and/or close binaries.
As an arbitrary example, other stars in the sample with $|v_{\rm rad}|>130$\kmsec\ include 
GLX\,J19111--1406, 
GLX\,J14258--0432, EC\,05242--2900, 
Ton\,S\,148, 
LB\,1630, 
and
Ton\,S\,415. 
They are indicated by black circles surrounding the spectral group symbol in Fig.\,\ref{f:lights}.
It will be interesting to investigate the space motions of these stars.

\begin{figure*}
\includegraphics[width=0.85\linewidth]{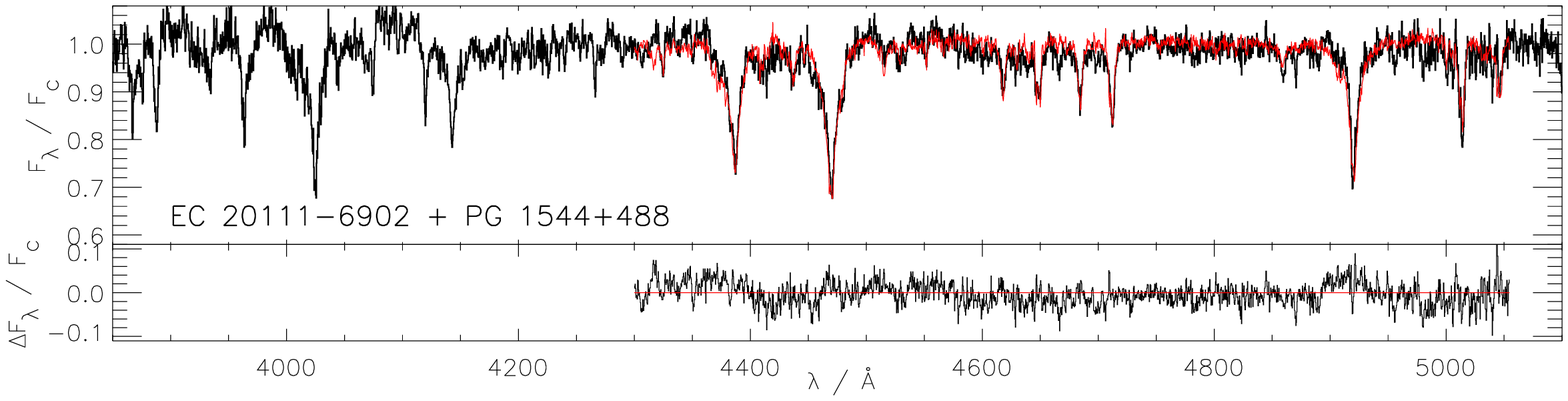}\\
\includegraphics[width=0.85\linewidth]{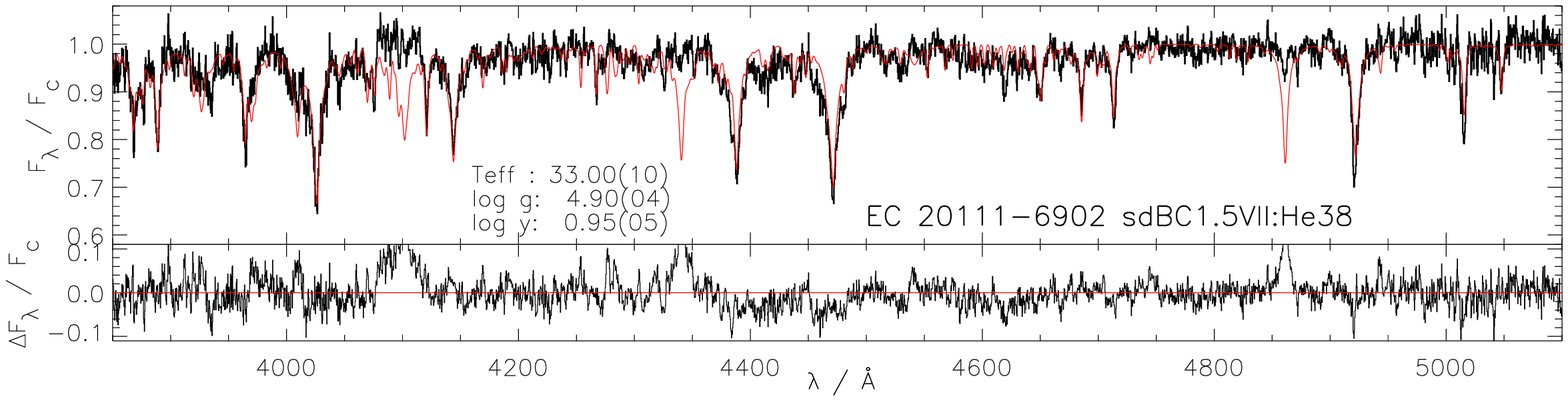}\\
\includegraphics[width=0.85\linewidth]{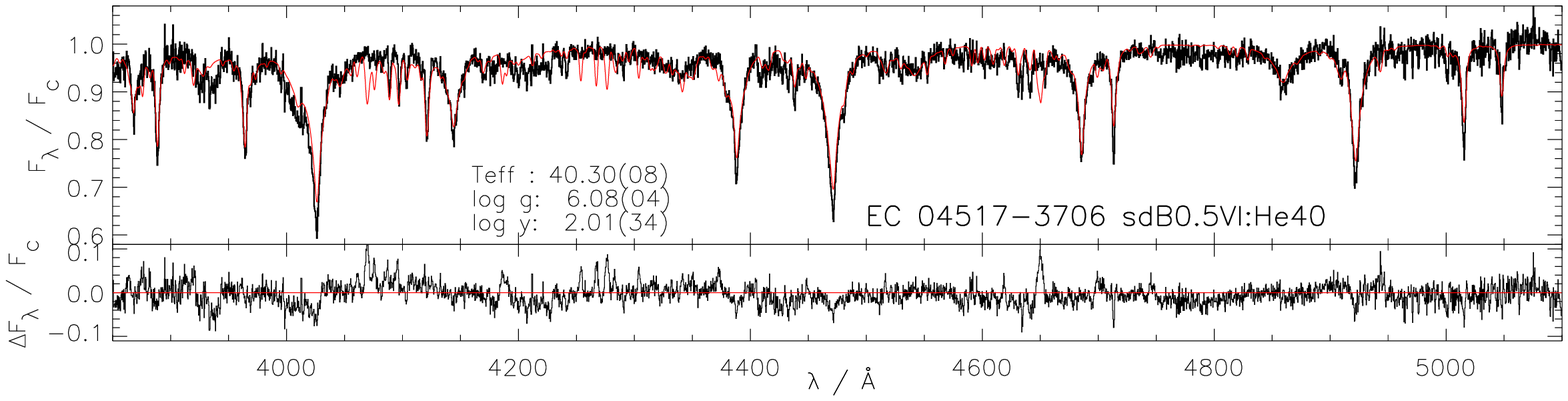}
\includegraphics[width=0.85\linewidth]{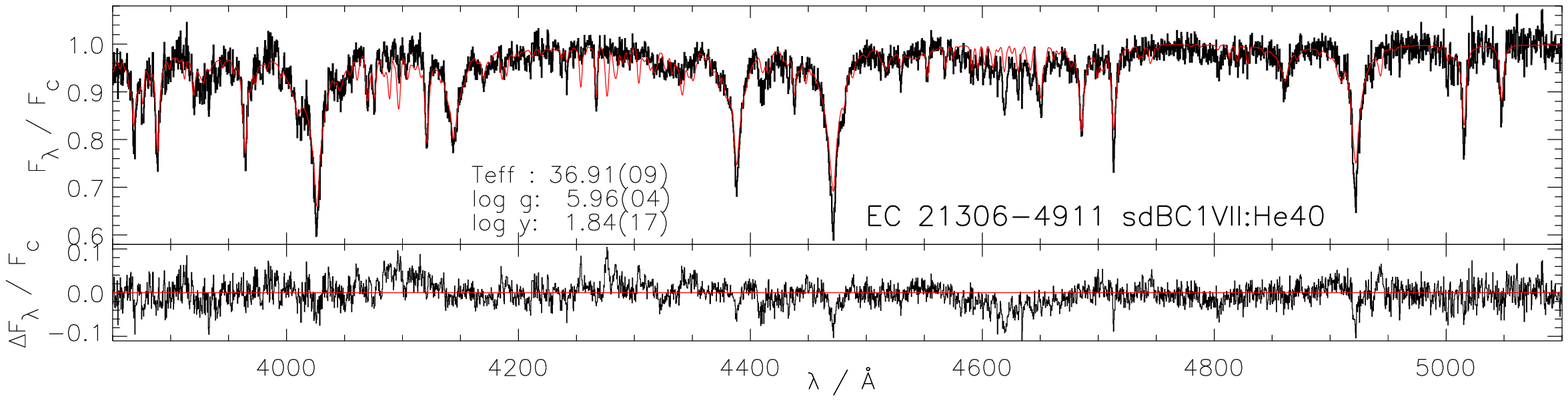}
\caption{Stars with spectra similar to the double helium-rich subdwarf PG\,1544+488. 
Top: Comparison of the SALT/RSS spectrum of EC\,20111--6902 (black) with a William Herschel Telescope spectrum (r746709) of the double helium white dwarf PG\,1544+488 (red) \citep[][Table 3]{sener14}. The latter has been shifted in radial velocity to match. 
The residual (EC\,20111 -- PG\,1544) is shown beneath at the same scale.  
Bottom three panels: As Fig.\,\ref{f:fits} for EC\,20111--6902, EC\,04571--3706 and EC\,21306--4911. } 
\label{f:ec20111}
\end{figure*}

\subsection{Broad lines: $v_{\rm wid} > 150 \kmsec$}
\label{s:hivrot}

The mean line width for RSS spectra in Table\,\ref{t:saltpars} is $\langle v_{\rm wid} \rangle = 85\pm21\kmsec$.
Excluding very hot stars $\Teff\gtrsim50$\,kK, where hydrogen-helium blends cannot be resolved,   $\langle v_{\rm wid} \rangle = 80\pm16$. 
Stars with $v_{\rm wid} \gtrsim \langle v_{\rm wid} \rangle  + 2 \sigma $ are of interest, since these indicate either a higher than average rotation velocity, a variable velocity spectrum used to construct the mean, or a spectrum originating in two or more similar stars with different velocities.  
Examples are  indicated by black bowties superimposed on  the spectral group symbol  in Fig.\,\ref{f:lights}.
Table\,\ref{t:saltpars} shows four stars  with $v_{\rm wid} > 112 \kmsec$ and $\Teff < 50$\,kK.

\paragraph*{EC\,20111--6902 (sdBC1.5VII:He38)}  (34.1, 5.7, 2.0).
The {\sc sfit} solution to the hydrogen-deficient spectrum of EC\,20111--6902 ($\Teff=34$\,kK) shows a well-above average value for the line width ($v_{\rm wid}=153\kmsec$). 
Using {\it XTgrid} yielded $v_{\rm wid}=204\kmsec$ and so the high value is not a consequence of using LTE rather than non-LTE models. 
The spectrum is well-exposed (S/N$\approx52$), being the sum of observations made on 4 separate nights.
The individual observations show a spread in {\it radial} velocity of 40\,\kmsec\ from cross-correlation with a model template and of 53\,\kmsec\ from shifts in the \ion{C}{ii} 4267\,\AA\ absorption line. 
Co-adding these spectra without correction will contribute substantially to the high value of  $v_{\rm wid}$. 
The cause of the variation requires further investigation. 
The spectrum bears a strong similarity to that of the double helium subdwarf binary PG\,1544+488 (sdBC1VII:He39p: D13) (Fig.\,\ref{f:ec20111}).
The latter has   a 12\,h orbital period with velocity semi-amplitudes of 87 and 95 \kmsec\ for each of the components, respectively \citep{ahmad04a,sener14}. 
It is proposed that  EC\,20111--6902 is very likely a spectroscopic binary containing at least one, if not two, helium-rich subdwarfs, and for which the velocity semi-amplitude is at least 50\,\kmsec.

\paragraph*{EC\,04517--3706 (sdB0.5VI:He40)} (40.3, 6.1, 2.0) has $v_{\rm wid}=118\kmsec$ on the 2$\sigma$ boundary. 
It is  warmer and less carbon-rich than  EC\,20111--6902 (Fig.\,\ref{f:ec20111}).
 
\paragraph*{EC\,21306--4911 (sdBC1VII:He40)} (36.9, 6.0, 1.8) has a spectrum and parameters similar to EC\,20111--6902 (Fig.\,\ref{f:ec20111}).
Both have strong carbon lines.
$v_{\rm wid}=105\kmsec$ is high but lies within $2\sigma$ of the mean. 
While only a single RSS observation contributes to the spectrum, the S/N ratio is the same as that of the combined spectrum of  EC\,20111--6902.
Variable radial velocity is not a contributing factor, but the presence of two similar spectra with different velocities, as in PG\,1544+488, cannot be ruled out.  
Additional time-resolved high-resolution measurements are essential for all of these potential binary-star candidates. 

\paragraph*{EC\,00468--5440 (sdBC0VII:He25)} (38.4, 5.9, --0.7) )has $v_{\rm wid}=117\kmsec$ but a spectrum similar to the otherwise sharp-lined intermediate helium-rich subdwarfs (see \S\,5.3). 
A higher S/N spectrum is required.

\subsection{Sp $\lesssim$ sdO6}
\label{s:hot}

Thirteen stars have spectral types earlier than sdO6, including 
GLX J17051--7156 (sdOC6VII:He40 --- 54.9, 6.1, 0.6), 
PB\,5462	(sdOC5VII:He40 --- 61.4, 5.5, 0.4),
LB\,1630	(sdOC5VI:He38 --- 53.4, 4.4, 0.9),
EC\,12420--2732	(sdOC4VII:He40 --- 60.4, 6.0, 0.6),
PG\,1220--056	(sdO4VII:He39 --- 59.0, 6.0, 0.5),
GLX\,J16546+0318	(sdOC3VII:He40 --- 61.3, 6.1, 0.7),
EC\,10475--2703	(sdOC3VI:He39 --- 59.9, 5.5, 0.5),
GLX\,J07158--5407	(sdO3VII:He40 --- 61.0, 6.0, 0.3),
PG\,1537--046	(sdOC2VII:He40 --- 61.3, 5.5, 0.4),
GLX\,J20251--0804	(sdOC2VII:He39 --- 59.9, 5.0, 0.4),
GLX\,J20133--1201	(sdOC2VII:He37 --- 60.4, 4.8, 0.7),
EC\,11236--1945	(sdO2VII:He40 --- 61.8, 4.9, 0.6),
and 
PG\,2158+082	(sdO2VII:He40 --- 63.4, 6.2, 0.4).
These correspond to subdwarfs with $\Teff\gtrsim50$\,kK
and stretch both the boundaries and the physics of the model atmosphere grids. 
They are indicated by purple downward triangles in Fig.\,\ref{f:lights}.

 The majority have strong carbon lines, including emission around 4650\AA. 
As discussed already, the absence of neutral hydrogen at these temperatures makes it difficult to measure the hydrogen abundance at the resolution of the RSS spectra. 

\paragraph*{LB\,1630 (sdOC5VI:He38)} (53.4, 4.4, 0.9) has markedly narrower \ion{He}{ii} lines than the remainder of this group, hence its lower luminosity class and, indeed, surface gravity. 
It is possibly similar to the helium-rich subdwarfs LSE\,153, 259 and 263 \citep{husfeld89} or BD+37 442 and BD+37 1977 \citep{jeffery10} and hence the descendant of a helium-shell-burning giant rather than a helium-core-burning subdwarf.  Detailed fine analysis of this and the higher-gravity subdwarfs in this group would address important questions about their origin and fate.

\subsection{Sp = sdO6.5 -- sdB0.5, He $\gtrsim 35$ }
\label{s:rest}

The remaining members of the sample comprise what are most commonly understood to be `He-sdO' stars. 
These are indicated by small blue circles in Fig.\,\ref{f:lights}.
They all have low hydrogen abundance and a (rms) dispersion in surface gravity which is less than the estimated measurement error ($\langle \log g\rangle = 6.05 \pm 0.19$), although several of the solutions lie uncomfortably close to the {\it XTgrid} boundary. 
Nearly half (35) have a carbon-rich `C' classification and 4 have a nitrogen-rich `N' classification.  52 are concentrated in spectral types sdO7 to sdO9.

Detailed inspection of the spectra of these stars will surely yield additional surprises.
Since these stars are likely to have surfaces which provide a chemical record of previous evolution, further analysis to obtain precise hydrogen, carbon and nitrogen abundances, as well as data for other species,  will be invaluable \citep[cf.][]{zhang12a}. 

\section{Conclusion}

The current survey aims to characterize the properties of a substantial fraction of helium-rich subdwarfs in the southern hemisphere, to establish the existence and sizes of subgroups within that sample, and to provide evidence with which to explore connections between these subgroups and other classes of evolved star.
This paper has presented and validated the methods used to observe, classify and measure atmospheric parameters from intermediate dispersion ($R\approx3600$) spectroscopy obtained primarily the with the Robert Stobie spectrograph of the Southern African Large Telescope. 
It has presented spectral classifications on the MK-like Drilling system (D13) and atmospheric parameters $\Teff, \log g, \log y$ based on non-LTE zero-metallicity  ($\Teff > 41$\,kK) or LTE line-blanketed  ($\Teff < 41$\,kK) model atmospheres. 
Although the majority of the sample, especially for spectral types earlier than sdO8, are classified as being extremely helium rich on the basis of line depth ratios, the helium to hydrogen ratio $y$ is not well constrained by model atmospheres for $\Teff > 40$\,kK at the classification resolution. 
There are two reasons: one is that it is increasingly difficult to resolve hydrogen from the dominant \ion{He}{ii} lines as \Teff\ increases and the other is that, as the hydrogen abundance $\nH \rightarrow 0$, the error in the denominator ($y = \nHe/\nH$) dominates.  

It is clear that the generic term `helium-rich subdwarfs' as applied to low-resolution classification surveys includes stars with a wide range of properties. 
The majority (74/106) occupy a  tight volume in parameter space with
$41 \lesssim \Teff/{\,\rm kK} \lesssim 52$, 
$5.9 \lesssim \log g/{\rm cm\,s^{-2}} \lesssim 6.4$, and $\log y > 0.5$. 
Of the remainder distinct groups include:
very hot stars with spectral types sdO6 or earlier (13),
cool low-gravity ($\log g<5$) extremely helium-rich stars (5), 
and stars with intermediate helium abundances and spectral types sdO8 -- sdB1 (14), of which up to 6 may have surfaces heavily enriched in s-process elements.

Several remarkable individual stars have been identified.   
A few have been reported previously, e.g. GLX\,J18459--4138, EC\,22536--5305 \citep{jeffery17a,jeffery19b}. 
At least one star (EC\,20111--6902) is a radial-velocity variable and bears a strong resemblance to the double helium subdwarf PG\,1544+488. 
Other binaries are likely to lie undetected within the sample.  
One star (EC\,19529--4430) at the extreme cool end of the sample is remarkable for the absence or weakness of its metal lines. 
LB\,1630 appears to be a high luminosity extreme helium subdwarf. 

Immediate future work will include completion of the low-resolution survey with SALT/RSS and its extension to high-resolution for all sufficiently bright sample members. 
The sample must also be reviewed for radial-velocity variables, and followed up for positive detections.  
Classification and parameterisation should be carried out for the remainder of the sample on completion of the observations, and should include stars observed with other telescopes so as to establish a complete magnitude limited sample. 

Methods used for atmospheric analyses must be extended to include line-blanketed nLTE models of appropriate composition wherever practically possible, though appropriate LTE models will continue to be useful for low temperature stars $\Teff > 30$\,kK. 
Robust techniques that deliver reliable, self-consistent and precise fundamental quantities and abundances for large numbers of helium-rich subdwarfs are urgently required. 
 
With the imminent improvement of Gaia parallaxes and proper motions to $<0.001$\arcsec, spectroscopy should be supplemented with total-flux methods to establish precise angular diameters which will yield useful radii, luminosities and galactic orbits. 

Spectroscopic masses derived therefrom together with abundances for carbon, nitrogen and other species, will provide illuminating tests for evolution models that otherwise pass the tests of radius, luminosity and galactic location.  

\section*{Acknowledgments}

The Armagh Observatory and Planetarium is funded by direct grant from the Northern Ireland Dept for Communities. That funding has enabled the Armagh Observatory and Planetarium to participate in the Southern African Large Telescope (SALT) through member of the United Kingdom SALT consortium (UKSC). 
All of the observations reported in this paper were obtained with SALT following  generous awards of telescope time from the UKSC and South African SALT Time Allocation Committees under programmes 2016-1-SCI-045, 2016-2-SCI-008, 2017-1-SCI-004, 2017-2-SCI-007,  2018-1-SCI-038, 2018-2-SCI-033, and 2019-1-MLT-003
The authors acknowledge invaluable assistance from current and former SALT staff, particularly Christian Hettlage and Steve Crawford. 
The model atmospheres were computed on a machine purchased under grant ST/M000834/1 from the UK Science and Technology Facilities Council.
This research has made use of the SIMBAD database, operated at CDS, Strasbourg, France 

\section*{Data Availability}
The raw and pipeline reduced SALT observations are available from the SALT Data Archive ({\tt https://ssda.saao.ac.za}). 
The model atmospheres and spectra computed for this project are available on the Armagh Observatory and Planetarium web server 
({\tt https://armagh.space/$\sim$SJeffery/Data/}).

\section*{Supplementary Material}
The supplementary material contains five appendices as follows.\\
{\bf A:}  provides dates on which SALT obtained data with either RSS or HRS for each star classified in Table\,2 (i.e. Table\,\ref{t:salt_obs} in full).  \\
{\bf B:}  compares  automatic classifications obtained by applying the algorithms described in \S\,3.1 with the manual classifications given by D13. The observational data  are the same for both sets of classifications.\\
{\bf C:} compares theoretical spectra  selected from the grids described in \S\,4.1. \\
{\bf D:} compares effective temperature, surface gravity and helium-to-hydrogen ratio ($\Teff, \log g, \log y$) as determined in \S\,4 with the corresponding spectral type, luminosity and helium classes  as determined in \S\,3. \\
{\bf E.} shows the complete ensemble of reduced survey spectra, best-fit models and residuals.  

\bibliographystyle{mnras}
\bibliography{ehe}

\label{lastpage}
\end{document}

\maketitle

\label{firstpage}

\appendix

%\counterwithin{figure}{section}
\section[]{SALT Observation Dates}
\label{s:appA}
\label{s:dates}
Table \ref{t:salt_obs} provides dates on which SALT obtained data with either RSS or HRS for each star classified in Table\,2. Full details are available in the SALT archives.  

%\counterwithin{figure}{section}
\section[]{Classification Verification}
\label{s:appB}
Figure \ref{f:drclass} compares  automatic classifications obtained by applying the algorithms described in \S\,3.1 with the manual classifications given by \citet{drilling13} (D13). The observational data  are the same for both sets of classification.

\section[]{Model Atmospheres}
\label{s:appC}
\label{s:modelfigs}
Models selected from the grids described in \S\,4.1 are presented and compared.
Figures\, \ref{f:mod_m99} and \ref{f:mod_m00} demonstrate the transition of the theoretical spectrum from helium-poor to helium-rich
($-2 \leq \log y \leq 2$ and from $25 \leq \Teff/{\rm kK} \leq 55$. 
Metal-poor and solar-metallicity {\sc sterne/spectrum} LTE models  are compared with the  {\sc tlusty/synspec} non-LTE zero-metal models of \citet{nemeth12}. All models are computed for a
surface gravity $\log g = 5.75$ characterstic of the majority of stars analysed in this paper.

\section[]{Classification Calibration}
\label{s:appD}
\label{s:class_pars}
D13 provides approximate calibrations between spectral type and \Teff\ and between luminosity class and $\log g$ for restricted subsets of their sample. 
Figure \ref{f:classpars} compares effective temperature, surface gravity and helium-to-hydrogen ratio ($\Teff, \log g, \log y$) as determined in \S\,4 with the corresponding spectral type, luminosity and helium classes  as determined in \S\,3. 

\section[]{Spectral Atlas}
\label{s:appE}
\label{s:atlas}
Figs.\,\ref{f:fit01}  - \ref{f:fit20}.7 show the complete ensemble of reduced survey spectra and best-fit models 
arranged by subsection in \S\,5 in the same format as Fig.\,6.

\bibliographystyle{mnras}
\bibliography{ehe}

\renewcommand\thefigure{A.\arabic{figure}} 
\renewcommand\thetable{A.\arabic{table}} 
\begin{table}[h]
    \caption{Observation dates in the form {\tt yyyymmdd}.  }
    \label{t:salt_obs}
    \setlength{\tabcolsep}{2pt}
    \centering
    \begin{tabular}{P{25mm}P{27mm}P{27mm}}
    \hline
Name & RSS Dates & HRS Dates \\
\hline
Ton S 144 &  20181101 &  20180611\\
Ton S 148 &  20191101 &  20180616 20180705 20180722 20190619 20190715\\
EC 00468$-$5440 &  20180801 20190617 20191114 &  \\
%% BPS 22946$-$0005 &  20180604 20190618 &  \\
SB 705 &  20180823 &  20170706\\
LB  3229 &  20180823 &  20170716\\
PG 0208+016 &  20180823 20190717 &  20170713\\
LB 1630 &  20190715 &  \\
PG 0240+046 &  20180729 20190803 &  20171106 20181118 20191103 20191104\\
LB 3289 &  20180729 20190714 &  20170827 20171028 20171105 20181118 20181119 20191103\\
EC 02523$-$6934 &  20190714 20191031 20191101 &  \\
EC 02527$-$7111 &  20190814 20191117 &  \\
PHL 1466 &  20190817 &  \\
EC 03505$-$6929 &  20190917 &  \\
EC 04013$-$4017 &  20180730 20190717 20191108 &  20181118 20191102 20191105\\
GLX J04111$-$0048 &  20180817 &  20170116 20170906 20190916 20191012\\
EC 04110$-$1348 &  20190727 &  \\
HE 0414$-$5429 &  20191108 &  \\
2M 0420+0120 &  20191103 &  20170113 20181117 20181118\\
HE 0421$-$5415 &  20190717 &  \\
EC 04271$-$2909 &  20191107 &  \\
EC 04281$-$4738 &  20190907 &  \\
EC 04299$-$1651 &  20180801 20190814 &  \\
LB 1741 &  20191004 &  \\
BPS 29520$-$0048 &  20190814 &  \\
HE 0440$-$3211 &  20190917 &  \\
EC 04517$-$3706 &  20181110 &  \\
%% EC 05048$-$2516 &  20180823 &  \\
GLX J05138$-$1944 &  20190814 &  \\
Ton S 415 &  20180823 20191102 &  20170316\\
EC 05242$-$2900 & 20200208 & \\
GLX J05580$-$2927 &  20191207  &  \\
EC 05593$-$5901 &  20180823 20190917 &  \\
GLX J06126$-$2712 &  20181216 &  20170909\\
GLX J07076$-$6222 &  20180915 20191010 &  20180414\\
GLX J07158$-$5407 &   &  20170124\\
GLX J07581$-$0432 &  20181024 20181122 &  20161115 20161126\\
UVO 0832$-$01 &  20191115 &  \\
GLX  J08454$-$1214 &  20191103 20191116 &  20170314 20170315\\
PG 0902+057 &  20181217 &  20170126\\
UVO 0904$-$02 &  20191210 &  \\
LSS 1274 &  20191127 &  \\
EC 09557$-$1551 &  20181123 &  \\
PG 0958$-$119 &  20191210 &  \\
EC 10475$-$2703 &  & 20170602 \\
EC 10479$-$2714 &  20181214 &  \\
EC 11236$-$1945 &  20181214 &  \\
PG 1127+019 &  20190105 20190420 &  20170119 20180109 20180128\\
PG 1220$-$056 &  20200111 &  \\
PG 1230+067 & 20200124 & 20170131 \\
EC 12349$-$2824 &  20190117 &  \\
EC 12420$-$2732 &  20190104 &  \\
PG 1318+062 &  20190330 20190620 &  20190206\\
EC 13290$-$1933 &  20180612 &  \\
GLX J14258$-$0432 &   &  20170416 20170514\\
PG 1455$-$069 &  20180628 20180824 &  \\
GLX J15235$-$1817 &  20180605 &  20170711\\

    \hline
    \end{tabular}
\end{table}
\addtocounter{table}{-1}

\begin{table}
    \caption{(contd.)}
    \setlength{\tabcolsep}{2pt}
    \centering
    \begin{tabular}{P{25mm}P{27mm}P{27mm}}
    \hline
Name & RSS Dates & HRS Dates \\
\hline
PG 1528+029 &  20180630 20190430 &  \\
EC 15348$-$1652 &  20180821 20190818 &  \\
PG 1537$-$046 &  20190501 &  \\
PG 1625$-$034 &  20190720 &  \\
GLX J16546+0318 &  20180630 &  \\
GLX J17051$-$7156 & 20200322 &  20170331 20180428 20180429\\
GLX J18325$-$4744 &   &  20170315 20170703\\
GLX J18372$-$3125 &   &  20170629\\
GLX J18387$-$5409 & 20180516 &  20170415 20170514\\
GLX J18459$-$4138 & 20180526 &  20170316 20170317 20170420 20170516\\
GLX J19059$-$4438 &   &  20170618\\
GLX J19108$-$4417 &  20180516 &  20170506\\
GLX J19111$-$1406 &   &  20170615\\
GLX J19150$-$4235 &  20180824 &  20170704\\
%% GLX J19188$-$3104 &   &  20170705\\
%% GLX J19307$-$3049 &  20190427 &  20180517 20190411\\
%% EC 19277$-$5829 &  20180801 &  \\
GLX J19333$-$2345 &  20180815 20190419 &  \\
GLX J19376$-$4303 &  20180513 &  20170510\\
BPS 22896$-$0128 &  20190615 20191025 20191102 &  \\
EC 19529$-$4430 &  20180516 &  20190323 20190329\\
GLX J20133$-$1201 &  20190709 &  \\
EC 20111$-$3724 &  20190427 &  \\
EC 20111$-$6902 &  20180628 20190424 20191103 20191104 &  \\
GLX J20204$-$1901 &  20180818 20181026 20190713 &  \\
EC 20184$-$3435 &  20180705 &  \\
EC 20187$-$4939 &  20180517 &  20160525\\
GLX J20251$-$0804 &  20180703 &  20170704\\
EC 20221$-$6249 &  20180604 &  20170403\\
EC 20236$-$5703 &  20190619 &  20190403\\
BPS 22940$-$0009 &  20160612 20180504 &  20160622 20160629 20180506 20190424\\
EC 20306$-$5127 &  20180816 &  20170404 20170416 20170614\\
EC 20337$-$2525 &  20190720 &  \\
EC 20450$-$1501 & 20180801  &  20160511\\
EC 20450$-$6947 & 20180727 20180801 & \\
EC 20481$-$5518 &  20190614 20191113 & \\
LS IV$-$14 116 &  20190427 20190501 &  \\
EC 20577$-$5641 &  20180822 &  \\
Ton S 14 &  20190617 &  \\
EC 21077$-$4815 &  20190817 &  \\
BPS 30319$-$0062 &  20180818 &  20190516 20190606\\
EC 21125$-$7013 &  20180627 &  \\
EC 21306$-$4911 &  20190709 &  \\
EC 21416$-$3645 &  20190617 20191109 20191111 &  \\
PHL 149 &  20180822 &  \\
PHL 178 &  20190617 &  \\
PG 2158+082 &  20190618 &  \\
BPS 22956$-$0090 &  20180628 20190521 &  20161031 20161101 20161106\\
BPS 22892$-$0051 &  20180822 &  \\
PB 7124 &  20180725 &  20170704\\
BPS 22875$-$0002 &  20180822 &  20170601\\
PG 2218+051  &  20191029 & \\
% EC 22211$-$2525 &  20180802 &  \\ % not in Tables 2 or 3 ??
EC 22332$-$6837 &  20190617 20191109 &  \\
BPS 22938$-$0044 &  20180604 20180612 20180628 20190514 &  \\
GLX J22565$-$5248 &  20180604 &  20170518 20181115 \\
BPS 22938$-$0073 &  20180821 &  20170519\\
PHL 540 &  20190618 &  \\
Ton S 103 &  20180608 &  20161106 20170923\\
HE 2347$-$4130 &  20190521 &  \\

    \hline
    \end{tabular}
\end{table}
\addtocounter{table}{-1}

\renewcommand\thefigure{B.\arabic{figure}} 
\renewcommand\thetable{B.\arabic{table}} 
\begin{figure*}
\begin{center}
\includegraphics[width=0.46\linewidth]{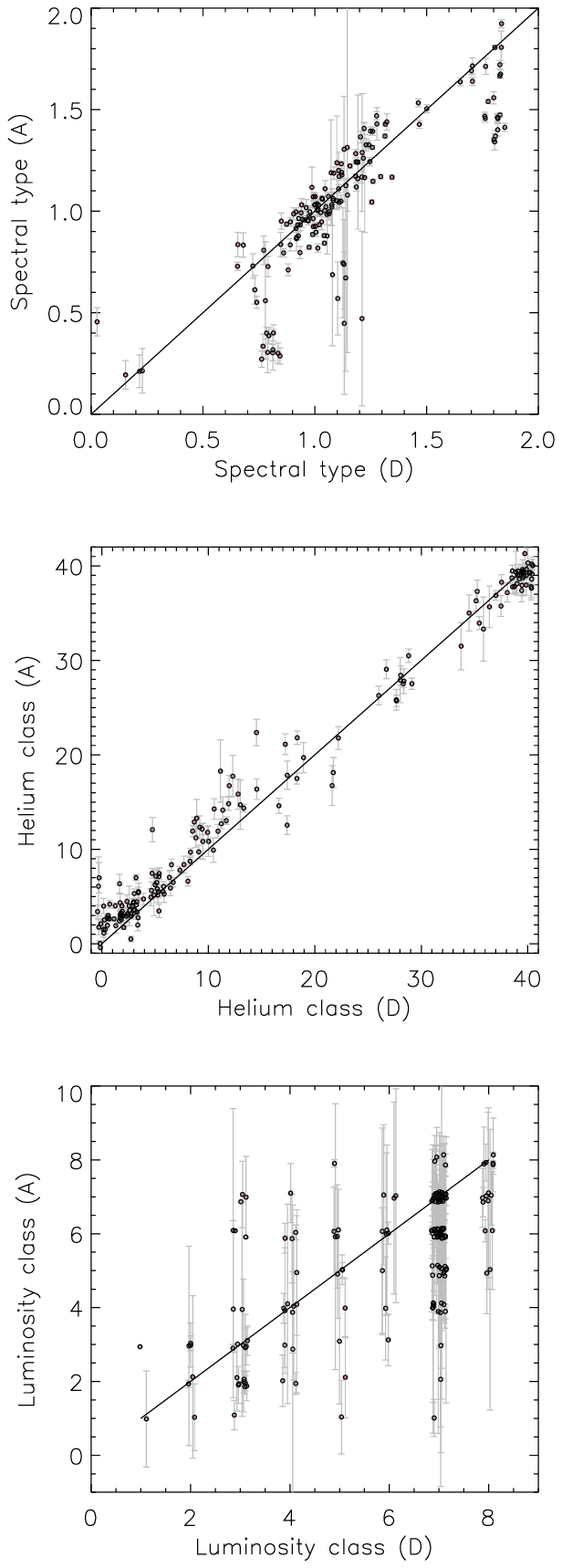}
\includegraphics[trim=0cm 6cm 0cm 0cm, clip, width=0.46\linewidth,]{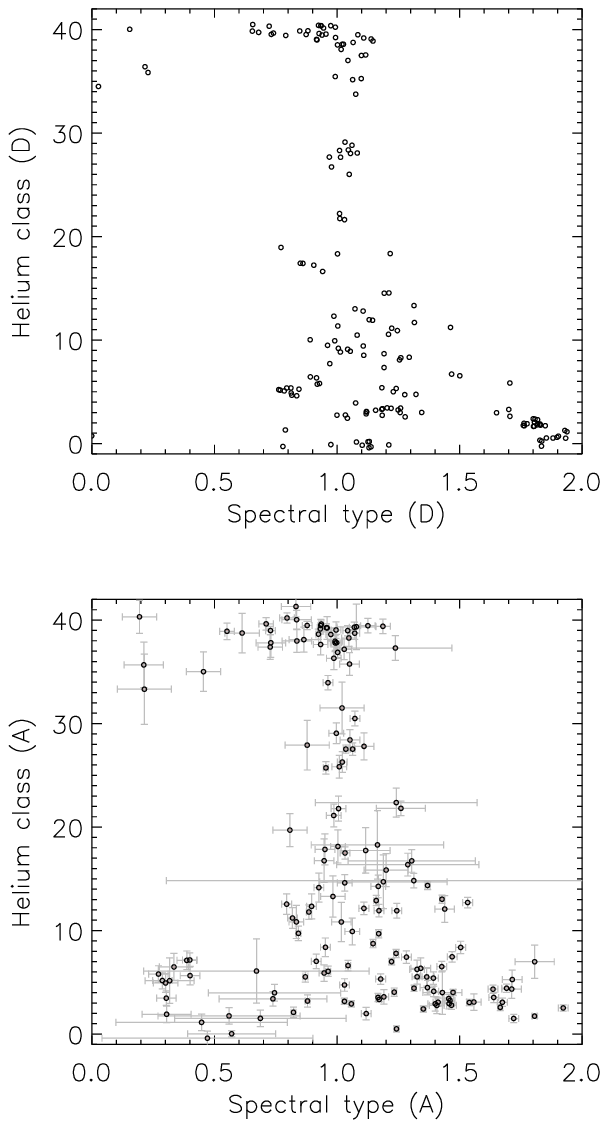}
\caption{Left: Comparison of automatic classifications (A) with D13 classifications (D) (Drilling sample). For visibility, all classification plots include a random jitter on intervals of $\pm0.05, \pm0.15$ and $\pm 0.5$ in SpT, LC and He respectively. 
Right: Comparison of the SpT -- He diagram  for D13 (top: D) and automatic (bottom: A) classifications (Drilling sample). If multiple spectra exist for the same star, each has been classified separately in the current exercise. } 
\label{f:drclass}
\end{center}
\end{figure*}
\addtocounter{figure}{-1}

\renewcommand\thefigure{C.\arabic{figure}} 
\renewcommand\thetable{C.\arabic{table}} 
\begin{landscape}
\begin{figure}
\includegraphics[width=0.98\linewidth]{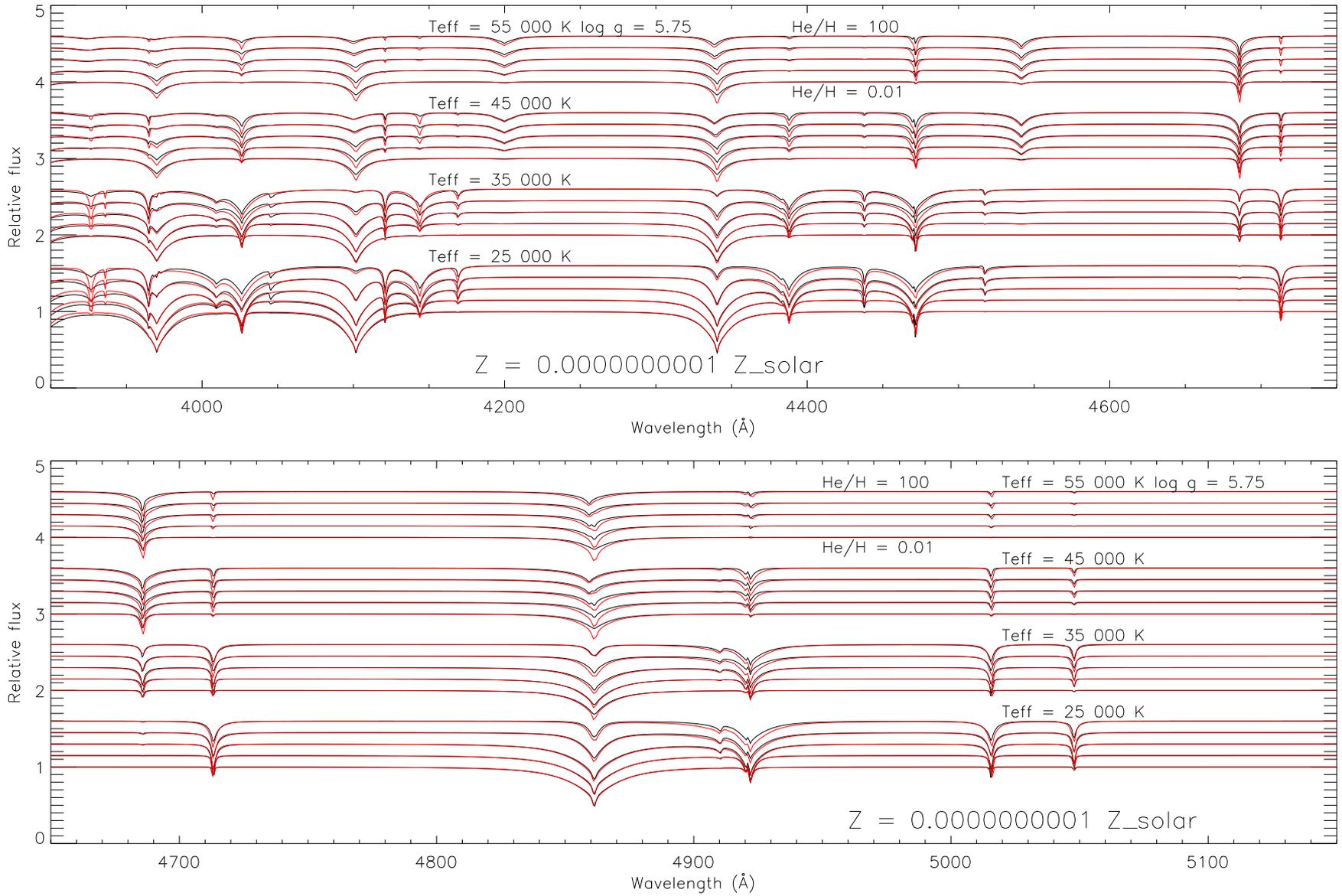}
\caption{A comparison of theoretical spectra for selected model atmospheres extracted from the grids described in \S\,4.1. {\sc sterne/spectrum} LTE models with negligible abundance of elements heavier than helium ($\log Z / Z_{\odot} = -10$: black) are compared with {\sc tlusty/synspec}
non-LTE zero-metal models (red).}
\label{f:mod_m99}
\end{figure}
\end{landscape}

\begin{landscape}
\begin{figure}
\includegraphics[width=0.98\linewidth]{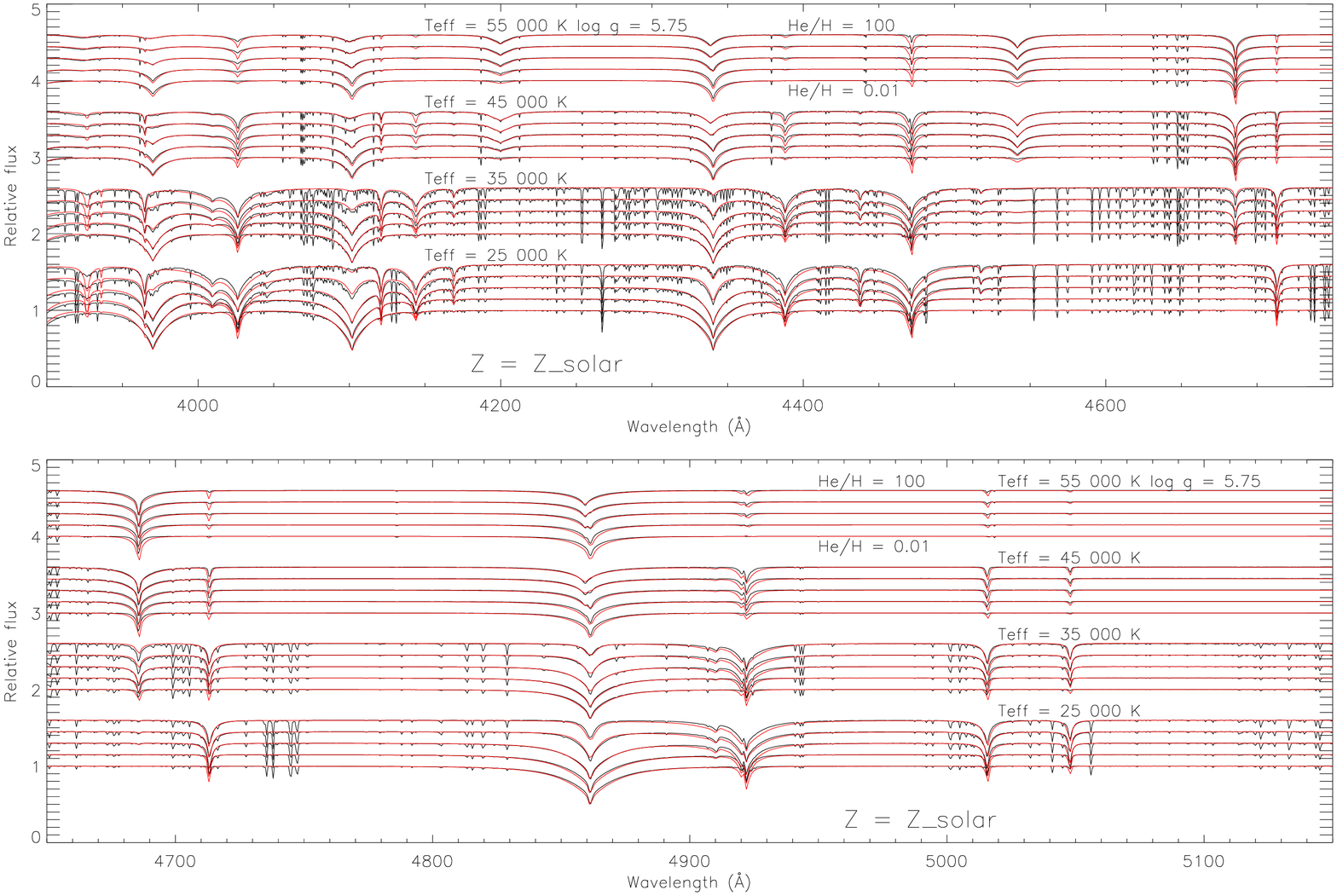}
\caption{As Fig.\,\ref{f:mod_m99} but for {\sc sterne/spectrum LTE} models with solar metallicity (black).}
\label{f:mod_m00}
\end{figure}
\end{landscape}
\addtocounter{figure}{-2}

\renewcommand\thefigure{D.\arabic{figure}} 
\renewcommand\thetable{D.\arabic{table}} 
\begin{figure*}
\begin{center}
\includegraphics[clip, width=0.48\linewidth]{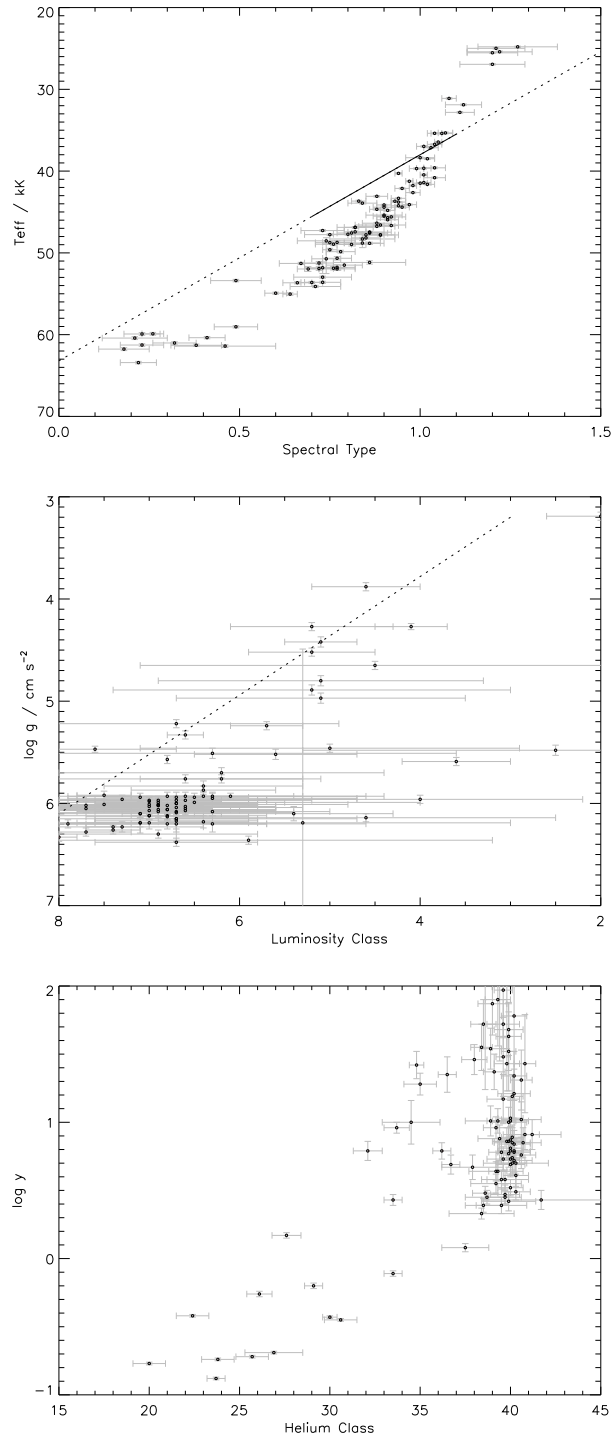}
\caption{\Teff, $\log g$ and $\log y$ from Fig.\,3 are compared with spectral type, luminosity and helium classes from Table\,2 (top to bottom).  
 The dotted lines approximately represent calibrations presented in D13, shown as solid over regions where they are considered valid. } 
\label{f:classpars}
\end{center}
\end{figure*}
\addtocounter{figure}{-1}

\renewcommand\thefigure{E.\arabic{figure}} 
\renewcommand\thetable{E.\arabic{table}} 

\begin{figure*}
\includegraphics[width=0.85\linewidth]{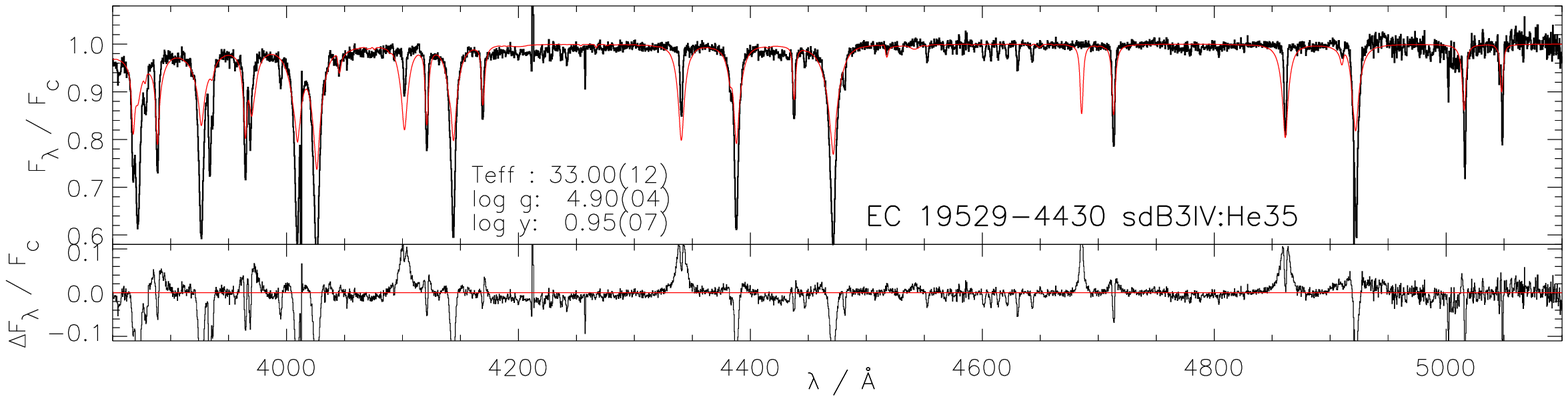}\\
\includegraphics[width=0.85\linewidth]{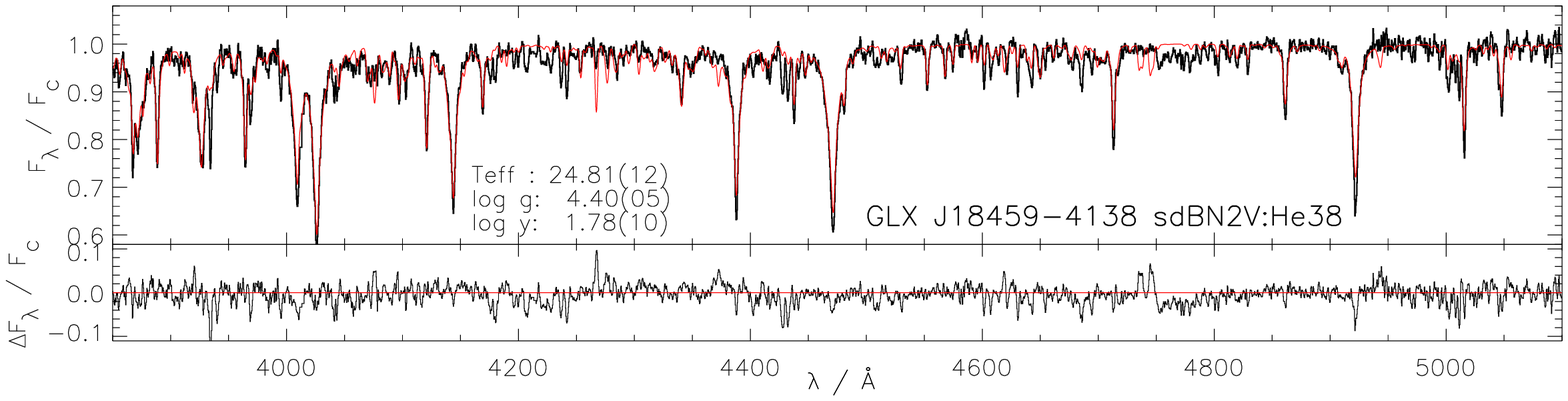}\\
\includegraphics[width=0.85\linewidth]{spectra/EC20236-5703.eps}\\
\includegraphics[width=0.85\linewidth]{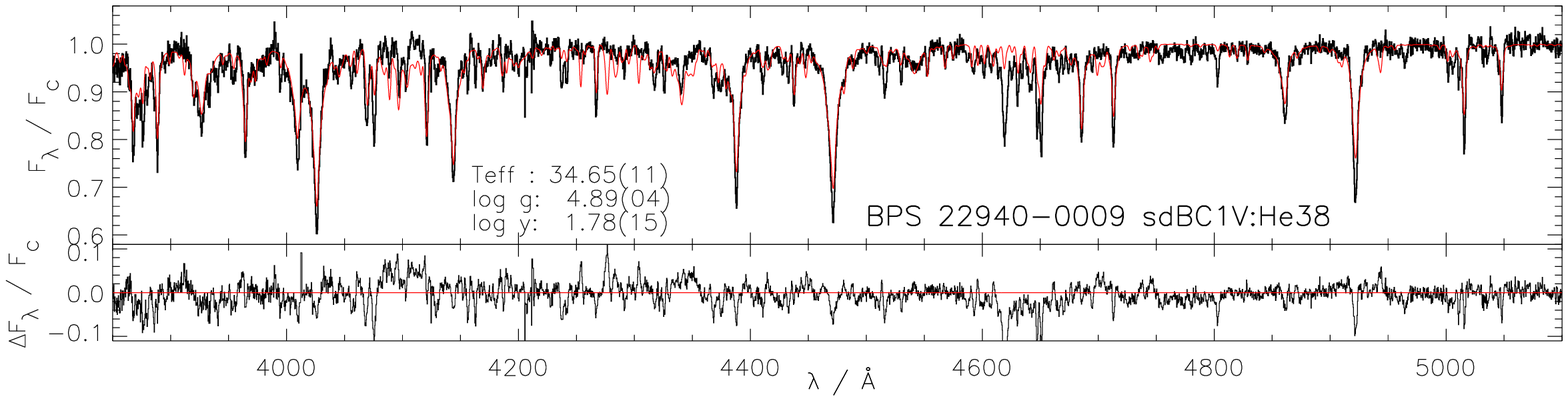}\\
\includegraphics[width=0.85\linewidth]{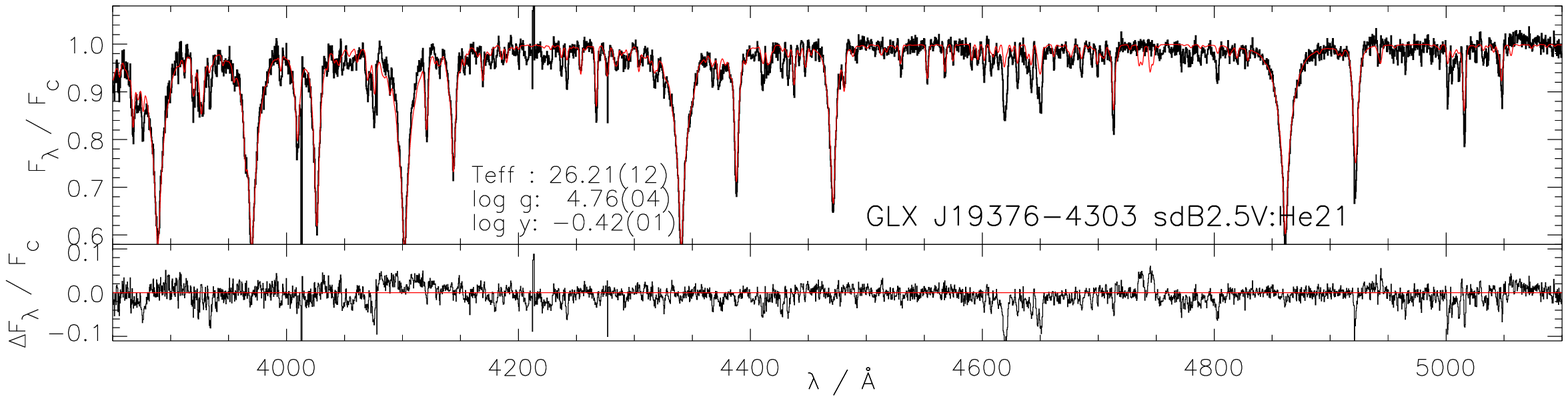}\\
\includegraphics[width=0.85\linewidth]{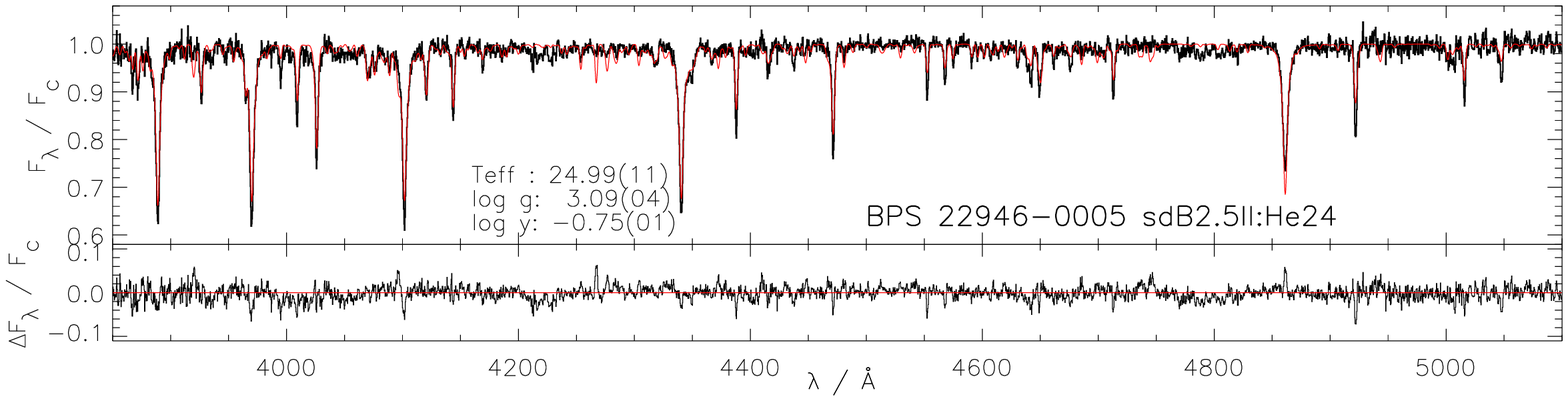}\\
\caption{As Fig.\,6 for the stars discussed in \S\,5.1: SpT = sdB1 -- sdB3, LC $\lesssim$V, He $>20$}
\label{f:fit01}
\end{figure*}

\begin{figure*}
\includegraphics[width=0.85\linewidth]{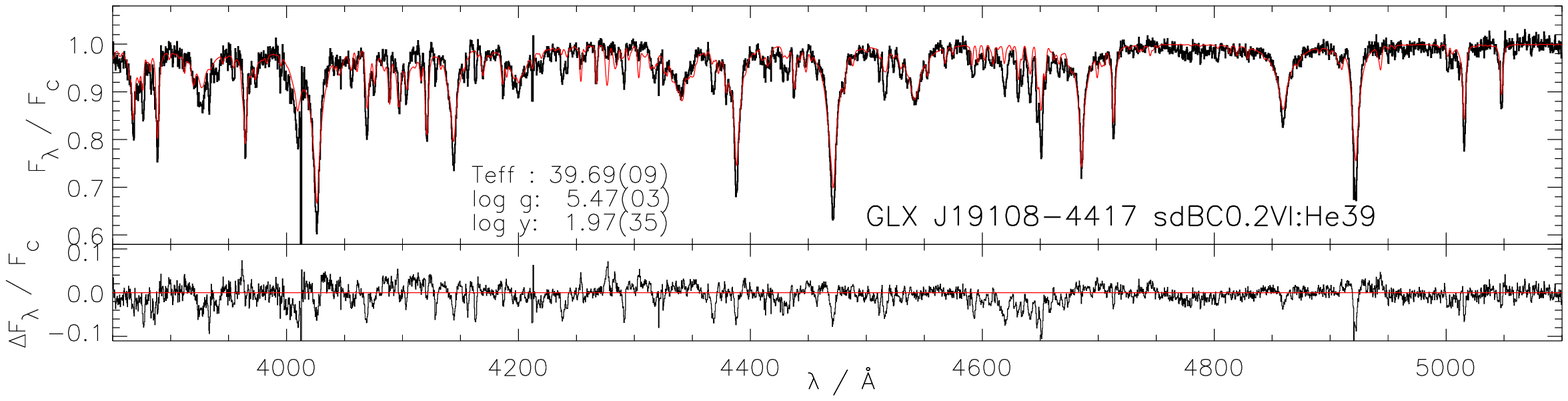}\\
\includegraphics[width=0.85\linewidth]{spectra/EC20111-6902.eps}\\
\includegraphics[width=0.85\linewidth]{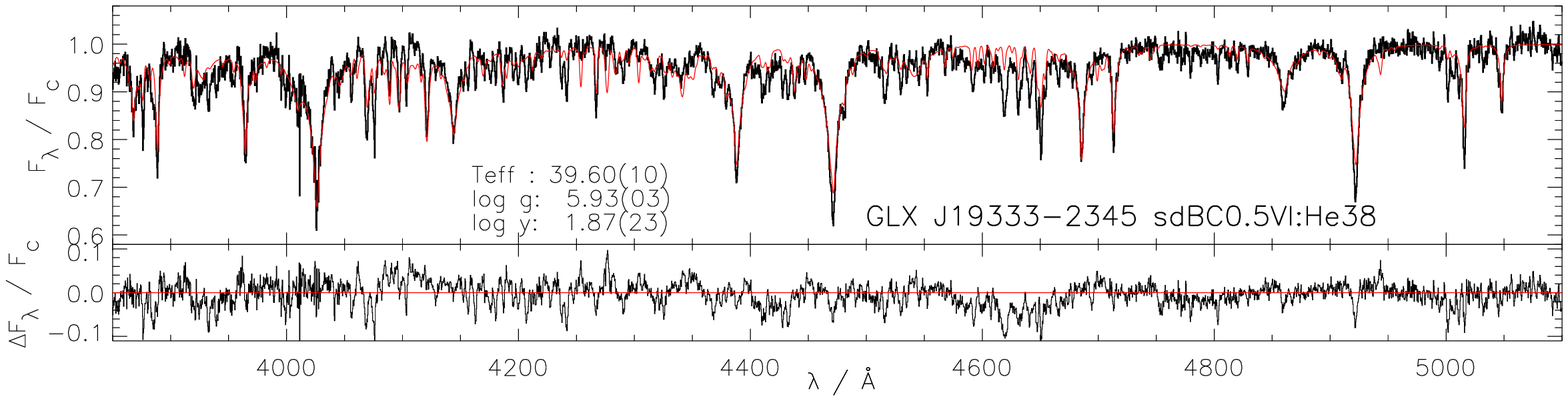}\\
\includegraphics[width=0.85\linewidth]{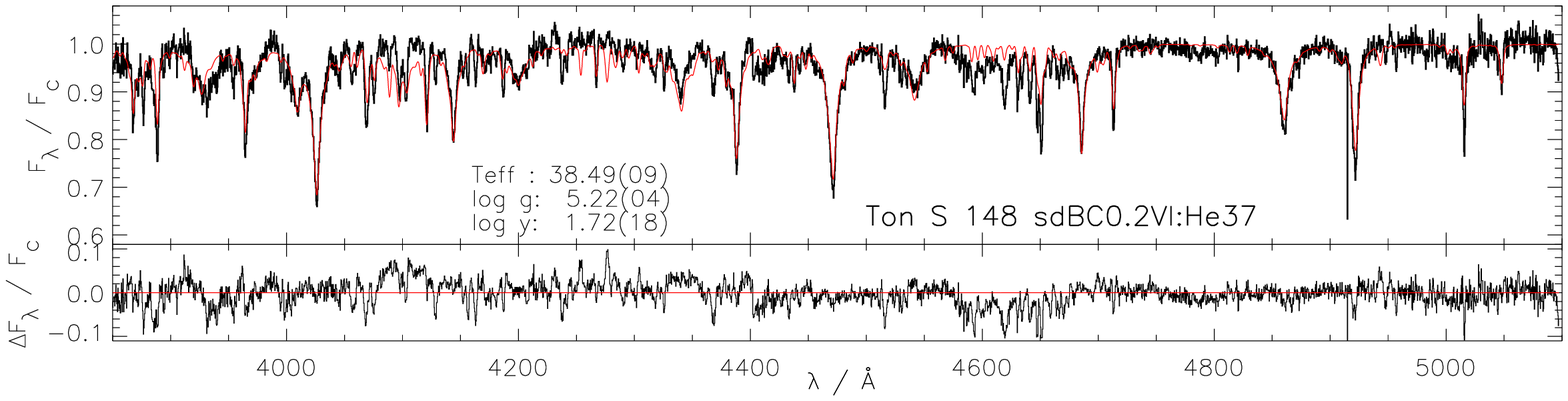}\\
\includegraphics[width=0.85\linewidth]{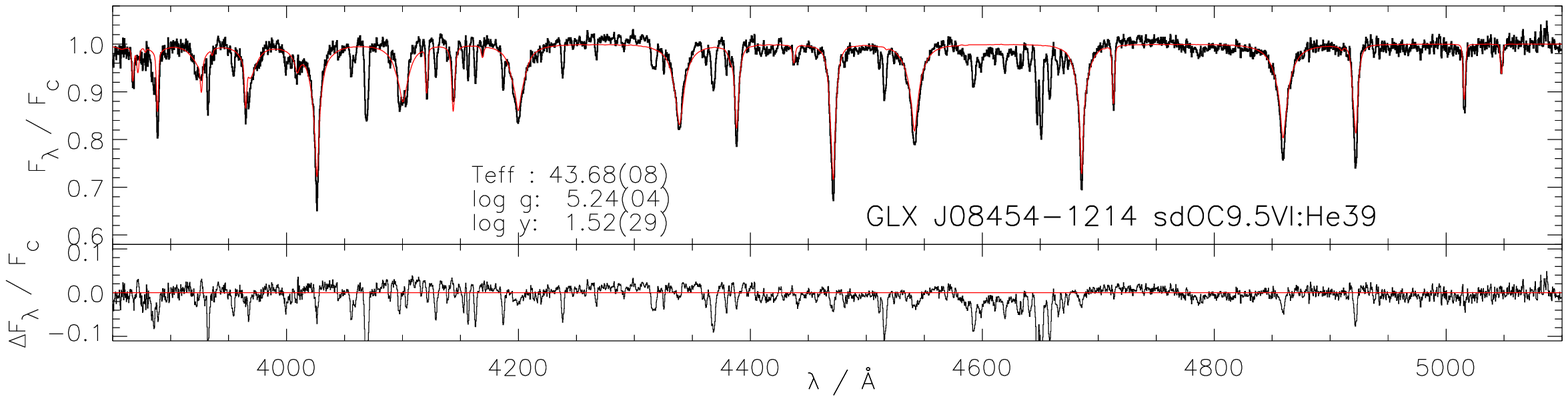}\\
\includegraphics[width=0.85\linewidth]{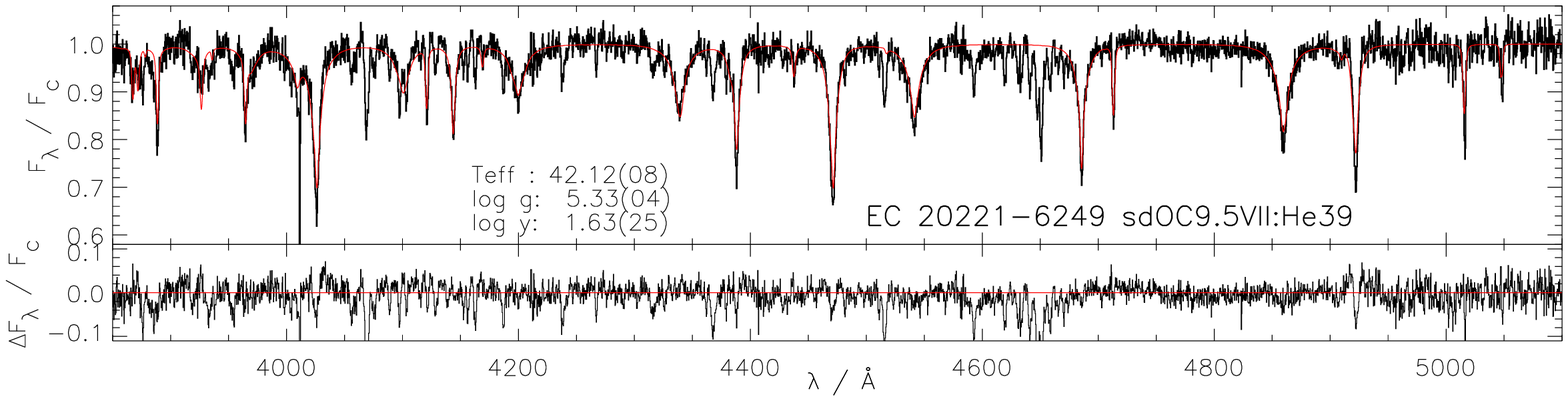}\\
\caption{As Fig.\,6 for the stars discussed in \S\,5.2: SpT = sdO9.5 -- sdB1, LC $\approx$ V -- VI, He $> 30$}
\label{f:fit02}
\end{figure*}

\begin{figure*}
\includegraphics[width=0.85\linewidth]{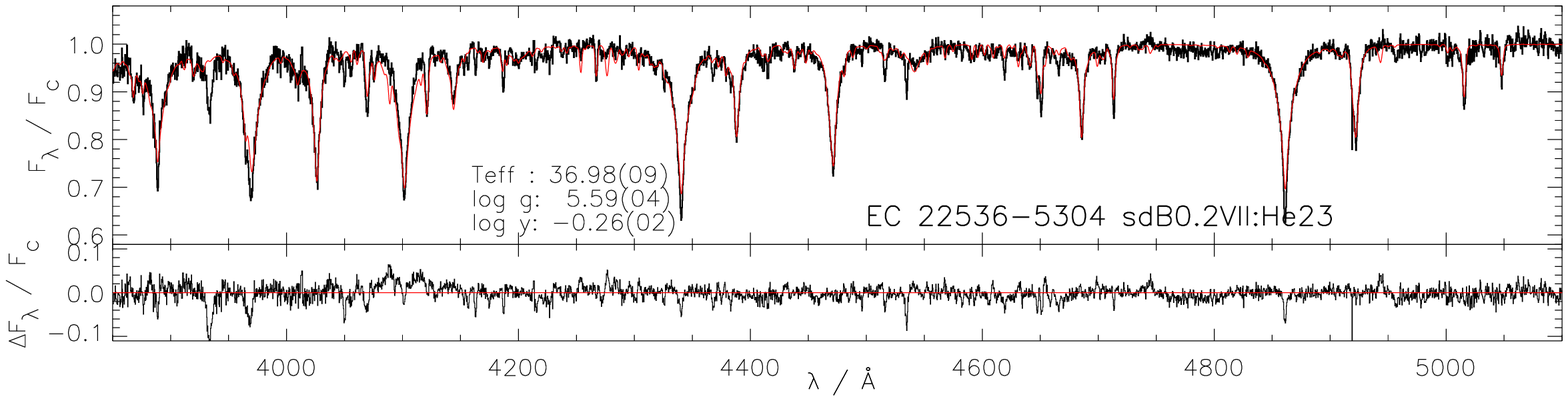}\\
\includegraphics[width=0.85\linewidth]{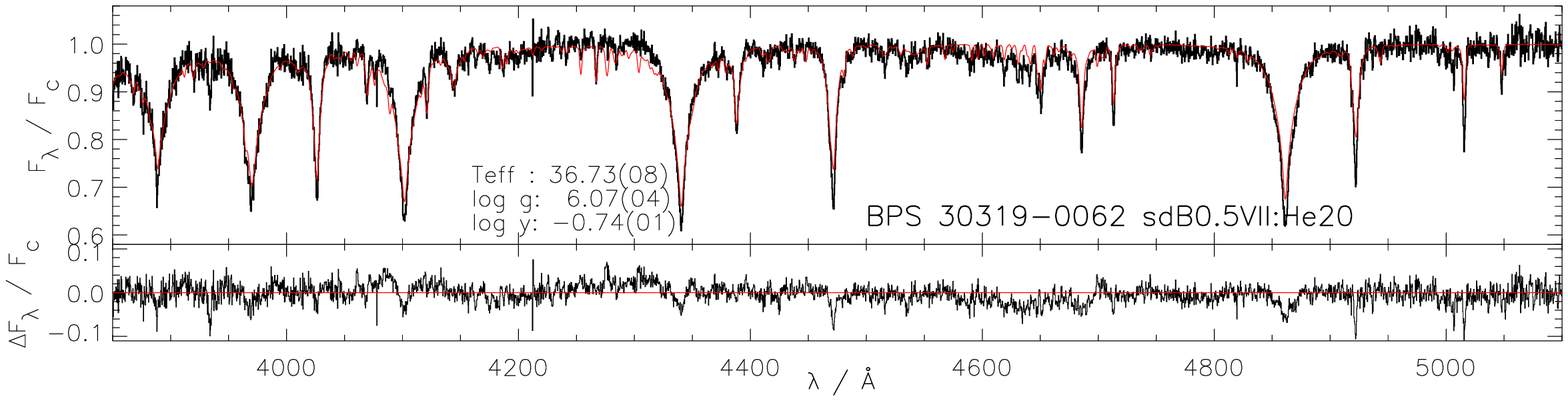}\\
\includegraphics[width=0.85\linewidth]{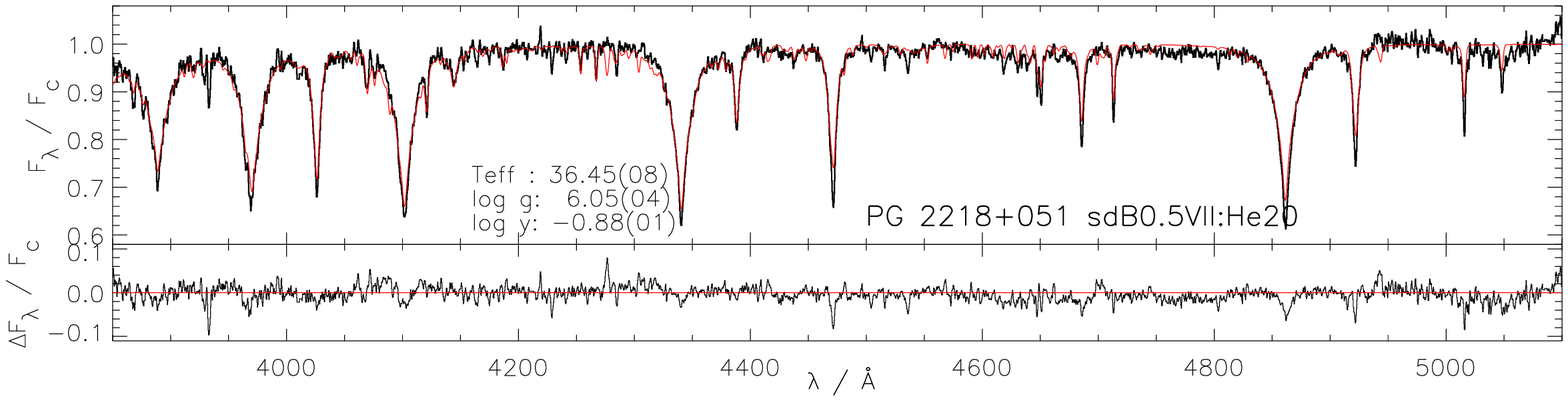}\\
\includegraphics[width=0.85\linewidth]{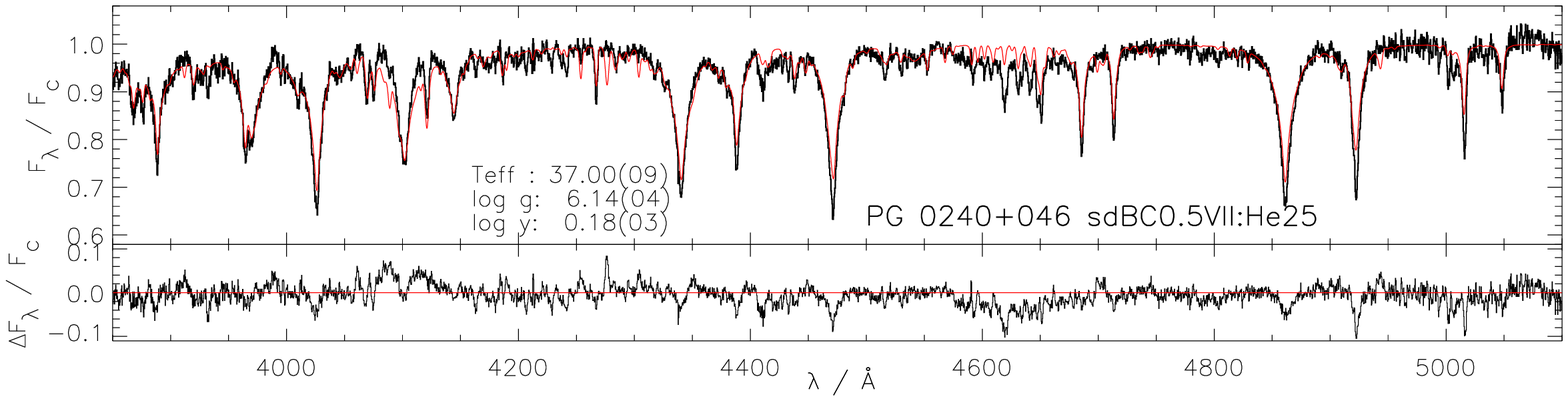}\\
\includegraphics[width=0.85\linewidth]{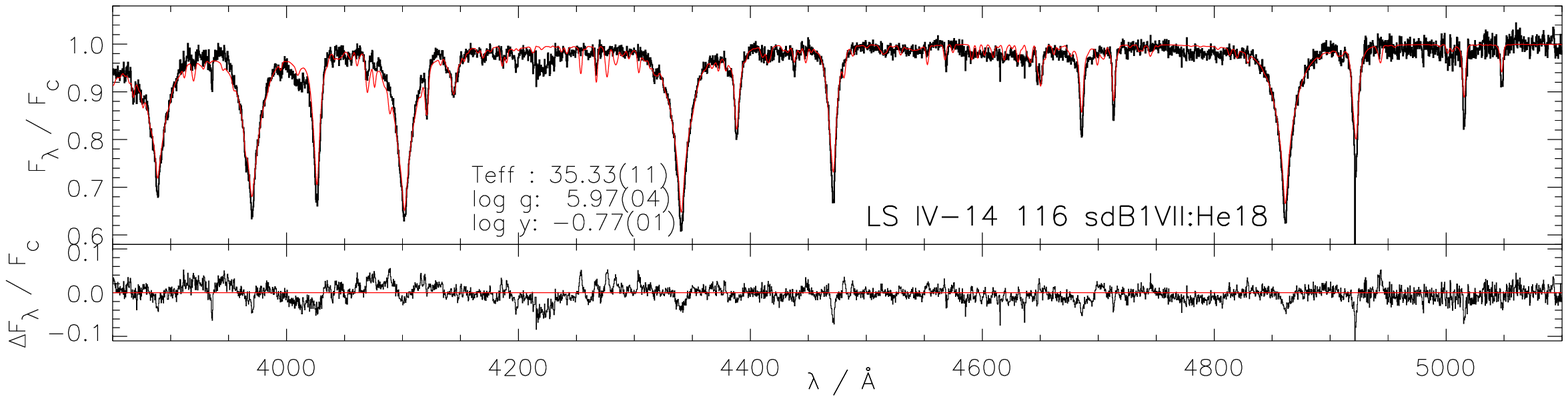}\\
\includegraphics[width=0.85\linewidth]{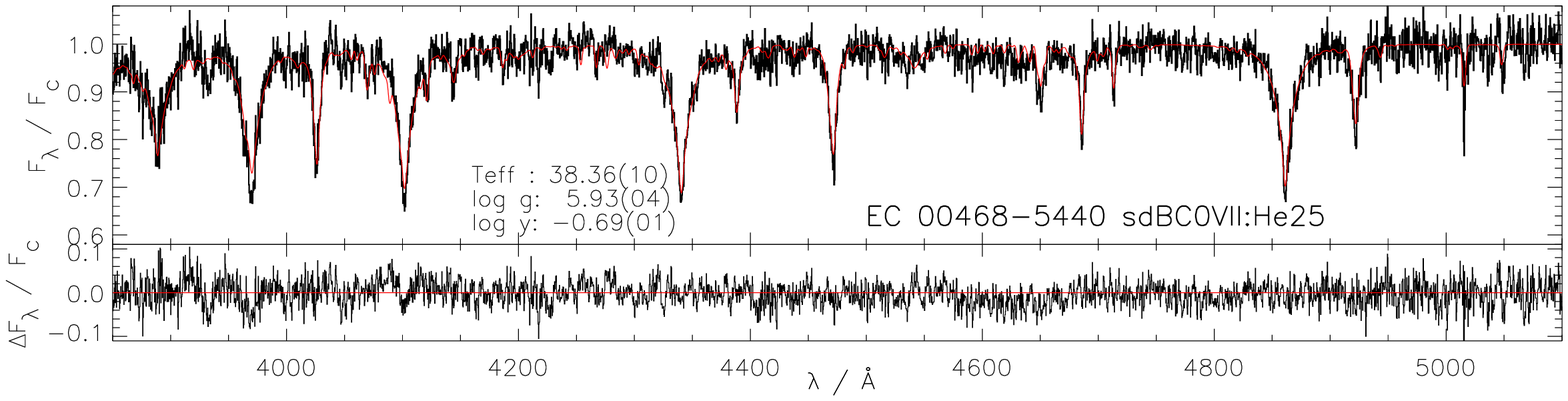}\\
\caption{As Fig.\,6 for the stars discussed in \S\,5.3: sdB0 $\lesssim$ Sp $\lesssim$ sdB1, He $<25$} 
\label{f:fit03}
\end{figure*}

\begin{figure*}
\includegraphics[width=0.85\linewidth]{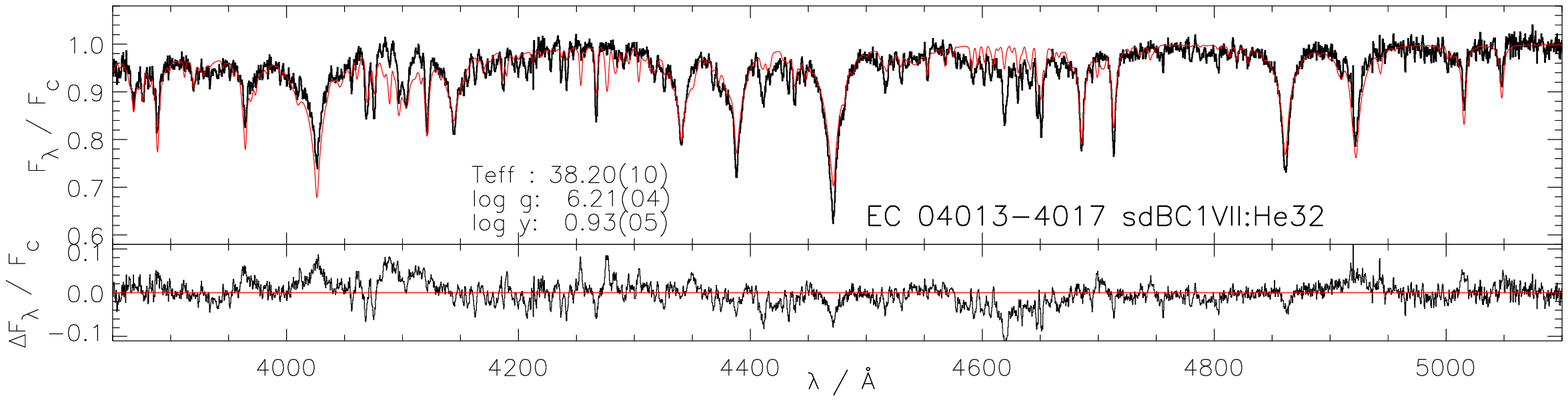}\\
\includegraphics[width=0.85\linewidth]{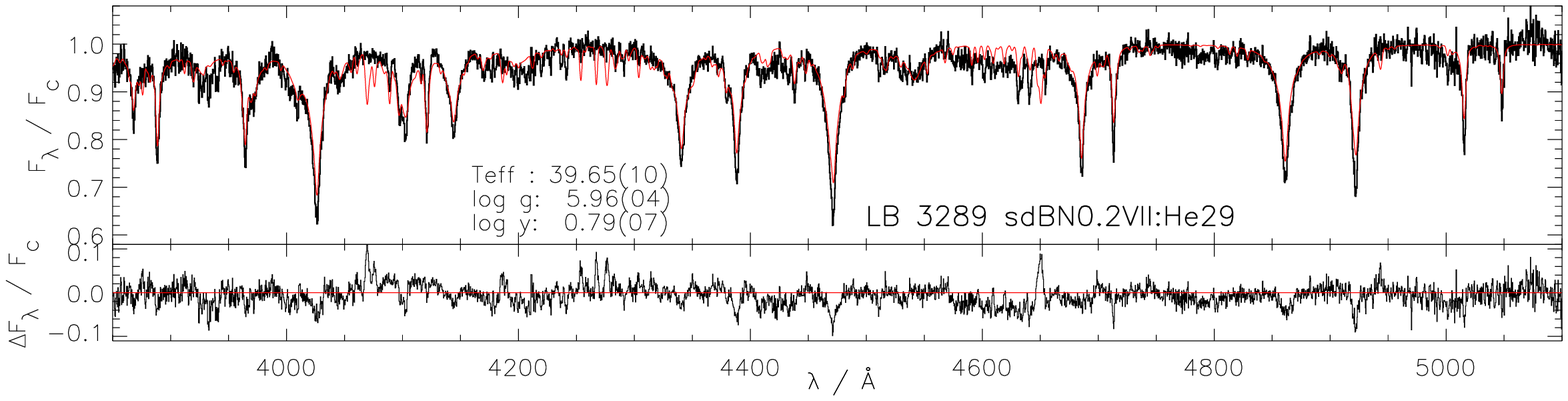}\\
\includegraphics[width=0.85\linewidth]{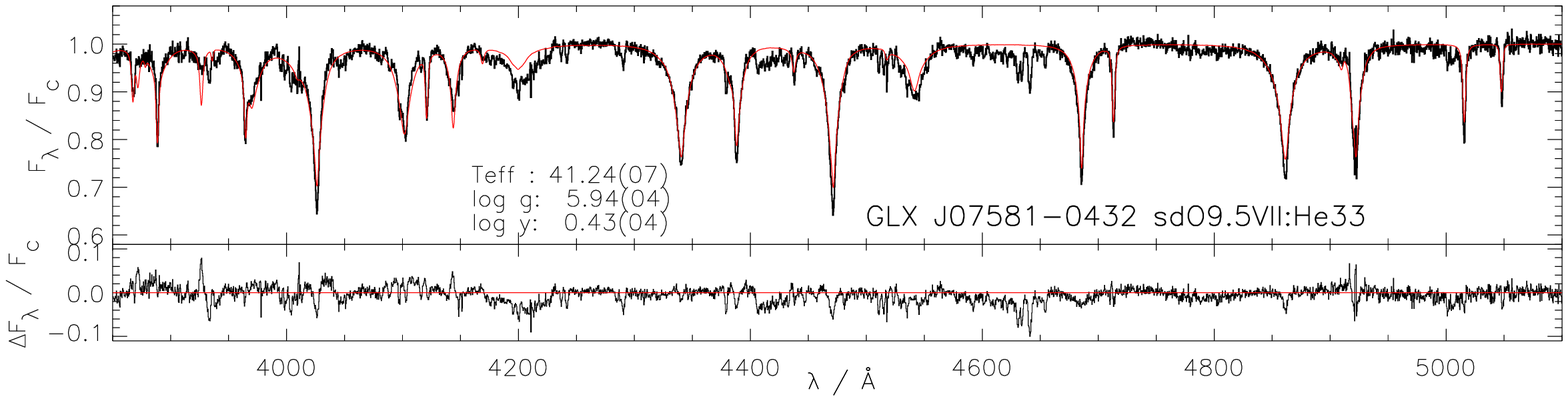}\\
\includegraphics[width=0.85\linewidth]{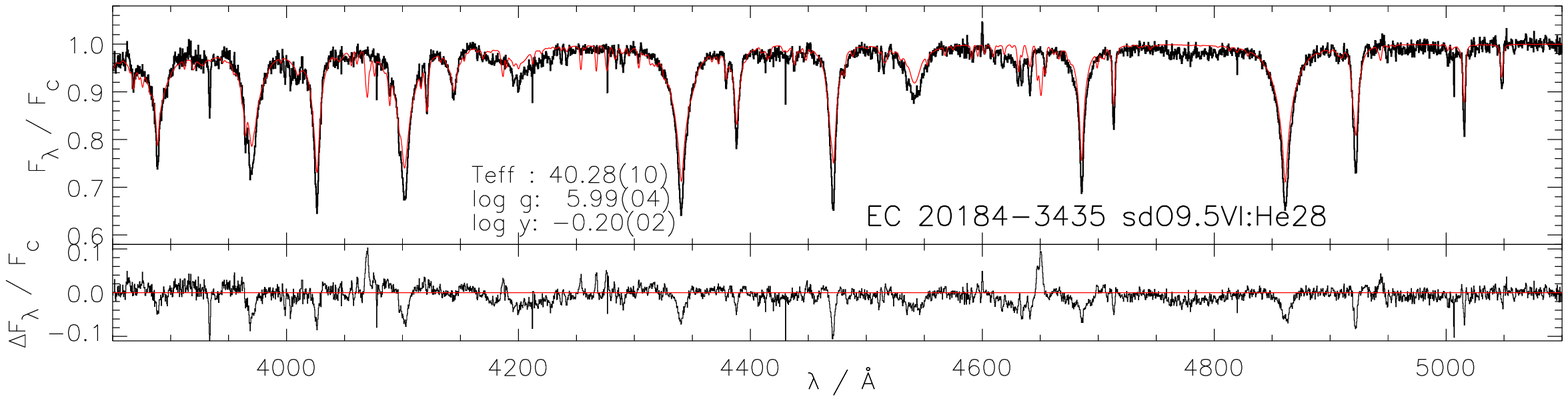}\\
\caption{As Fig.\,6 for the stars discussed in \S\,5.4: Sp = sdO8 -- sdB1, $25 <$ He $<35$ } 
\label{f:fit04}
\end{figure*}

\begin{figure*}
\includegraphics[width=0.85\linewidth]{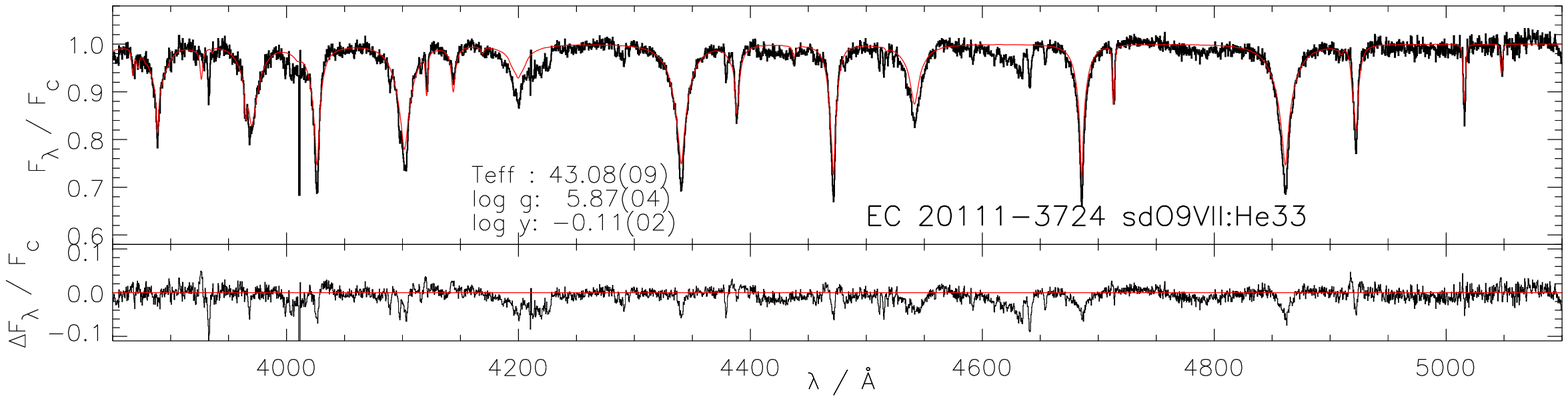}\\
\includegraphics[width=0.85\linewidth]{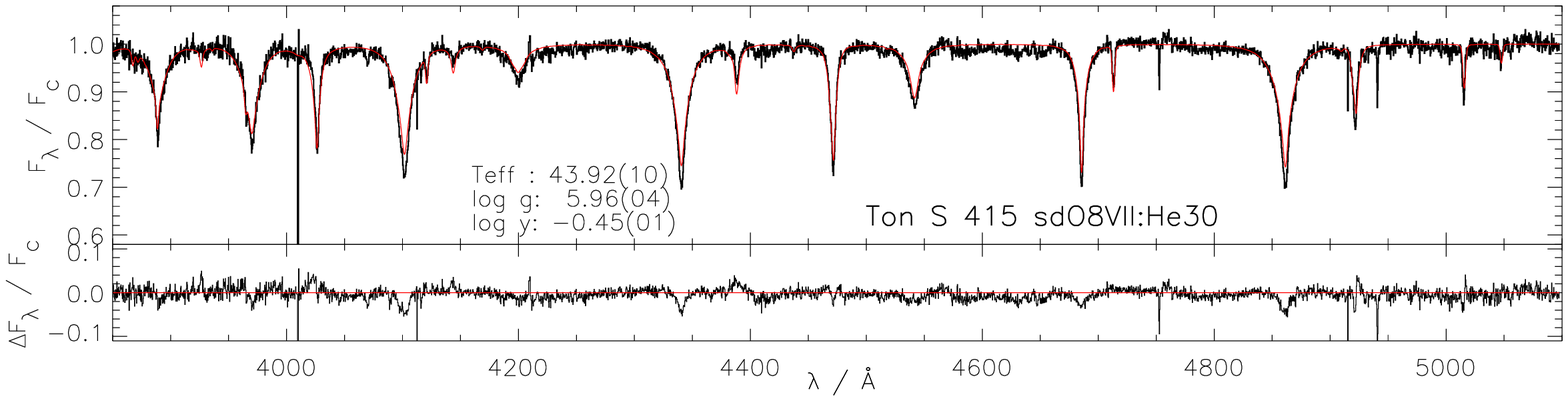}\\
\includegraphics[width=0.85\linewidth]{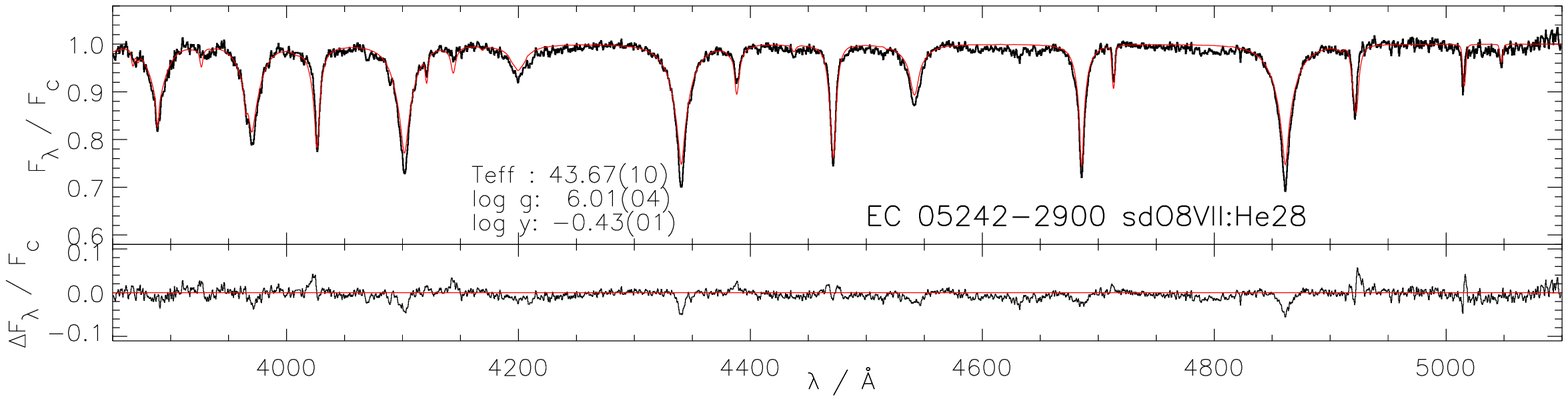}\\
\includegraphics[width=0.85\linewidth]{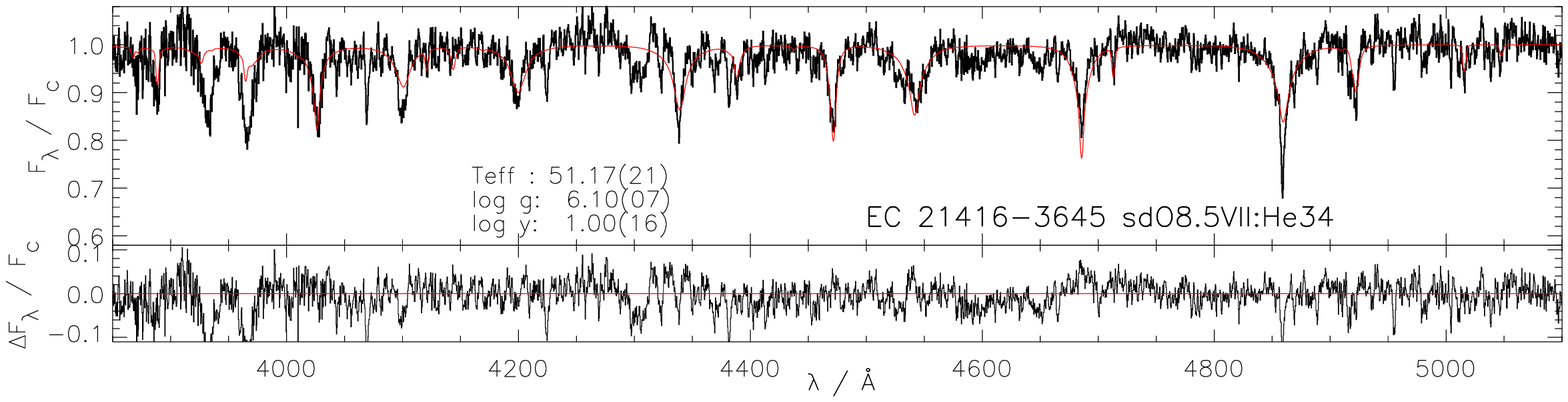}\\
\contcaption{}
\label{f:fit05}
\end{figure*}

\begin{figure*}
\includegraphics[width=0.85\linewidth]{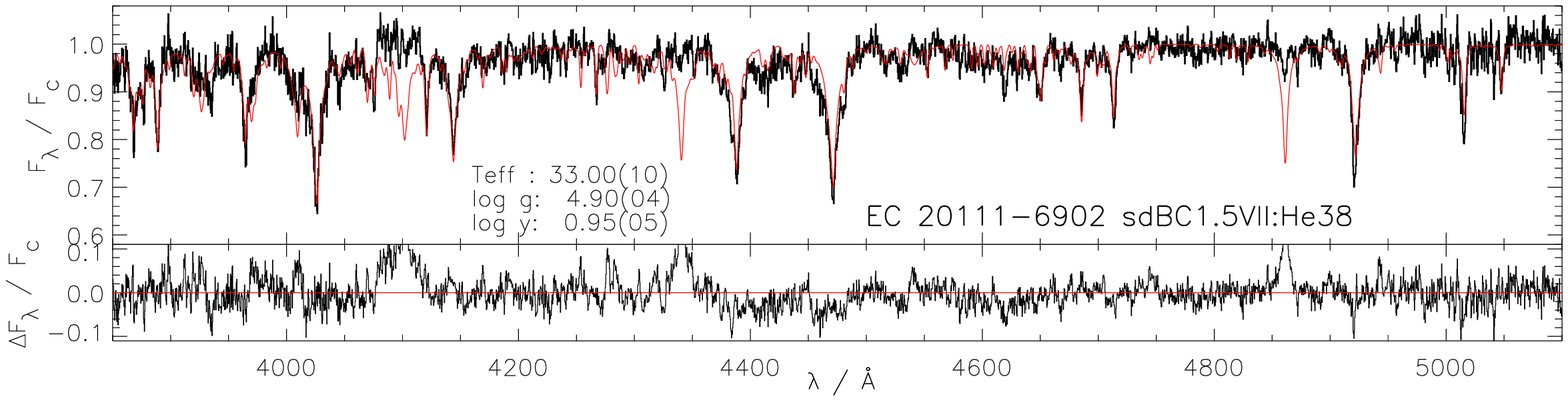}\\
\includegraphics[width=0.85\linewidth]{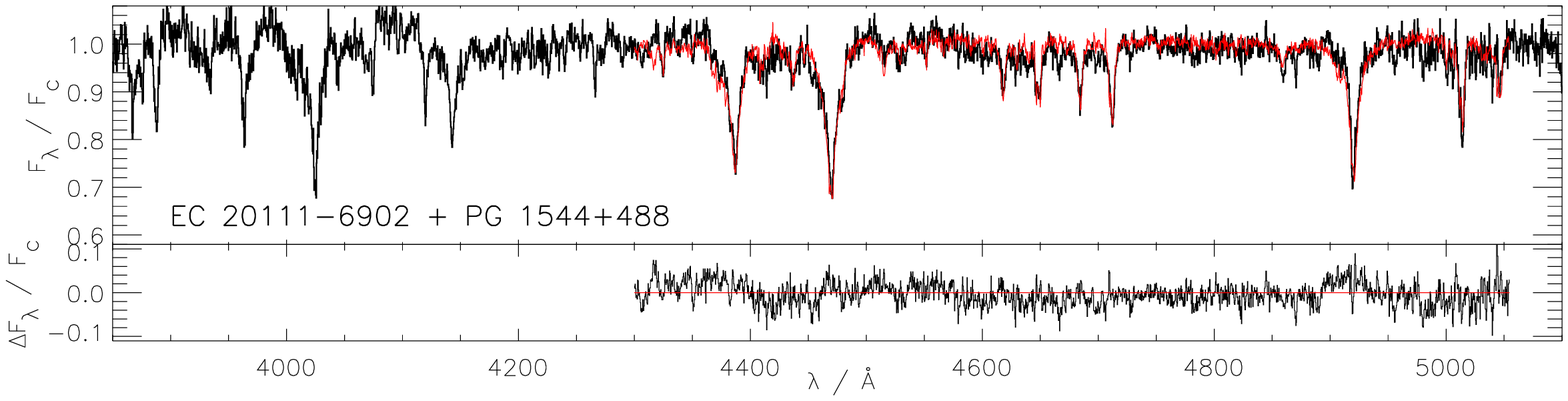}\\
\includegraphics[width=0.85\linewidth]{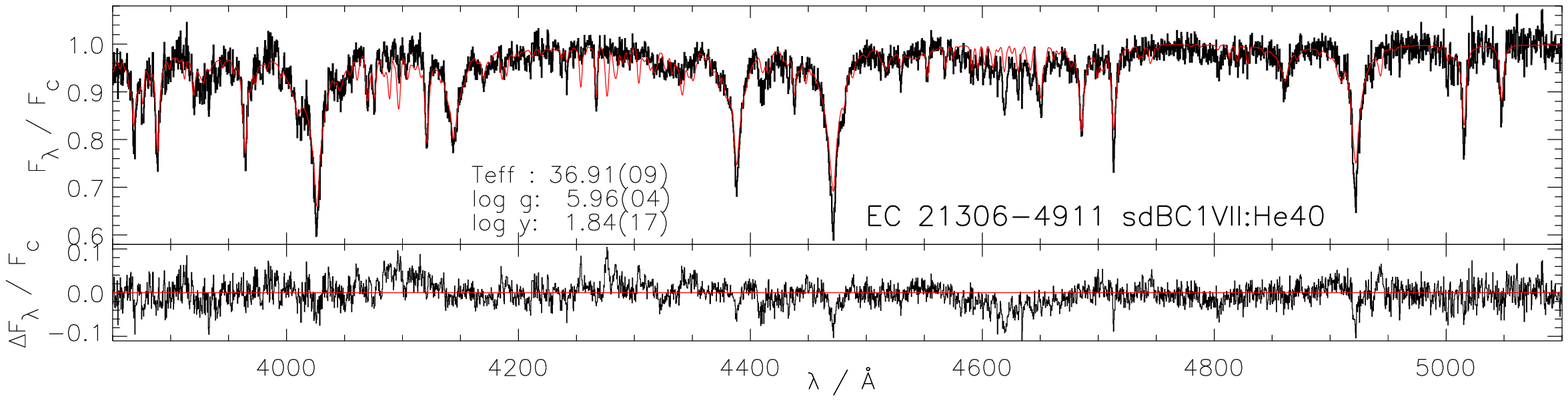}
\caption{As Fig.\,6 for the stars discussed in \S\,5.6: broad lines}
\label{f:fit06}
\end{figure*}

% hot stars

\begin{figure*}
\includegraphics[width=0.85\linewidth]{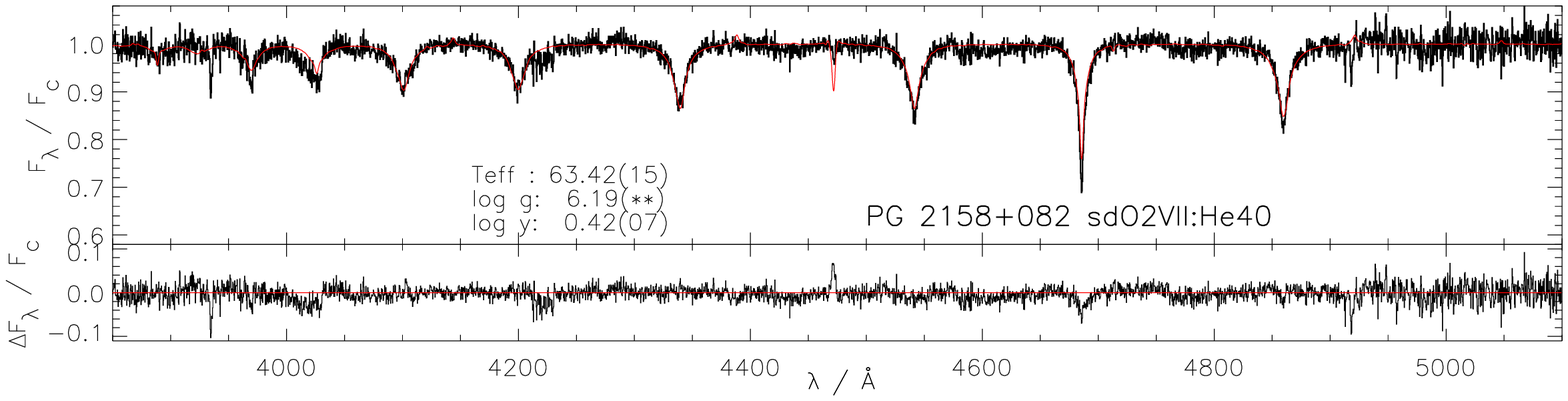}\\ 
\includegraphics[width=0.85\linewidth]{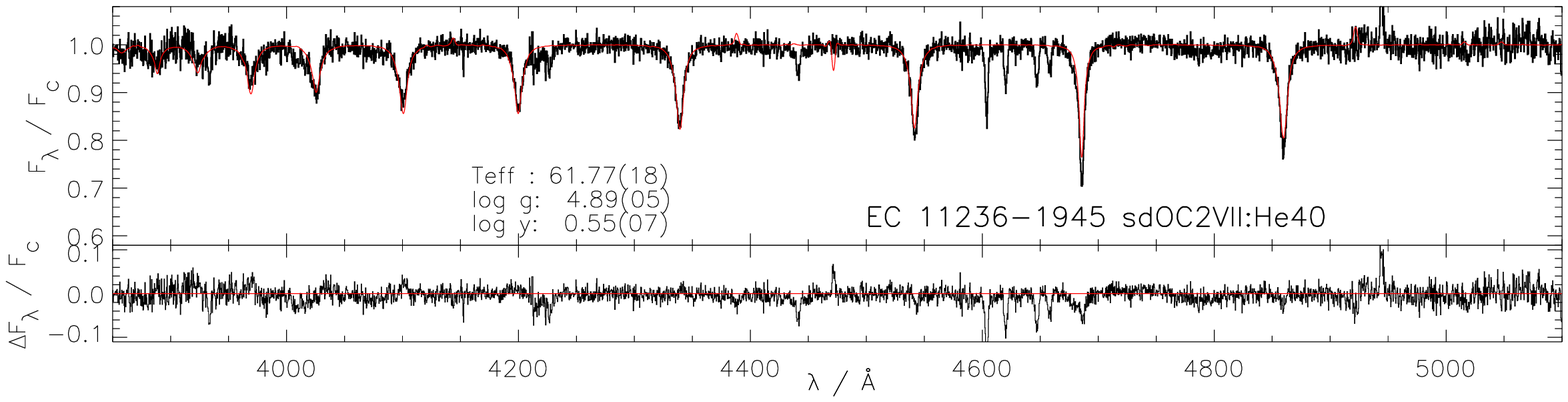}\\
\includegraphics[width=0.85\linewidth]{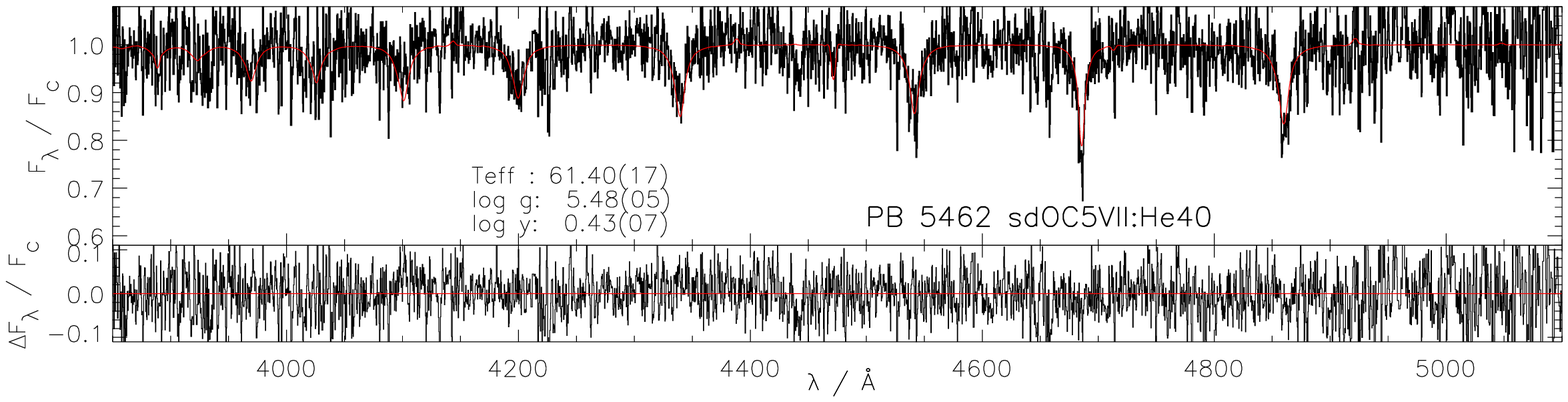}\\
\includegraphics[width=0.85\linewidth]{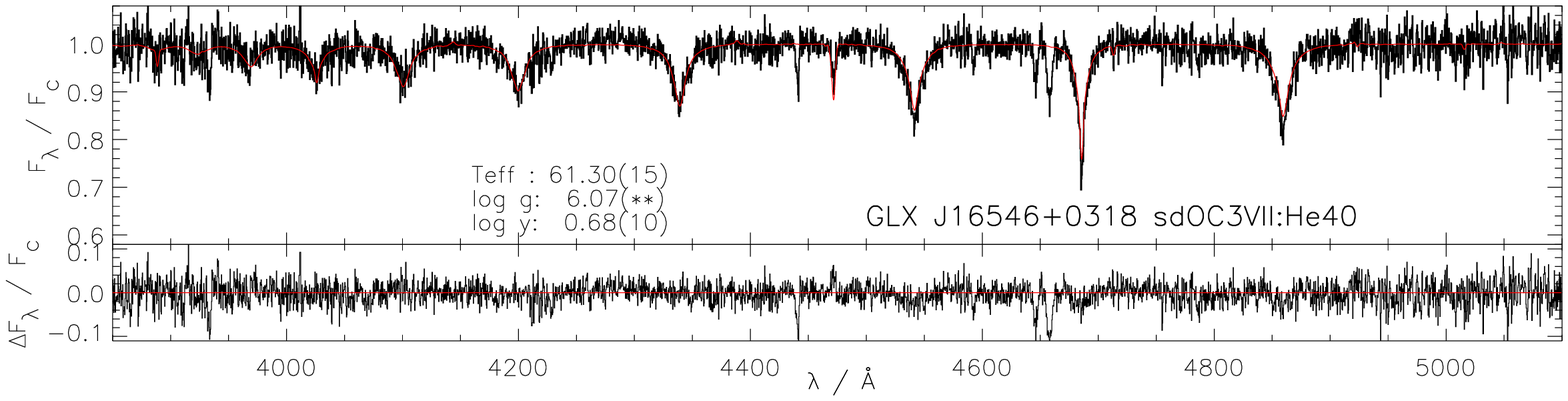}\\
\includegraphics[width=0.85\linewidth]{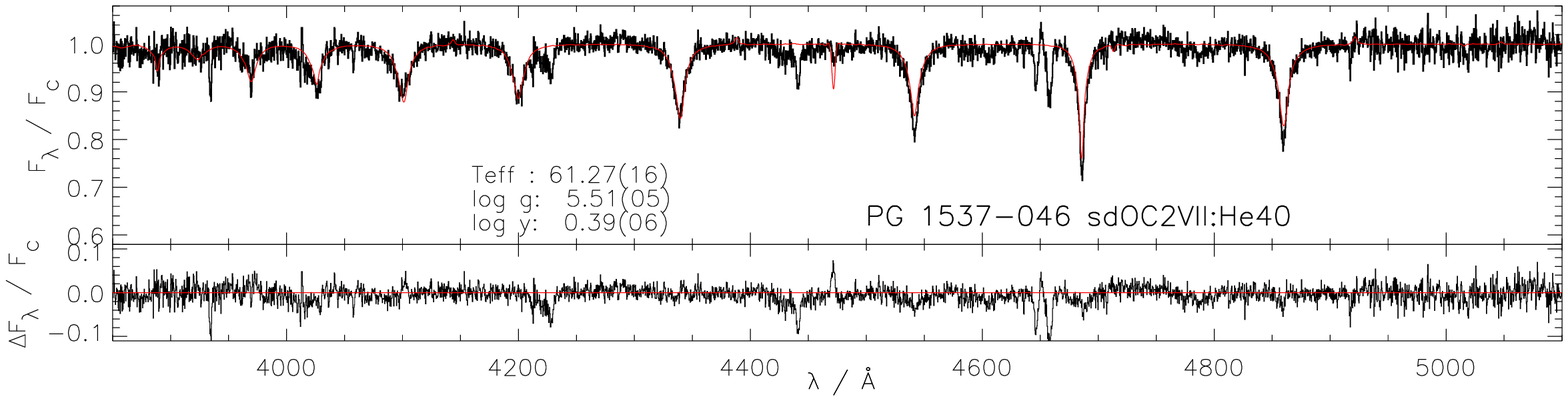}\\
\includegraphics[width=0.85\linewidth]{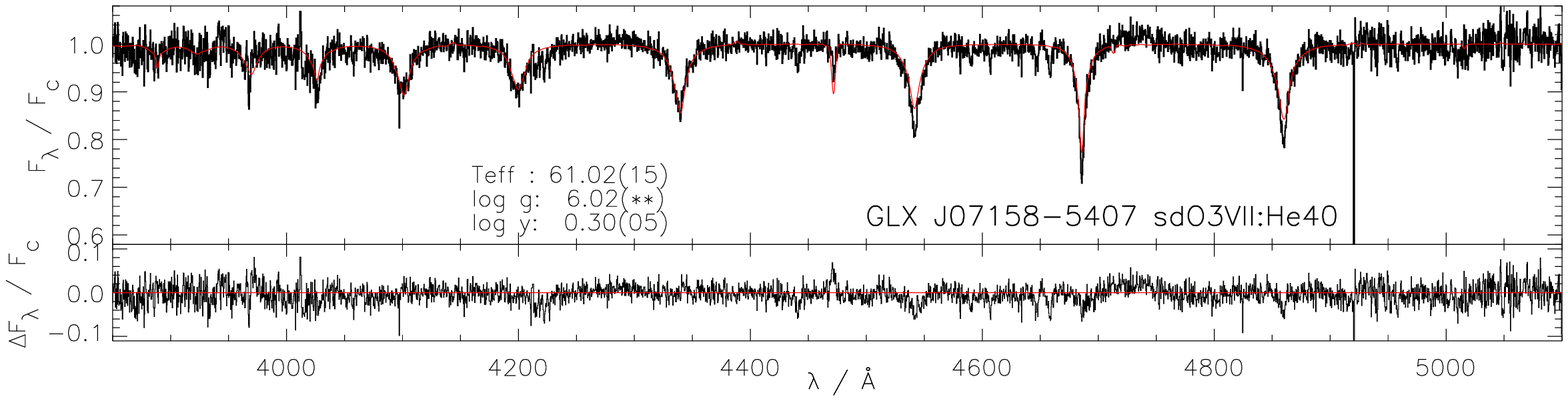}\\ % odd !
\caption{As Fig.\,6 for the stars discussed in \S\,5.7: Sp $\leq$ sdO6}
\label{f:fit07}
\end{figure*}

\begin{figure*}
\includegraphics[width=0.85\linewidth]{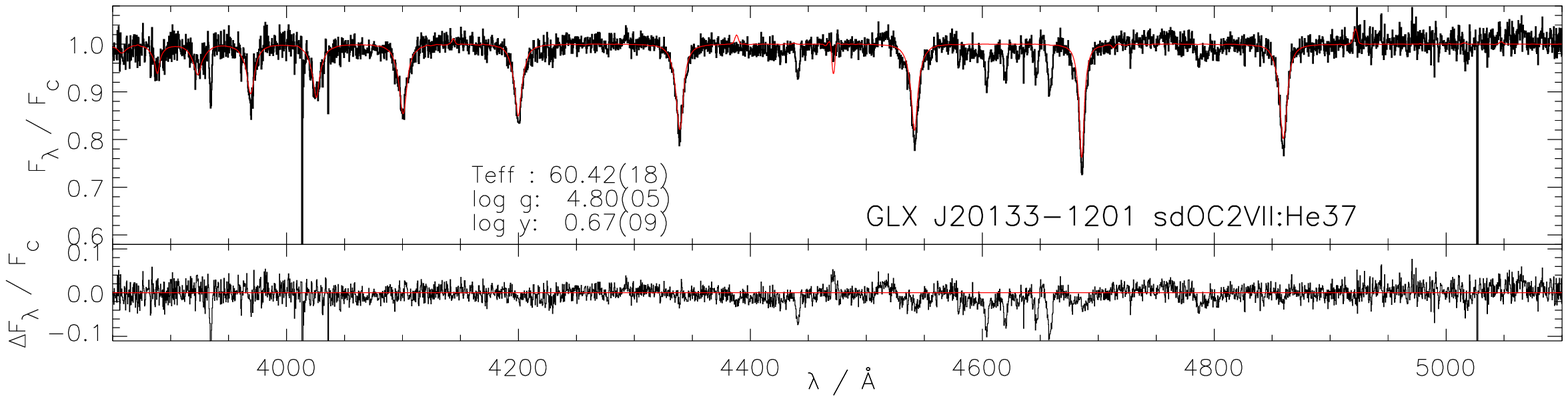}\\
\includegraphics[width=0.85\linewidth]{spectra/EC12420-2732.eps}\\
\includegraphics[width=0.85\linewidth]{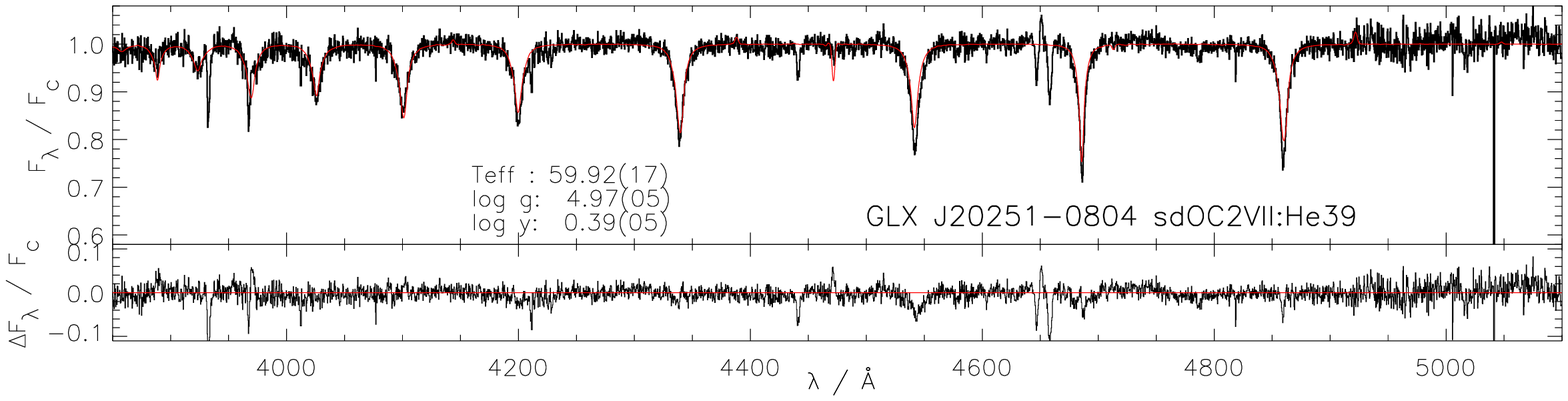}\\
\includegraphics[width=0.85\linewidth]{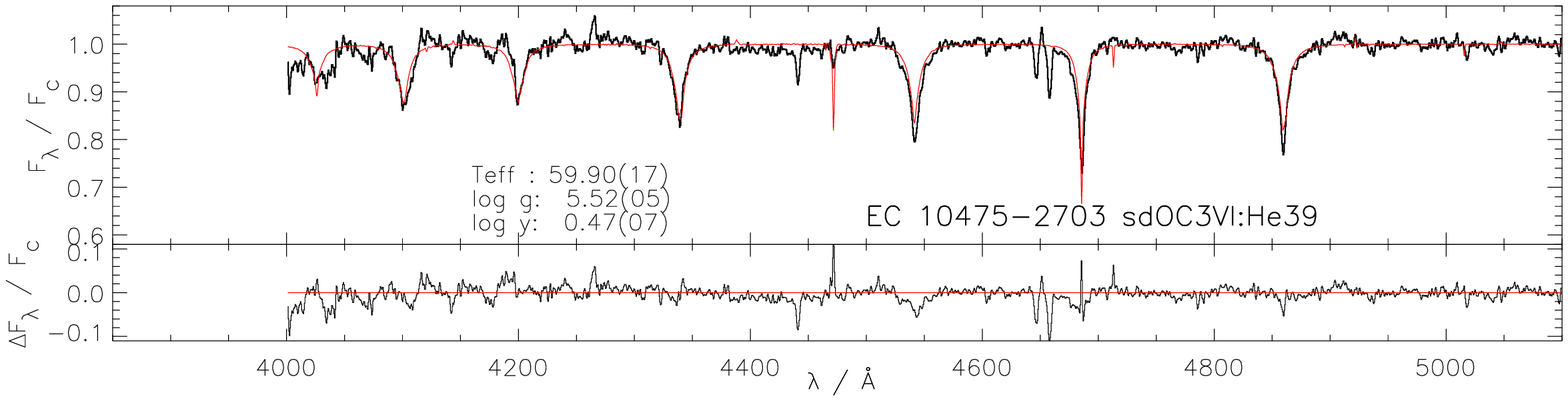}\\
\includegraphics[width=0.85\linewidth]{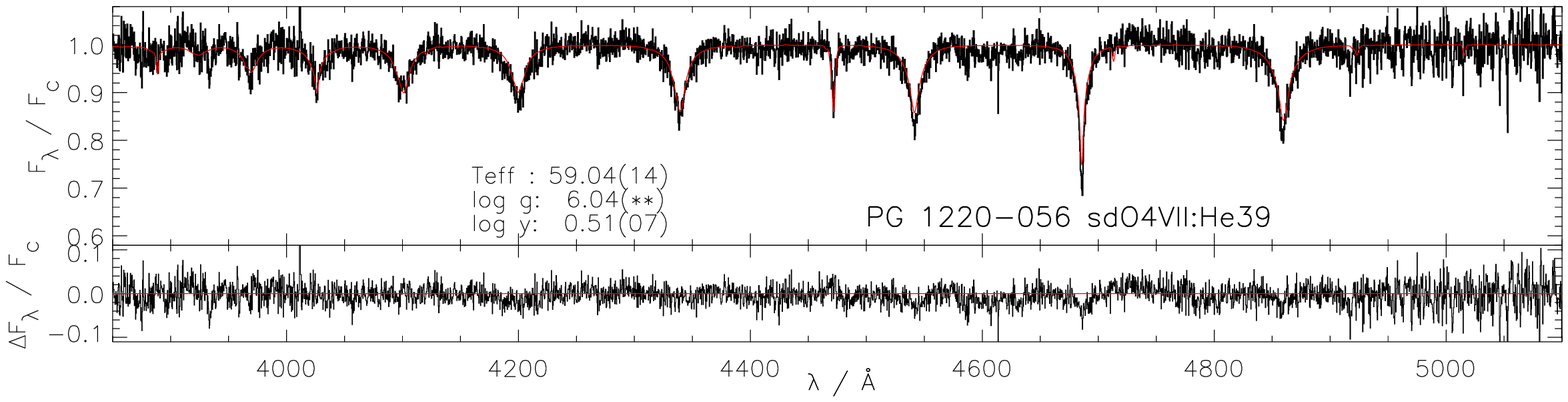}\\
\includegraphics[width=0.85\linewidth]{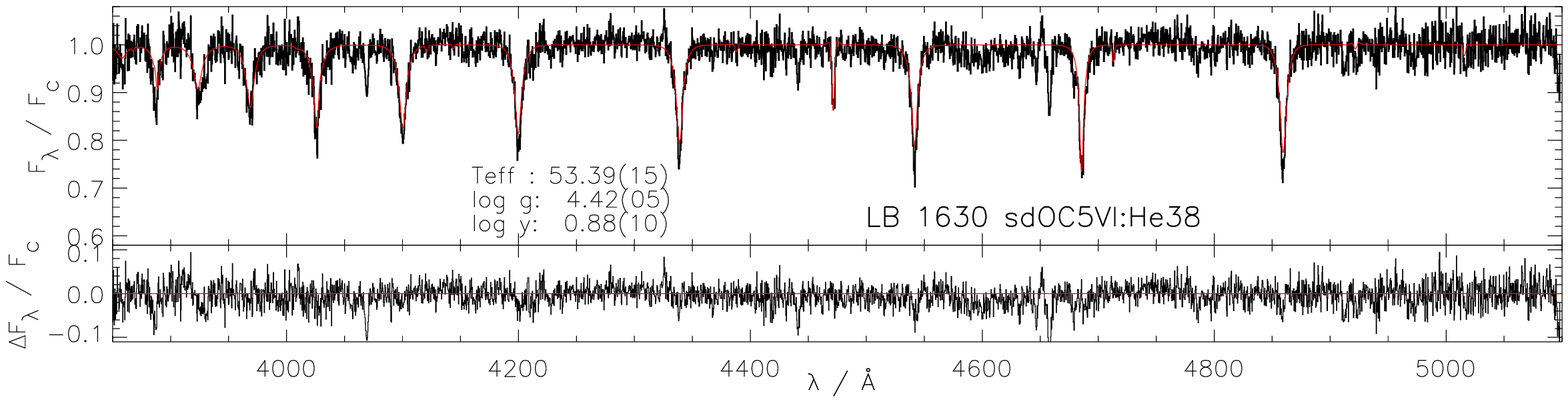}\\
\contcaption{}
\label{f:fit08}
\end{figure*}

% the rest

\begin{figure*}
\includegraphics[width=0.85\linewidth]{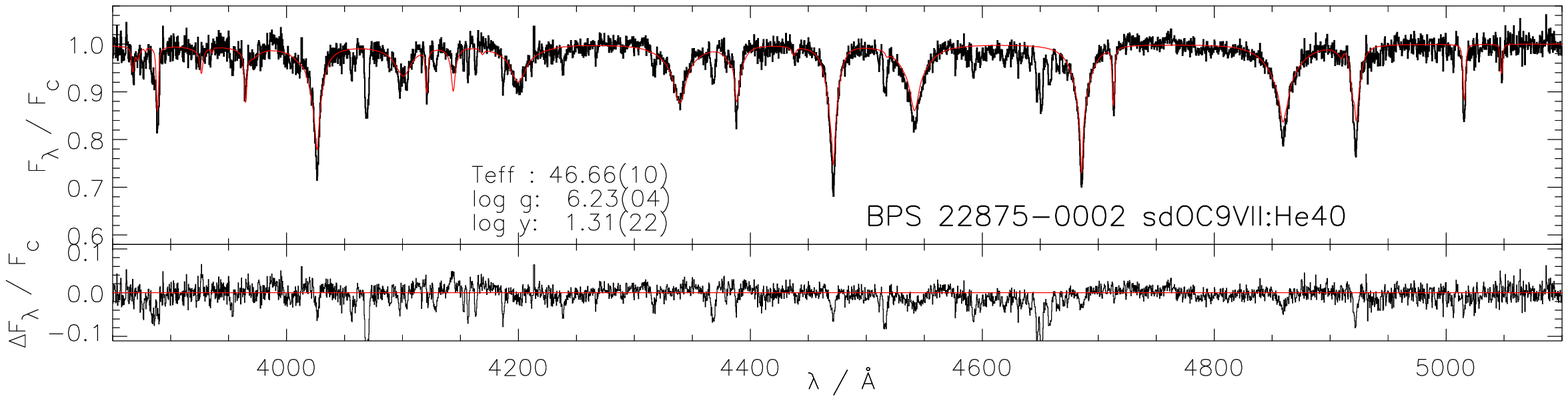}\\
\includegraphics[width=0.85\linewidth]{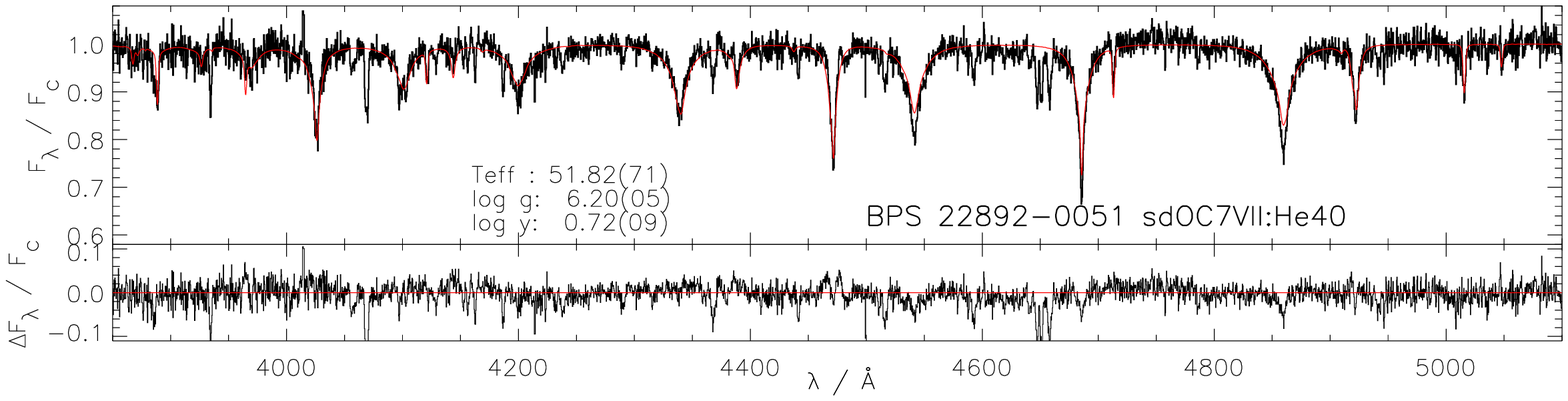}\\
\includegraphics[width=0.85\linewidth]{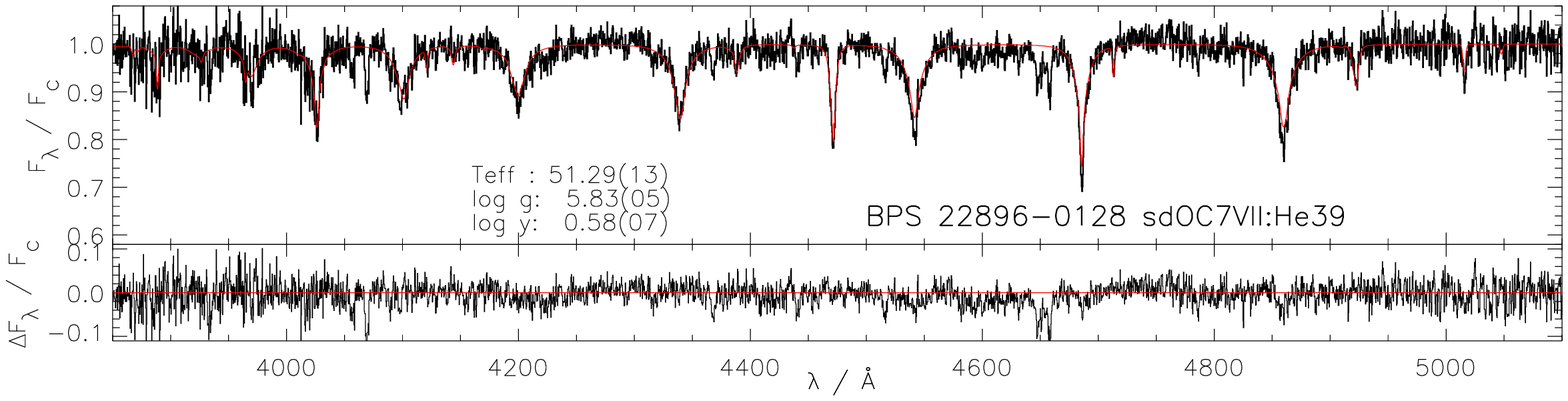}\\
\includegraphics[width=0.85\linewidth]{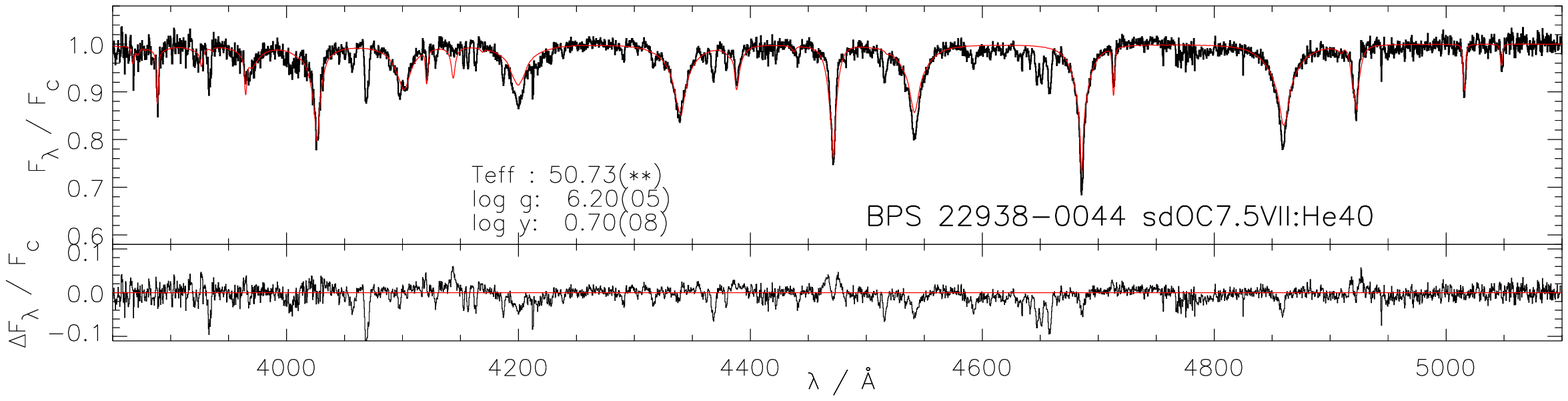}\\
\includegraphics[width=0.85\linewidth]{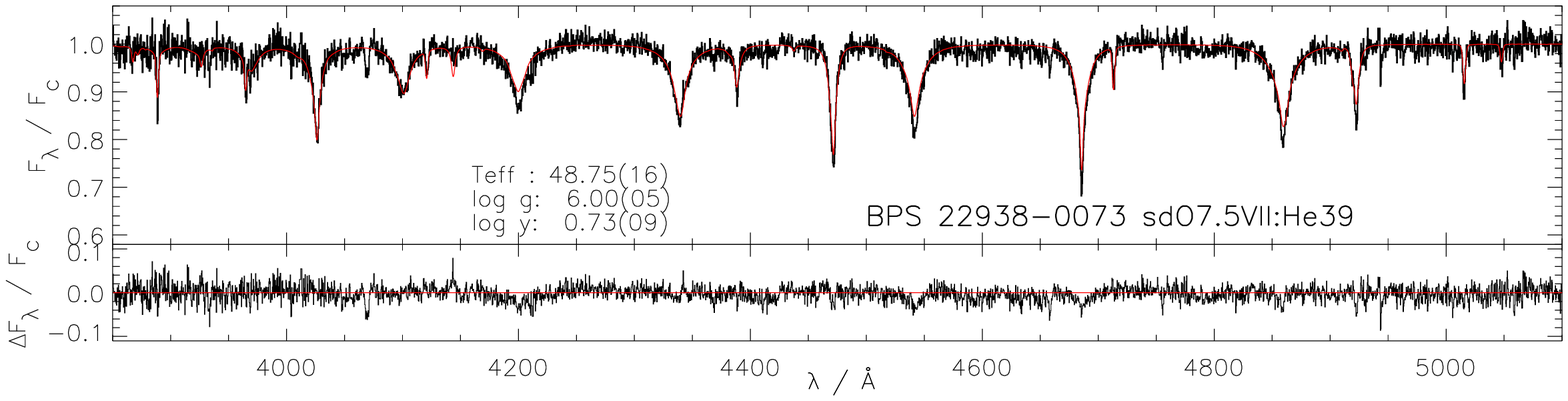}\\
\includegraphics[width=0.85\linewidth]{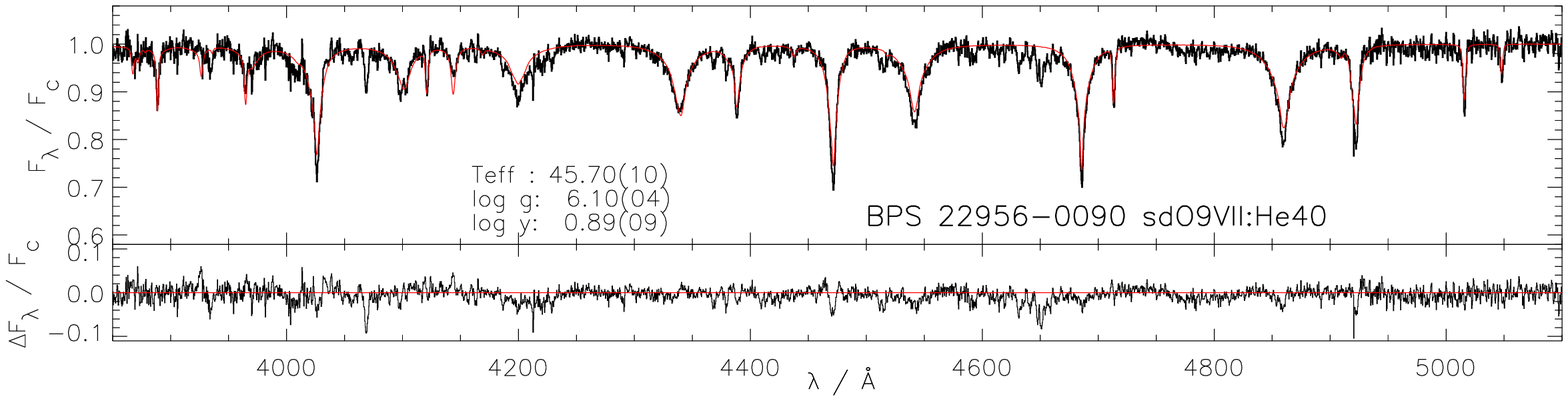}\\
\caption{As Fig.\,6 for the stars discussed in \S\,5.8: Sp = sdO6.5 -- sdB0.5, He $\gtrsim 35$}
\label{f:fit09}
\end{figure*}

\begin{figure*}
\includegraphics[width=0.85\linewidth]{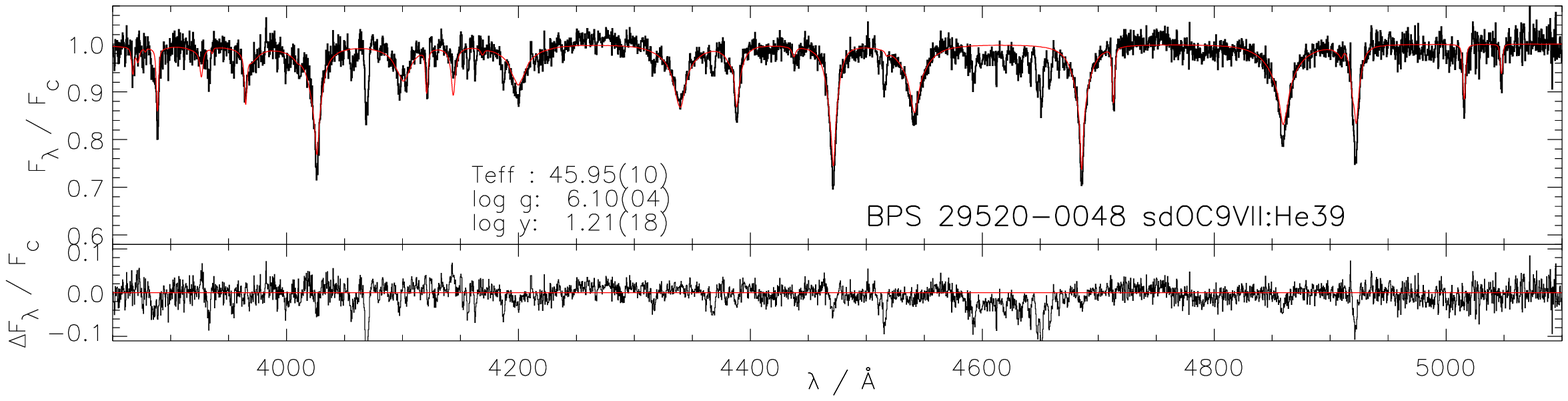}\\
\includegraphics[width=0.85\linewidth]{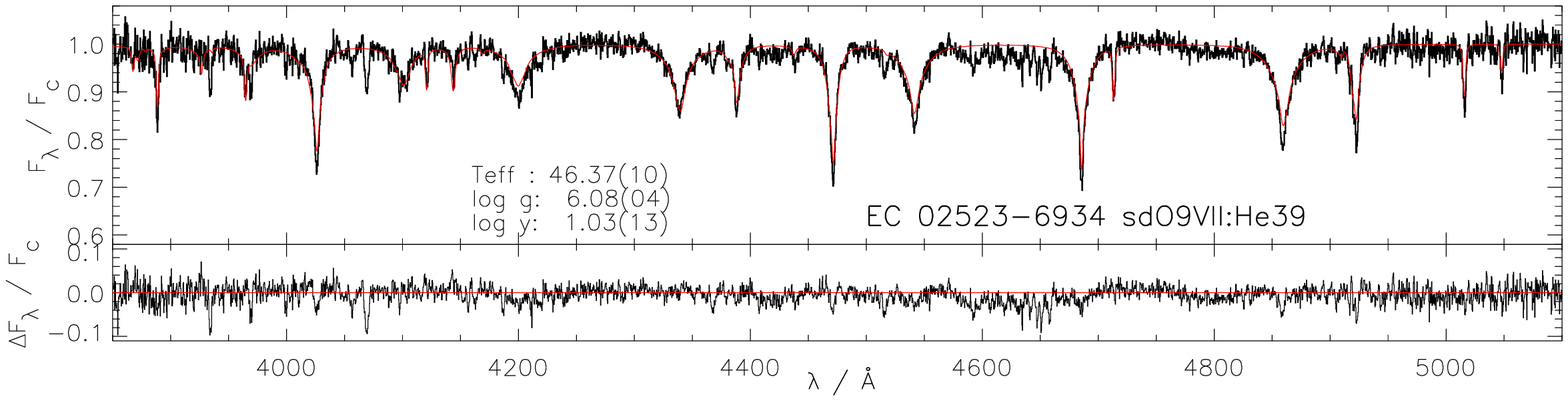}\\
\includegraphics[width=0.85\linewidth]{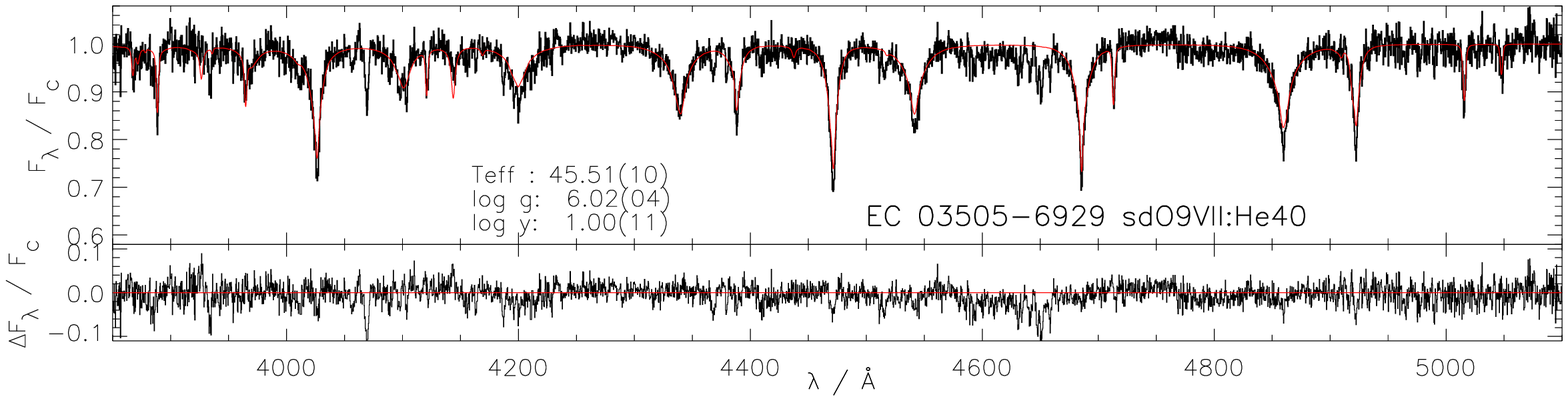}\\
\includegraphics[width=0.85\linewidth]{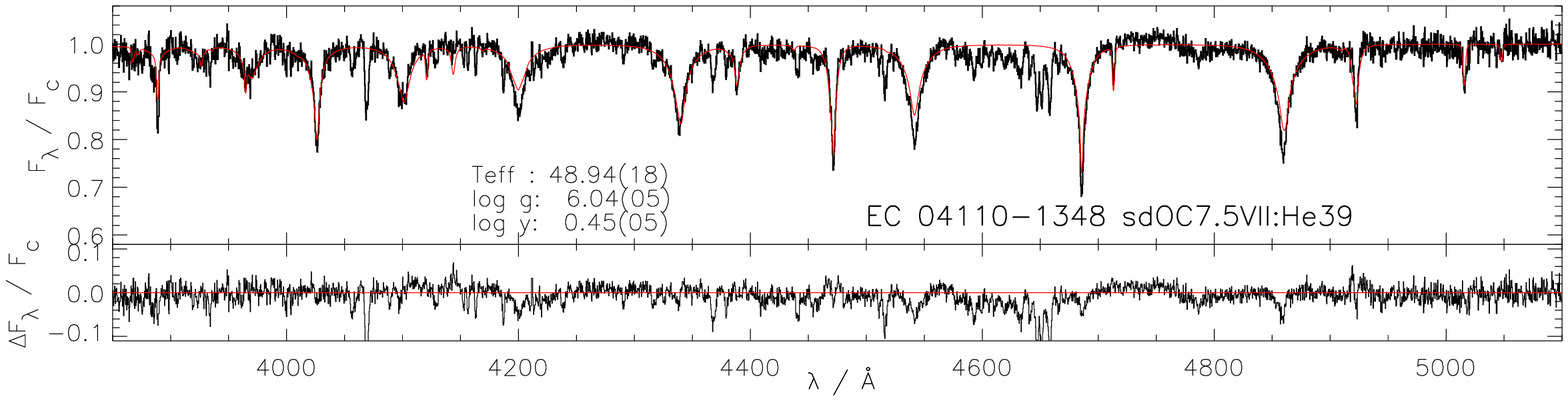}\\
\includegraphics[width=0.85\linewidth]{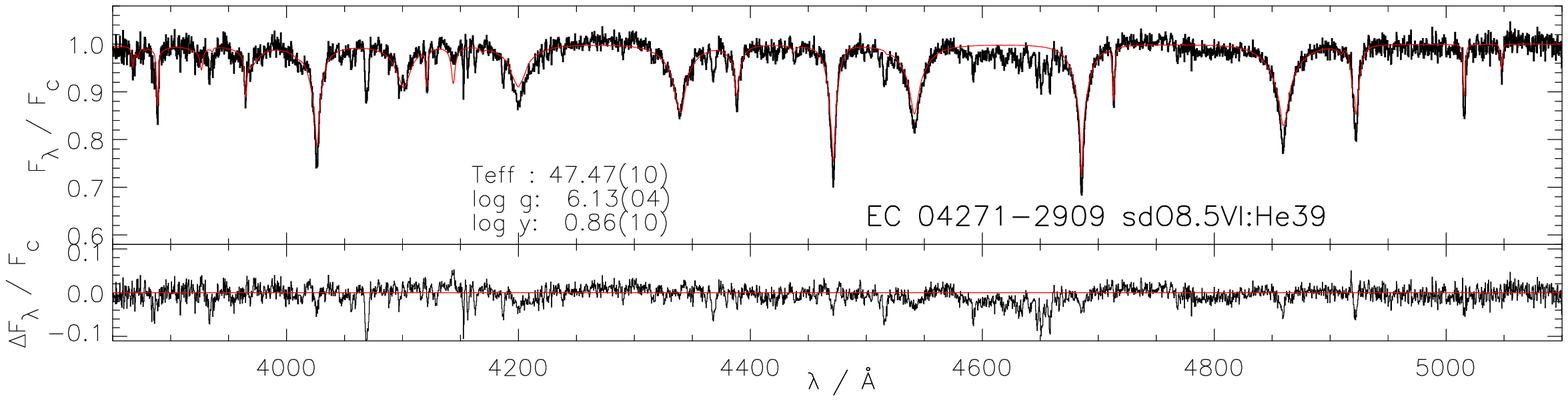}\\
\includegraphics[width=0.85\linewidth]{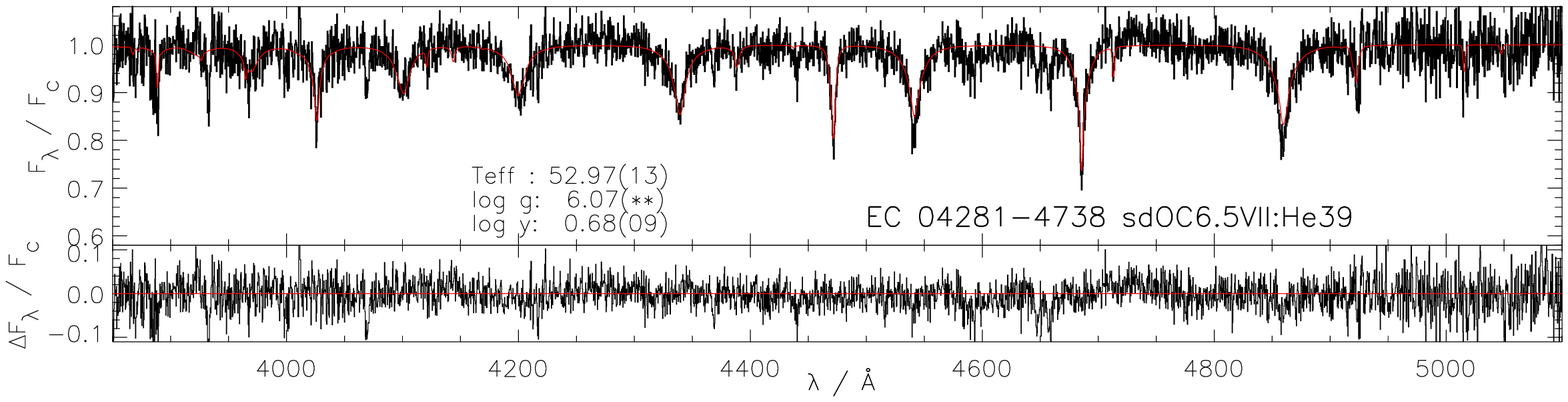}\\
\contcaption{}
\label{f:fit10}
\end{figure*}

\begin{figure*}
\includegraphics[width=0.85\linewidth]{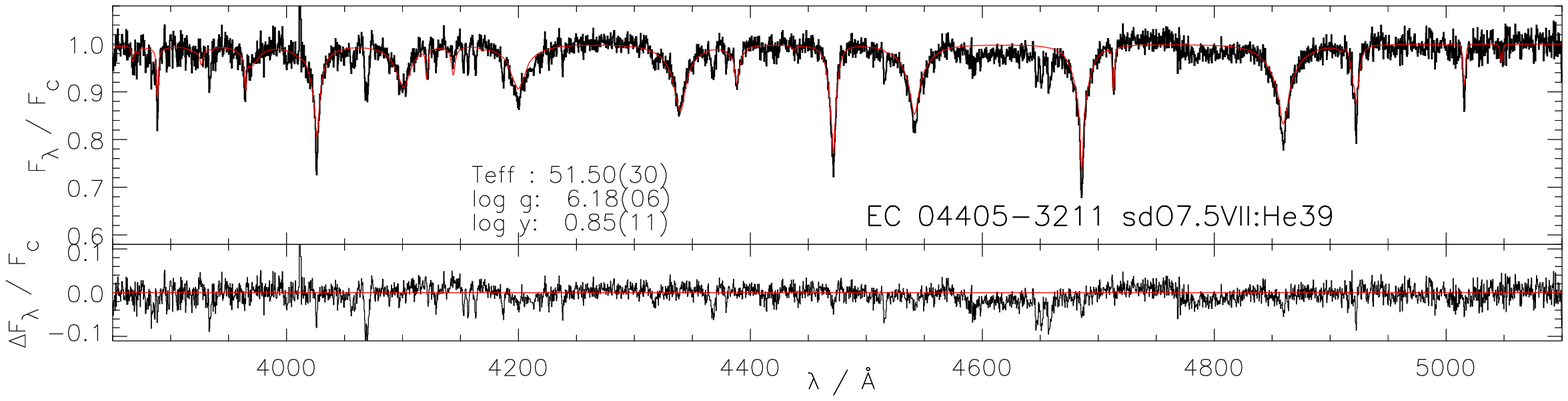}\\
\includegraphics[width=0.85\linewidth]{spectra/EC04517-3706.eps}\\
\includegraphics[width=0.85\linewidth]{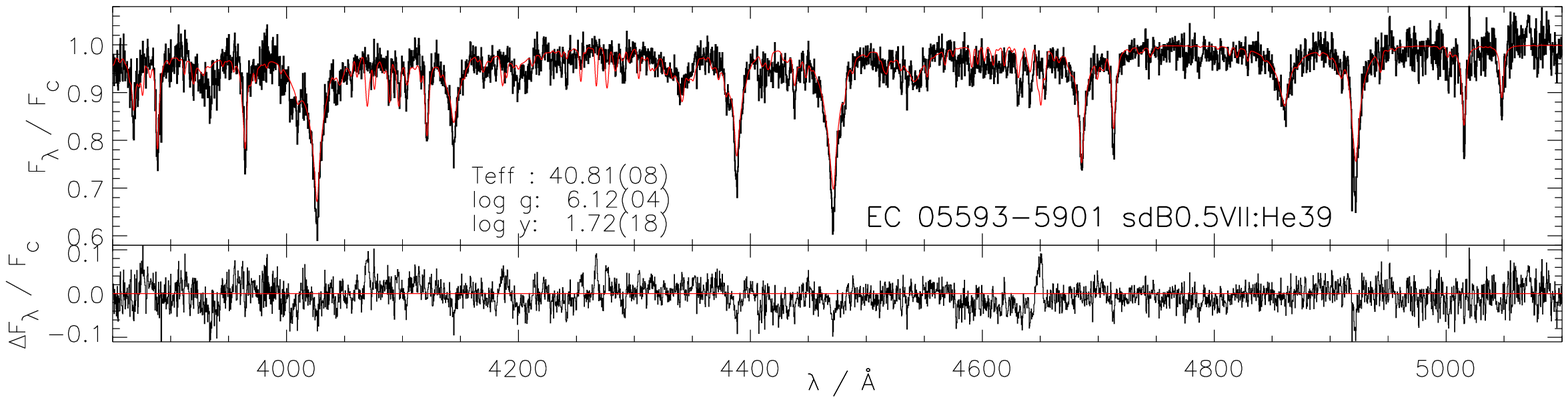}\\
\includegraphics[width=0.85\linewidth]{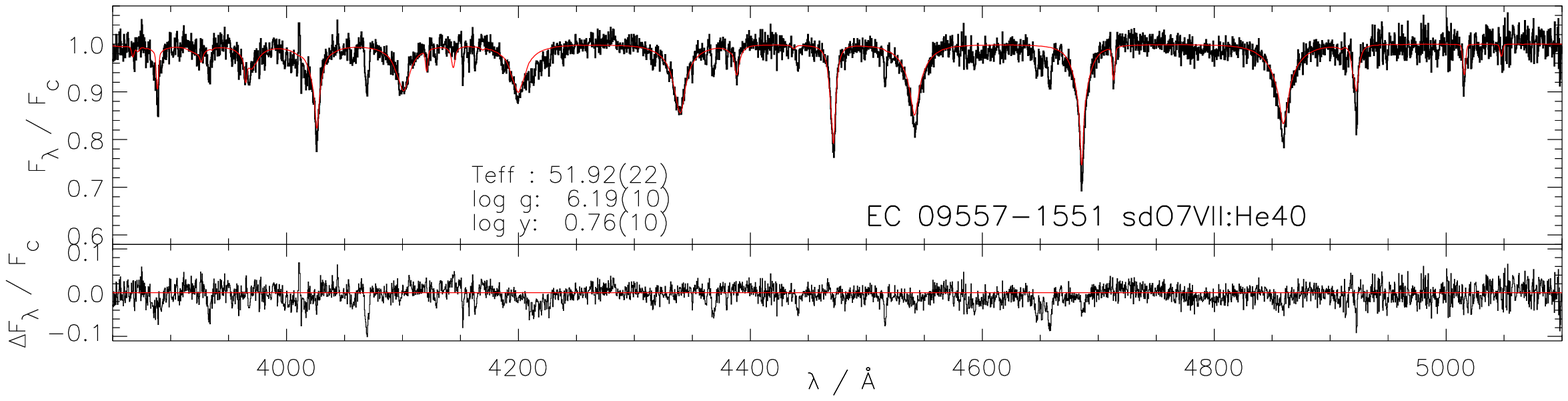}\\
\includegraphics[width=0.85\linewidth]{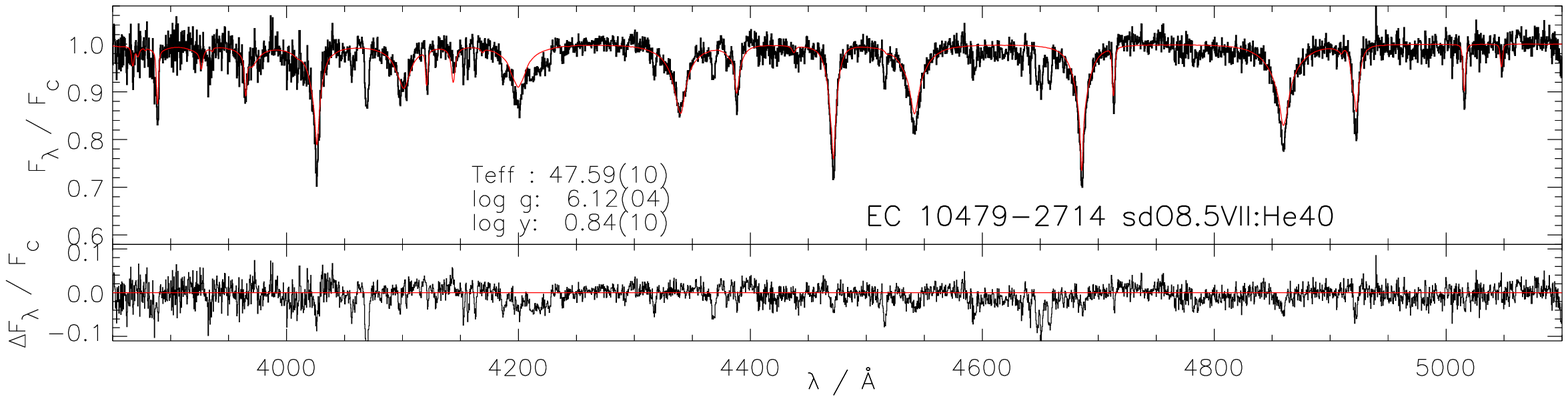}\\
\includegraphics[width=0.85\linewidth]{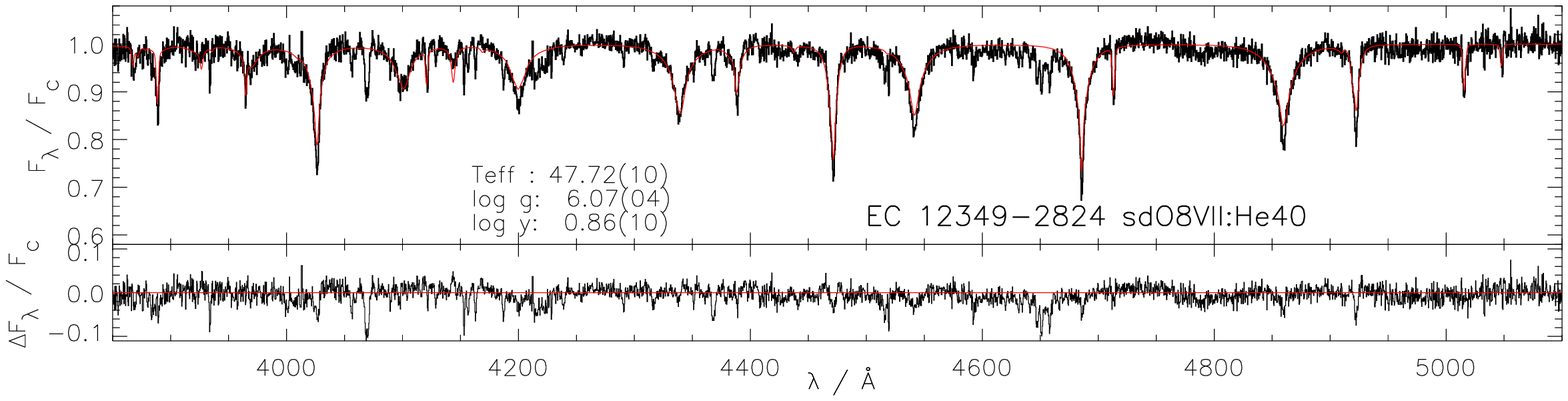}\\
\contcaption{}
\label{f:fit11}
\end{figure*}

\begin{figure*}
\includegraphics[width=0.85\linewidth]{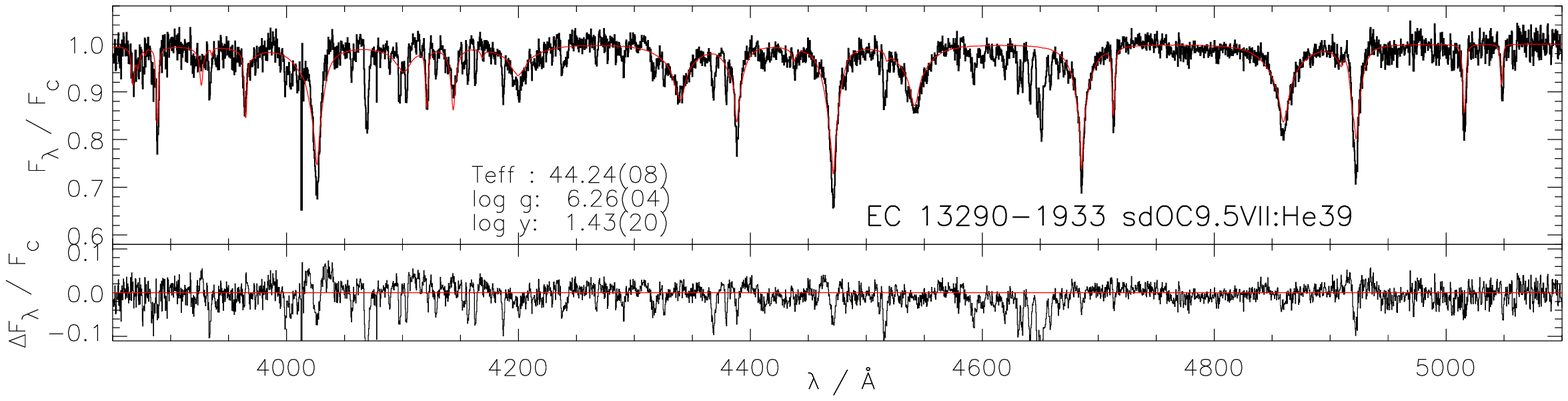}\\
\includegraphics[width=0.85\linewidth]{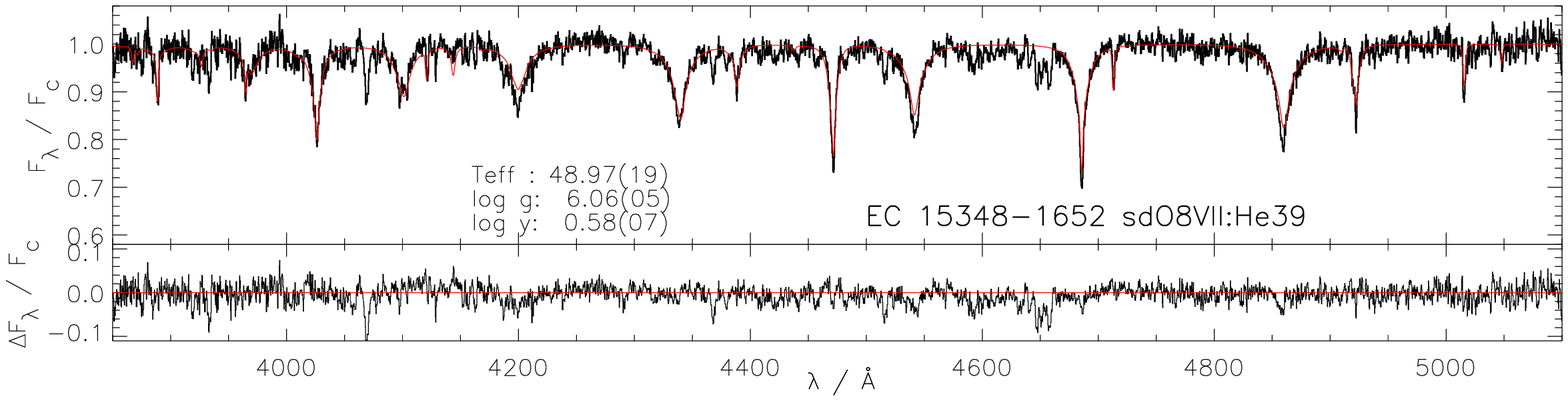}\\
\includegraphics[width=0.85\linewidth]{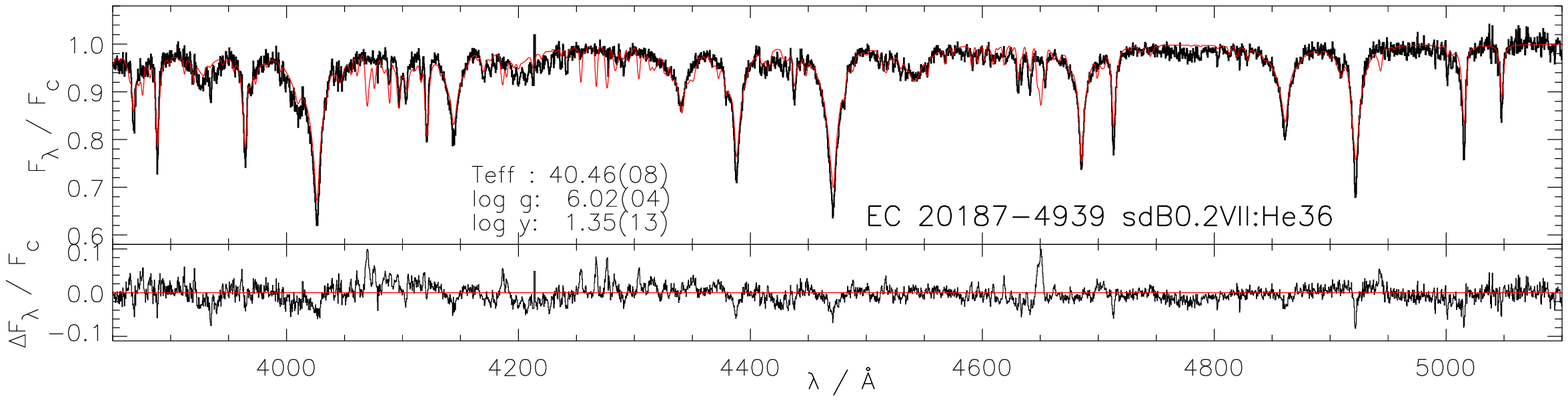}\\
\includegraphics[width=0.85\linewidth]{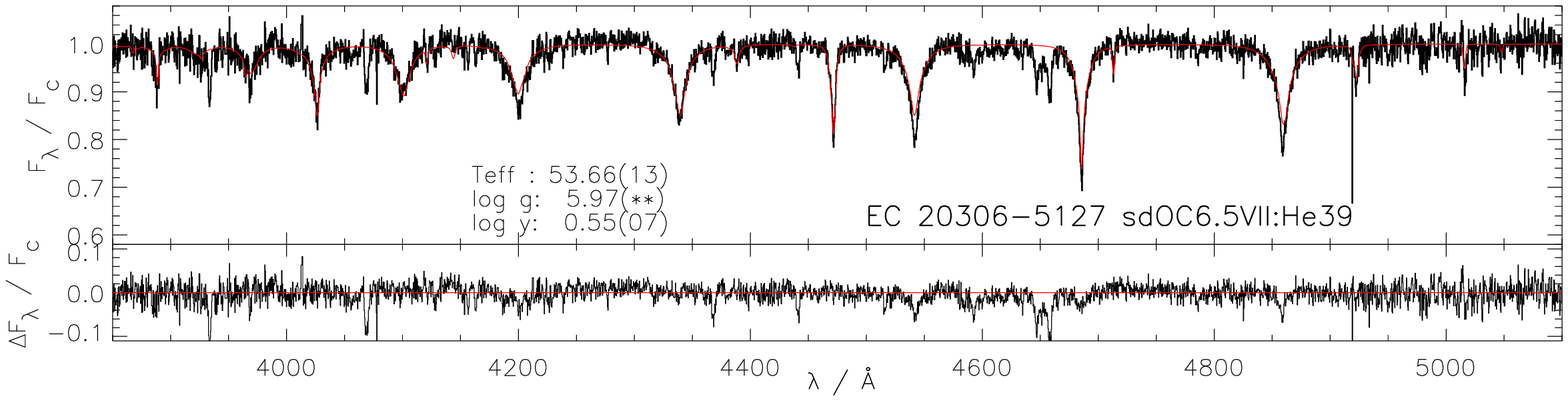}\\
\includegraphics[width=0.85\linewidth]{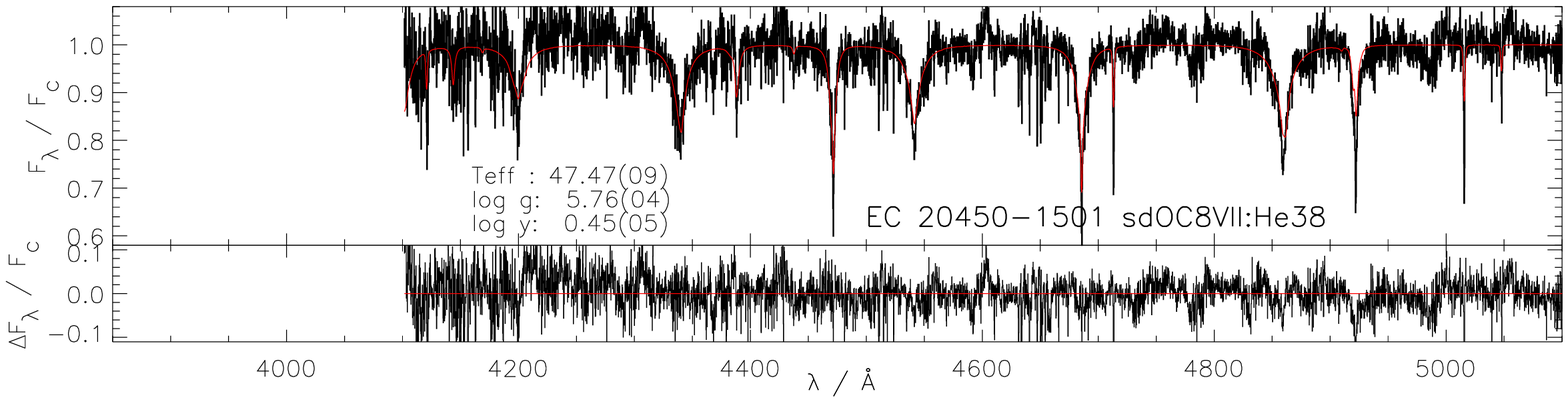}\\
\includegraphics[width=0.85\linewidth]{spectra/EC20450-6947.eps}\\
\contcaption{}
\label{f:fit12}
\end{figure*}

\begin{figure*}
\includegraphics[width=0.85\linewidth]{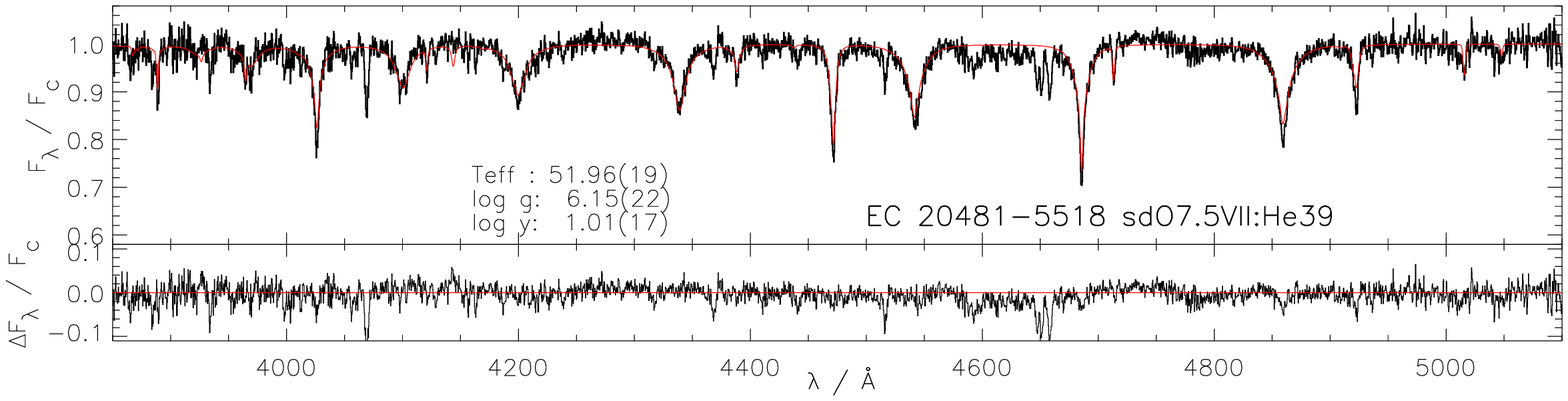}\\
\includegraphics[width=0.85\linewidth]{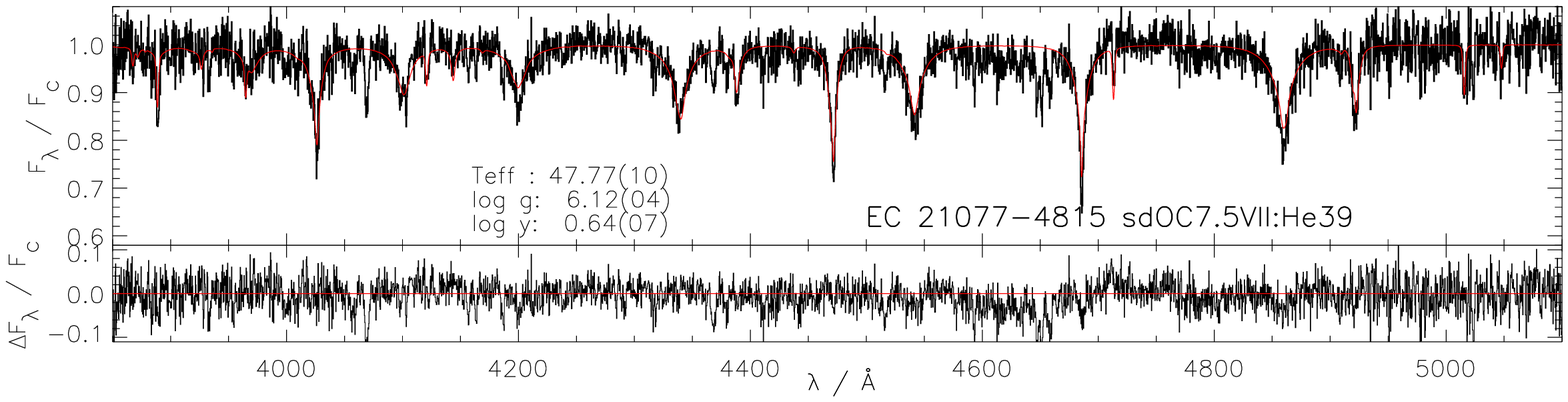}\\
\includegraphics[width=0.85\linewidth]{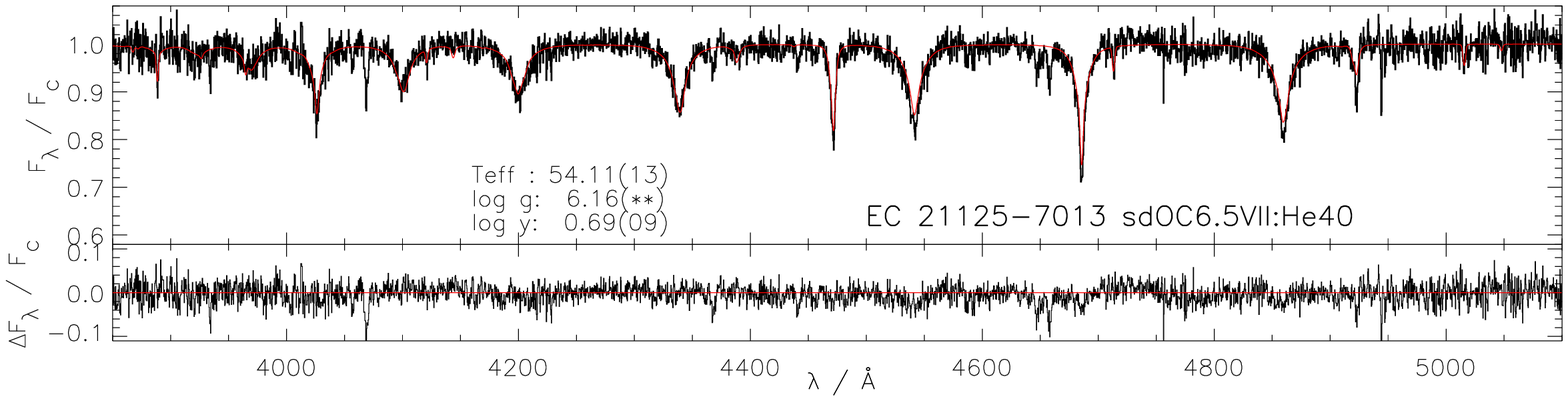}\\
\includegraphics[width=0.85\linewidth]{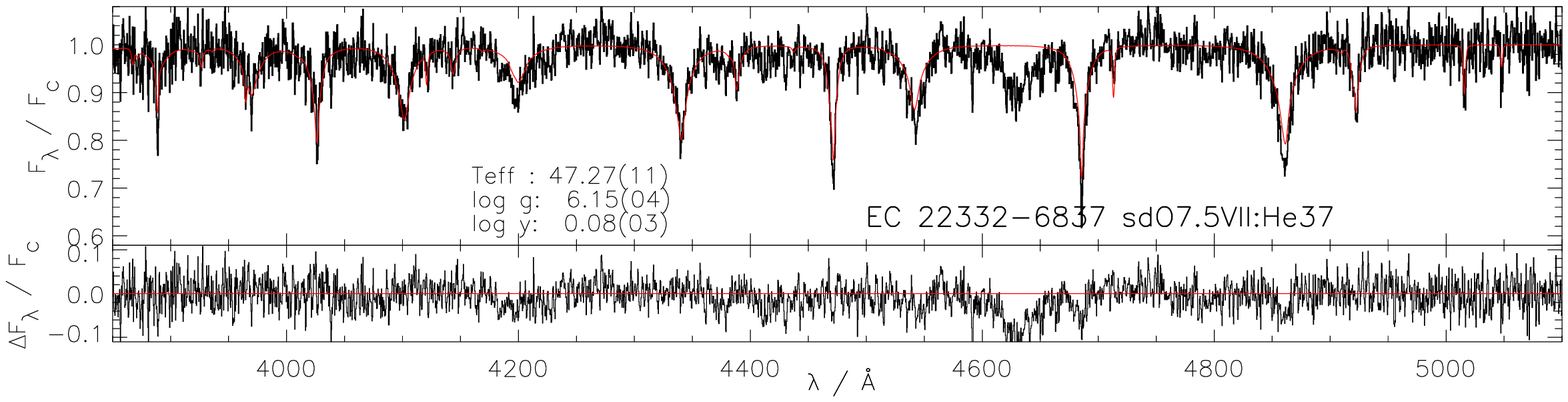}\\
\includegraphics[width=0.85\linewidth]{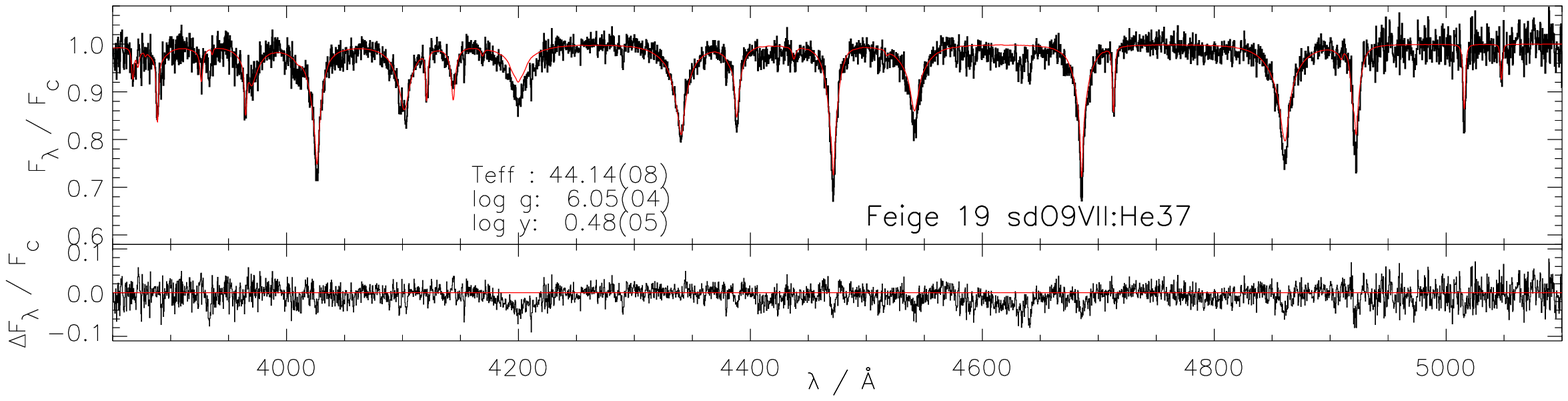}\\
\includegraphics[width=0.85\linewidth]{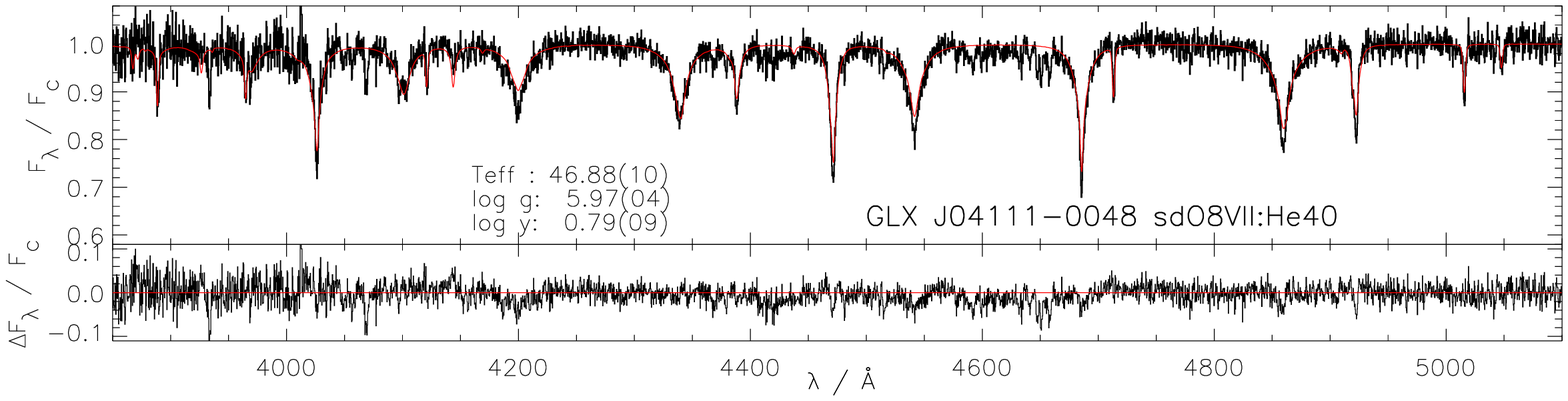}\\
\contcaption{}
\label{f:fit13}
\end{figure*}

\begin{figure*}
\includegraphics[width=0.85\linewidth]{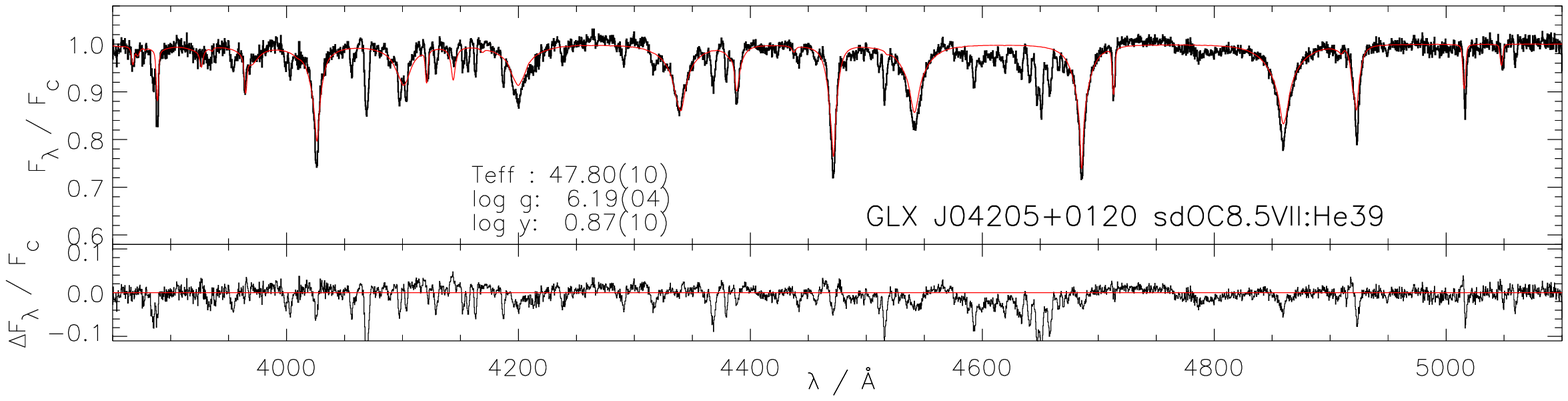}\\
\includegraphics[width=0.85\linewidth]{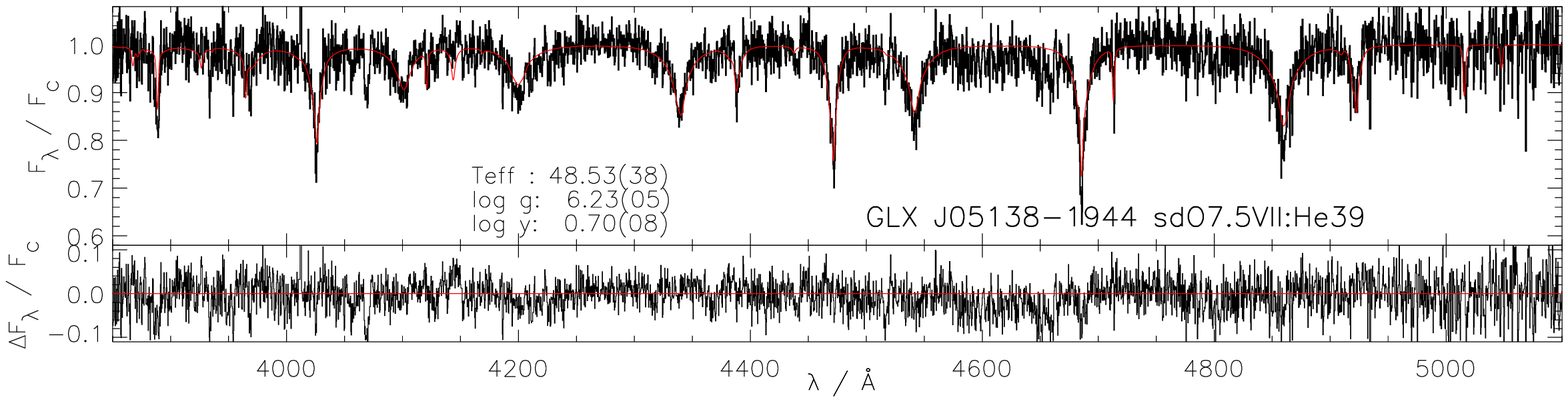}\\
\includegraphics[width=0.85\linewidth]{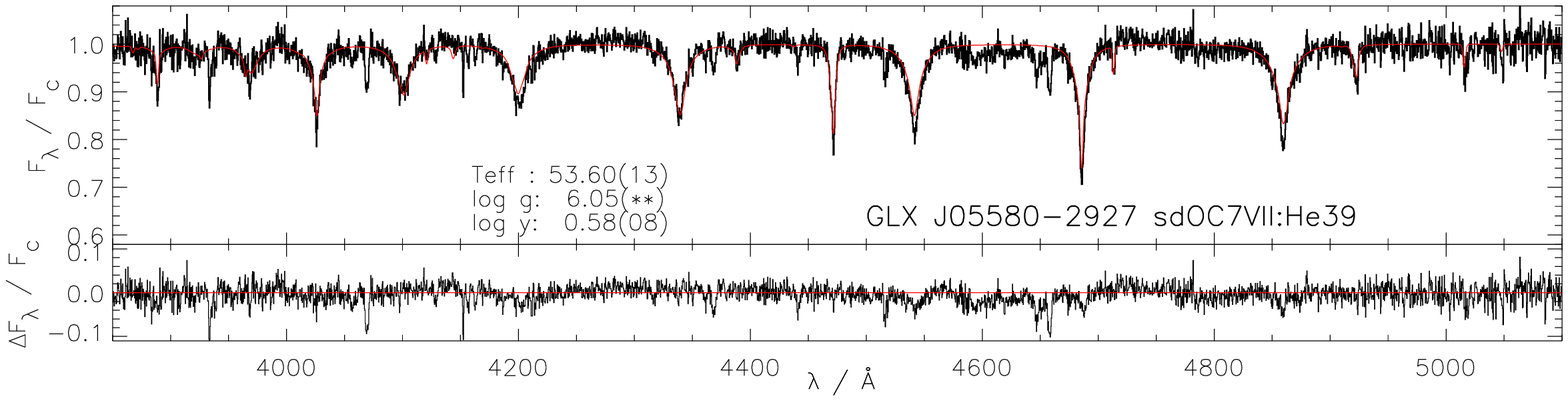}\\
\includegraphics[width=0.85\linewidth]{spectra/GLXJ06126-2712.eps}\\
\includegraphics[width=0.85\linewidth]{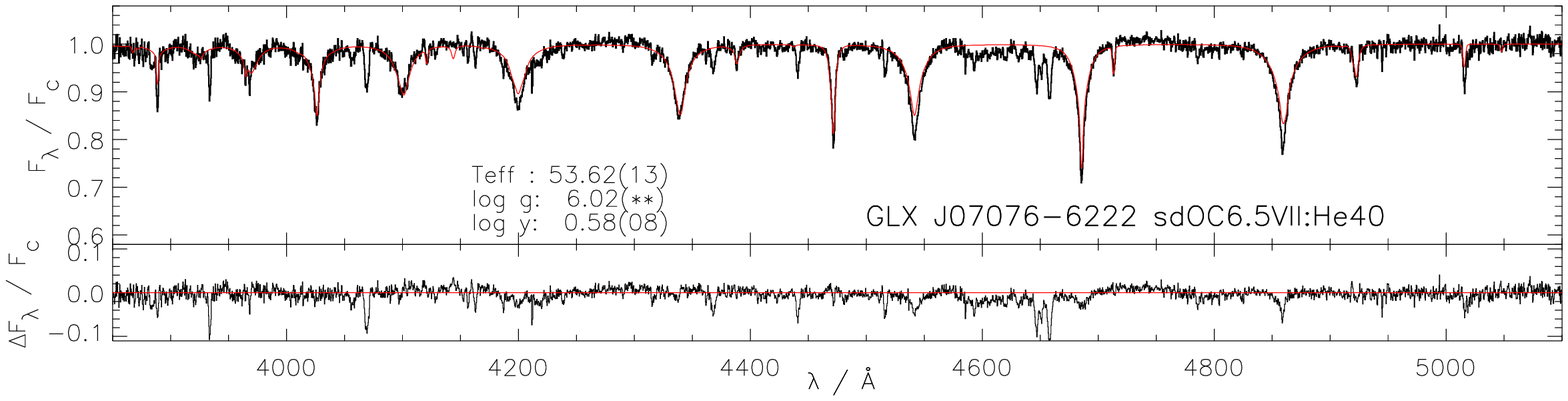}\\
\includegraphics[width=0.85\linewidth]{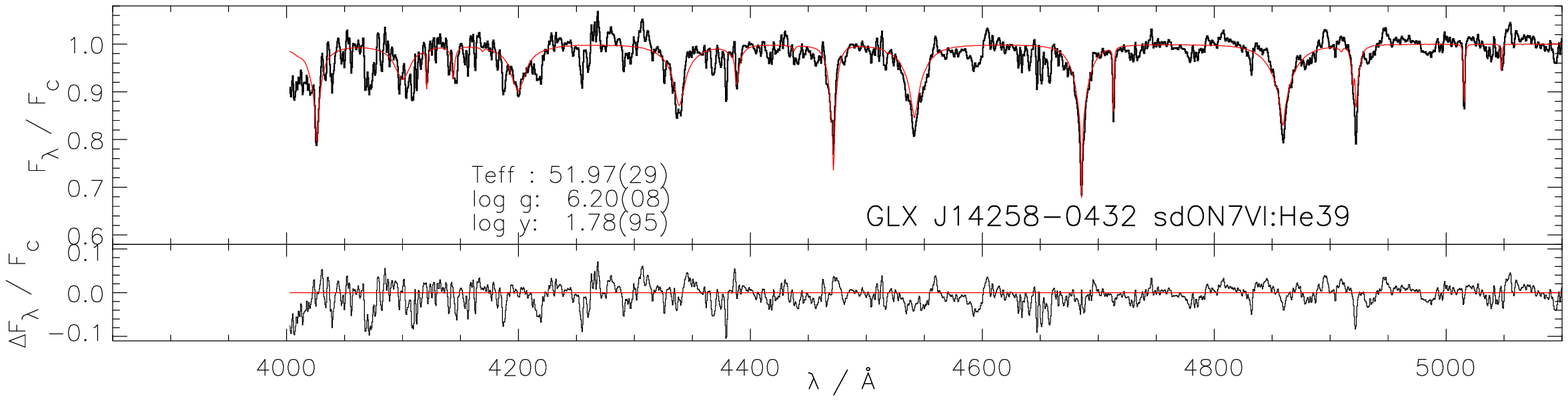}\\
\contcaption{}
\label{f:fit14}
\end{figure*}

\begin{figure*}
\includegraphics[width=0.85\linewidth]{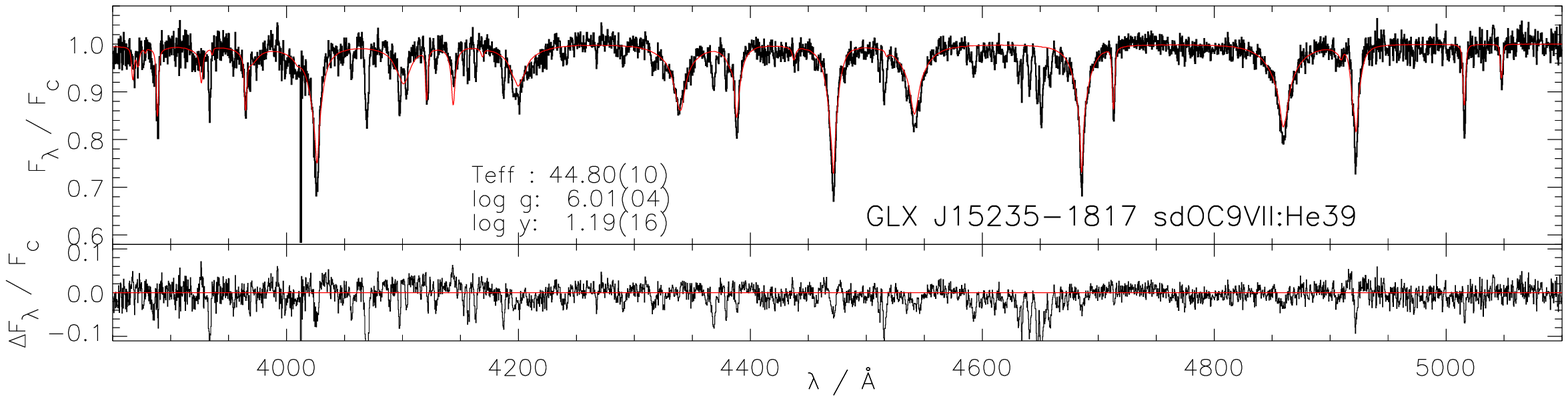}\\
\includegraphics[width=0.85\linewidth]{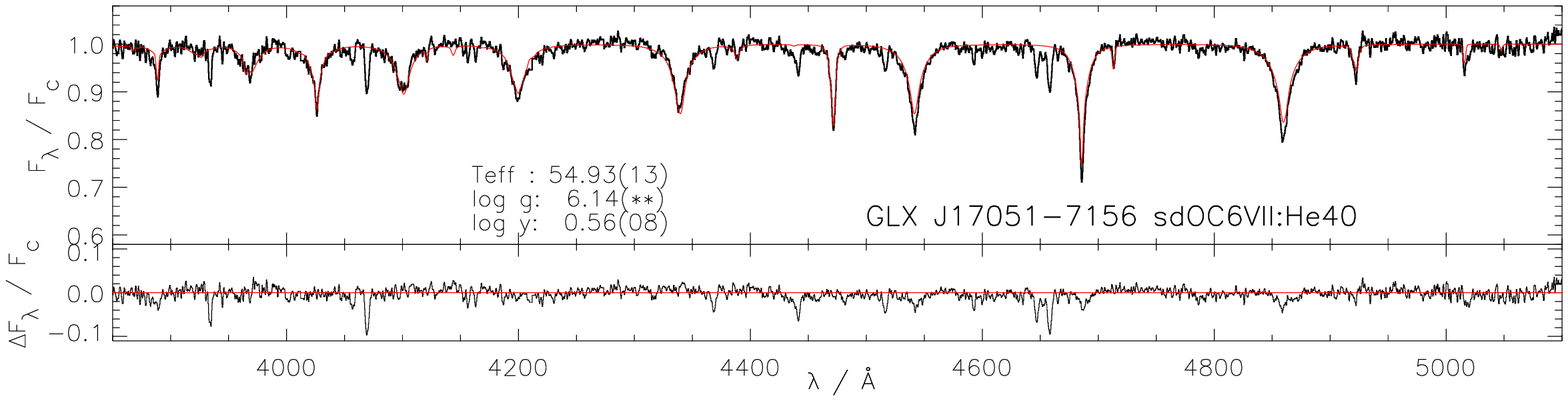}\\
\includegraphics[width=0.85\linewidth]{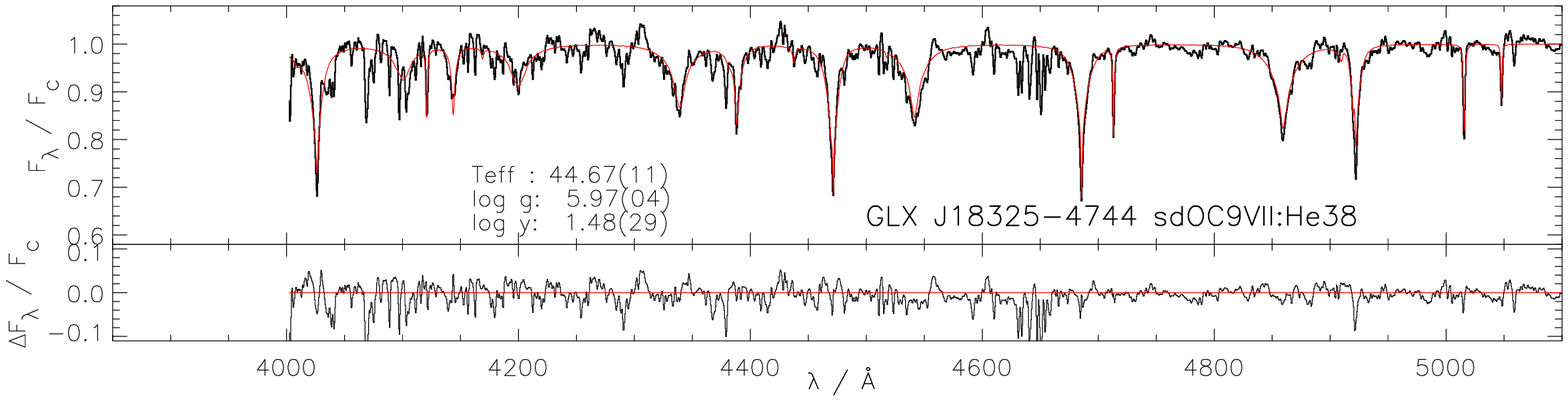}\\
\includegraphics[width=0.85\linewidth]{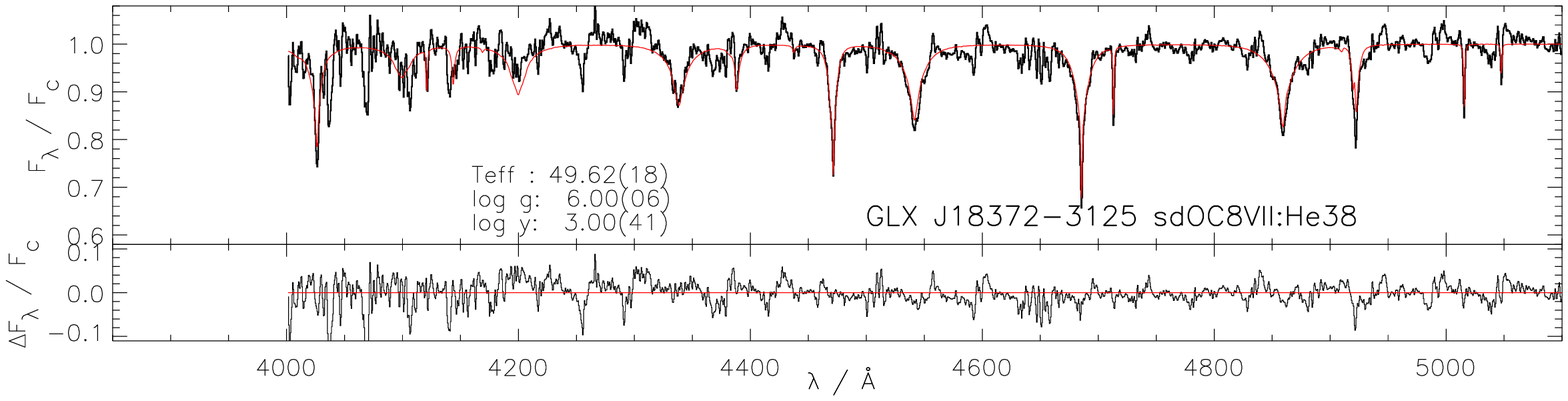}\\
\includegraphics[width=0.85\linewidth]{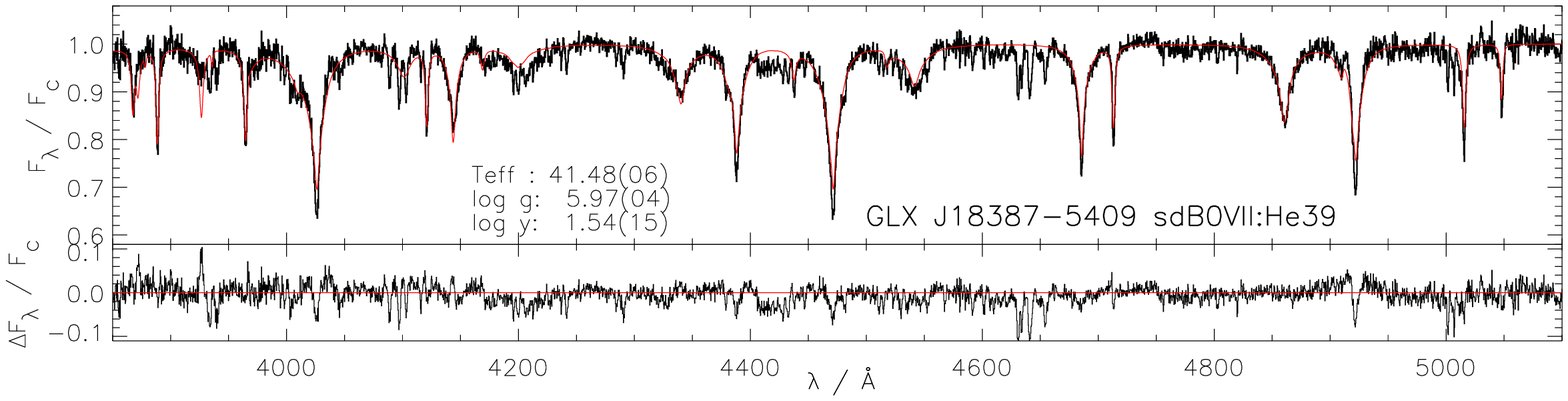}\\
\includegraphics[width=0.85\linewidth]{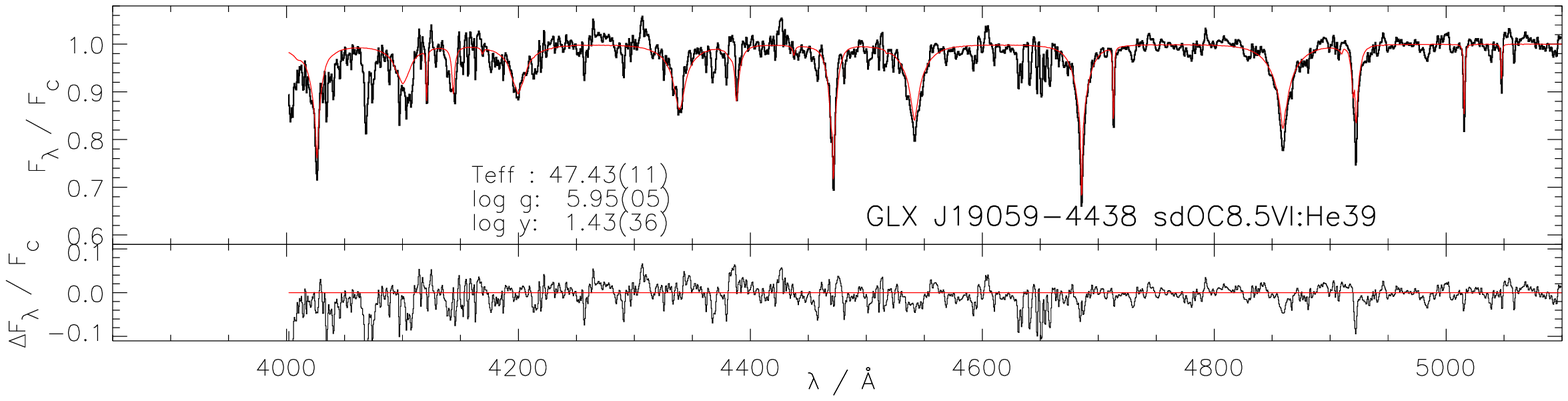}\\
\contcaption{}
\label{f:fit15}
\end{figure*}

\begin{figure*}

\includegraphics[width=0.85\linewidth]{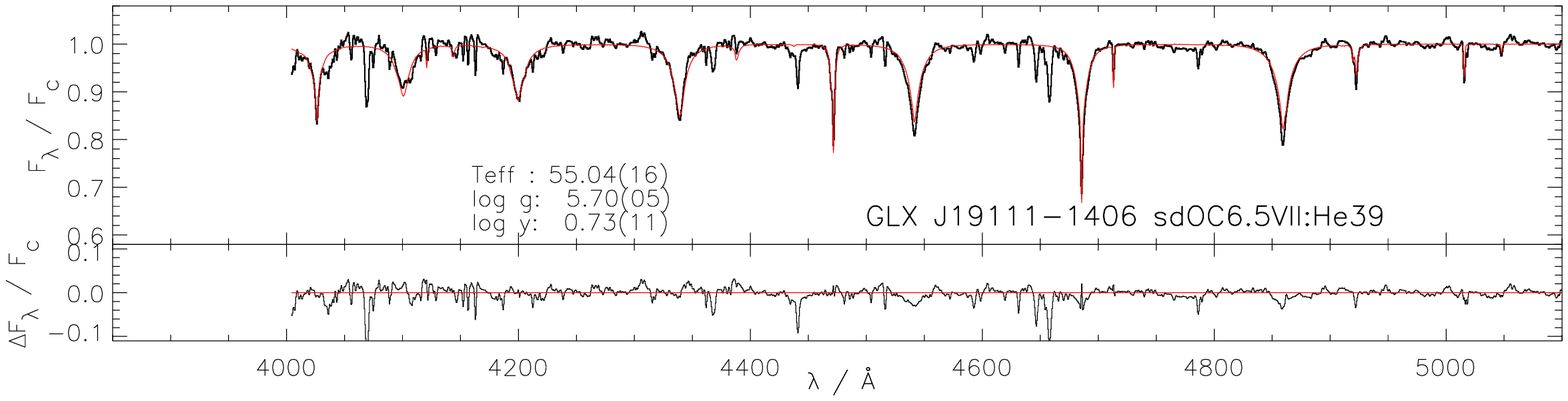}\\
\includegraphics[width=0.85\linewidth]{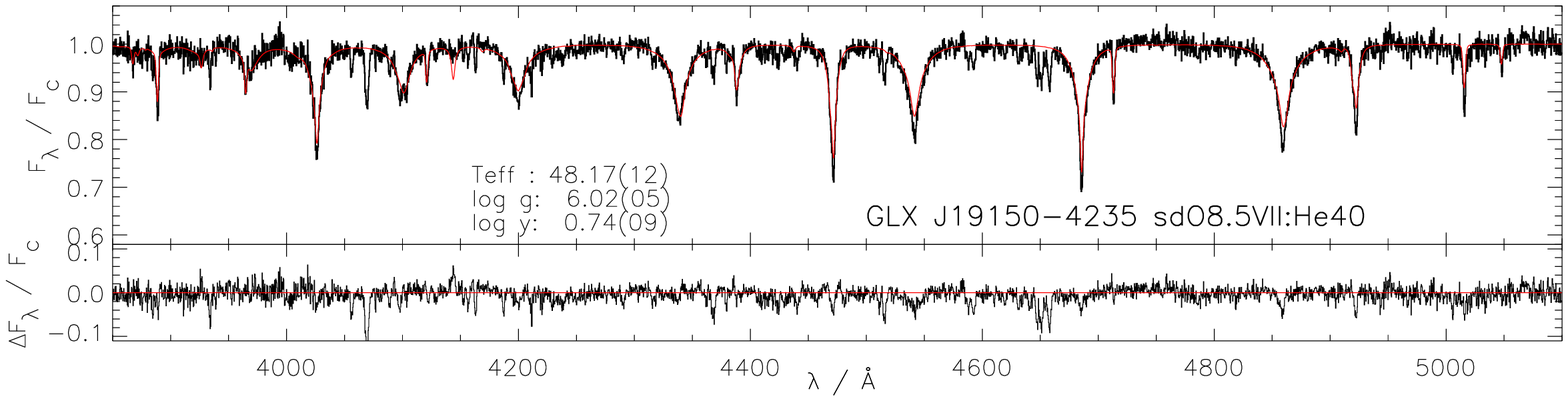}\\
\includegraphics[width=0.85\linewidth]{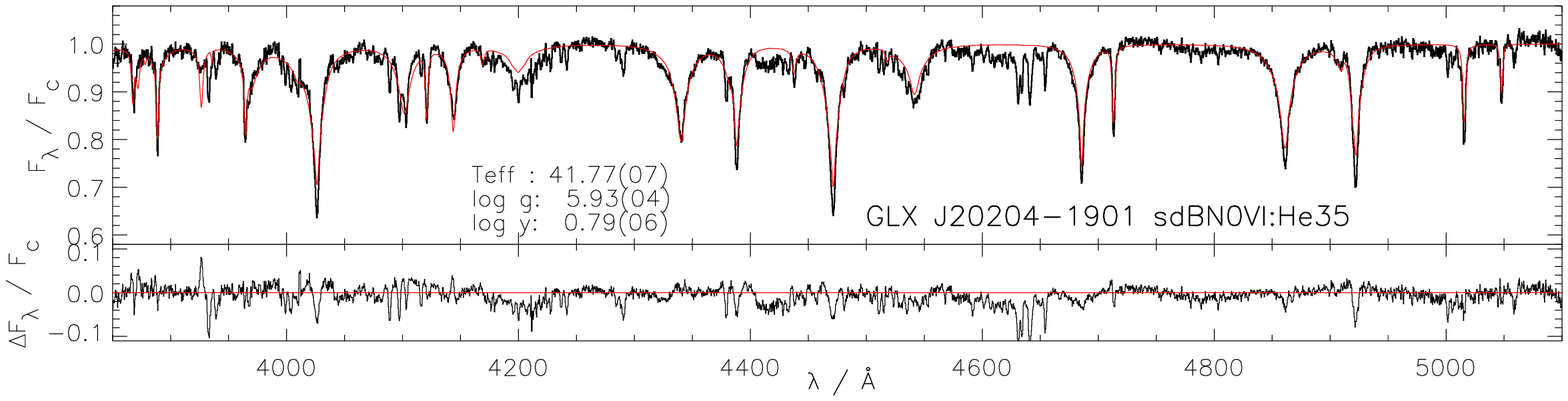}\\
\includegraphics[width=0.85\linewidth]{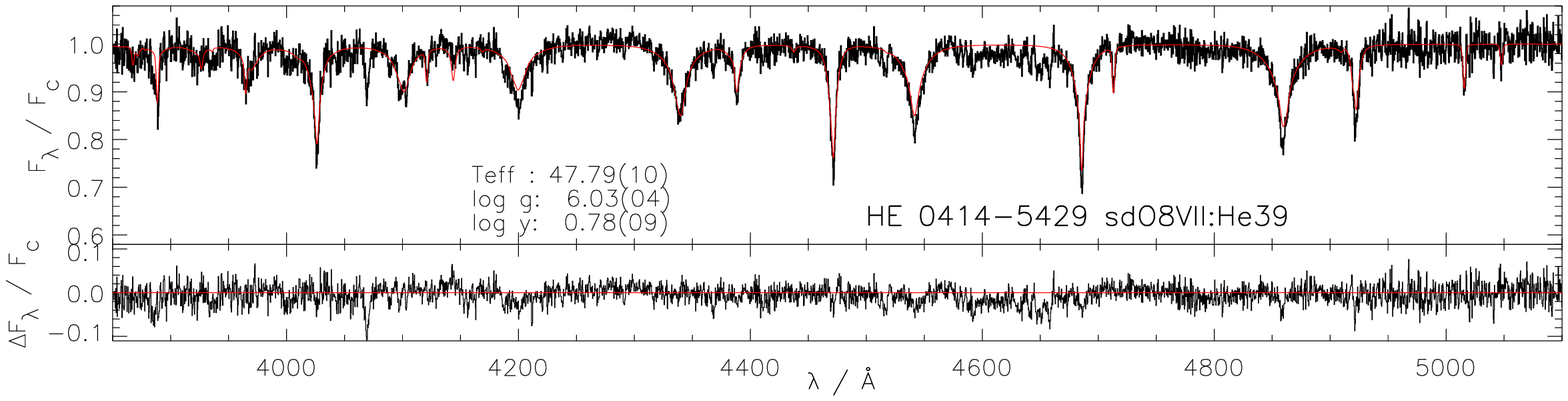}\\
\includegraphics[width=0.85\linewidth]{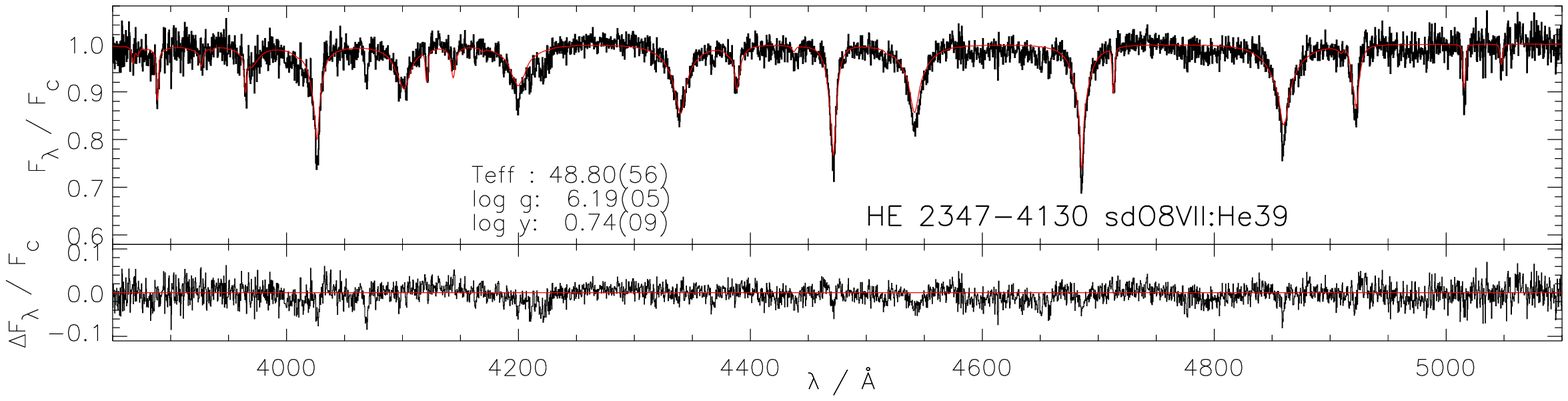}\\
\includegraphics[width=0.85\linewidth]{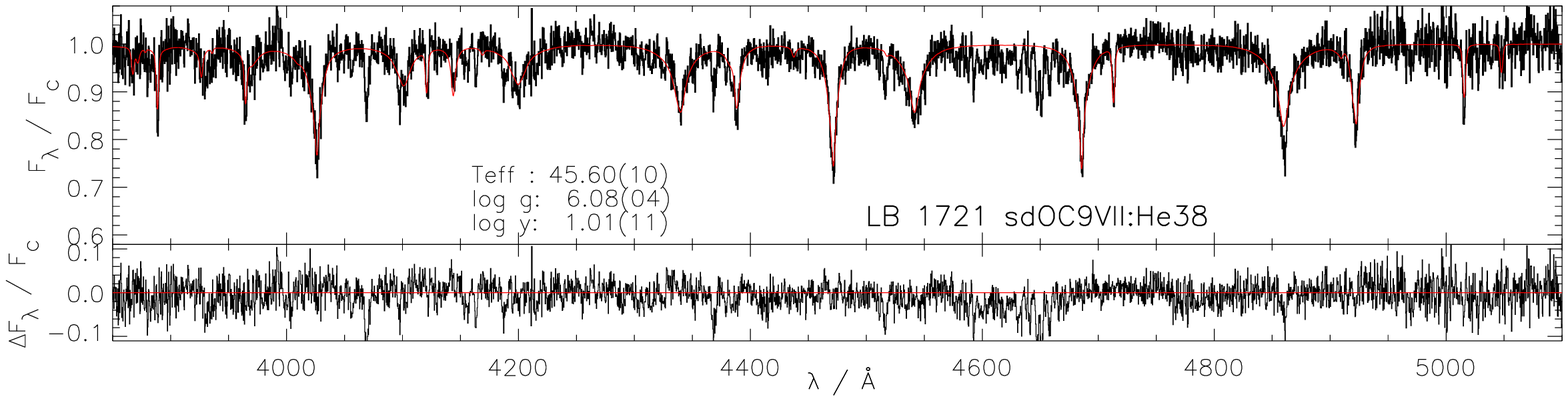}\\
\contcaption{}
\label{f:fit16}
\end{figure*}

\begin{figure*}
\includegraphics[width=0.85\linewidth]{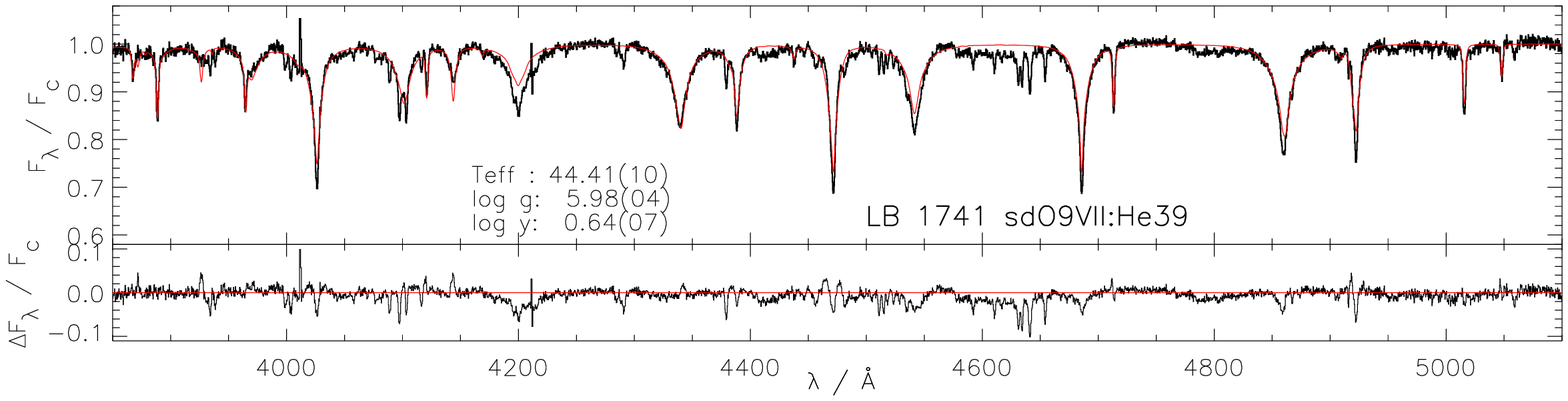}\\
\includegraphics[width=0.85\linewidth]{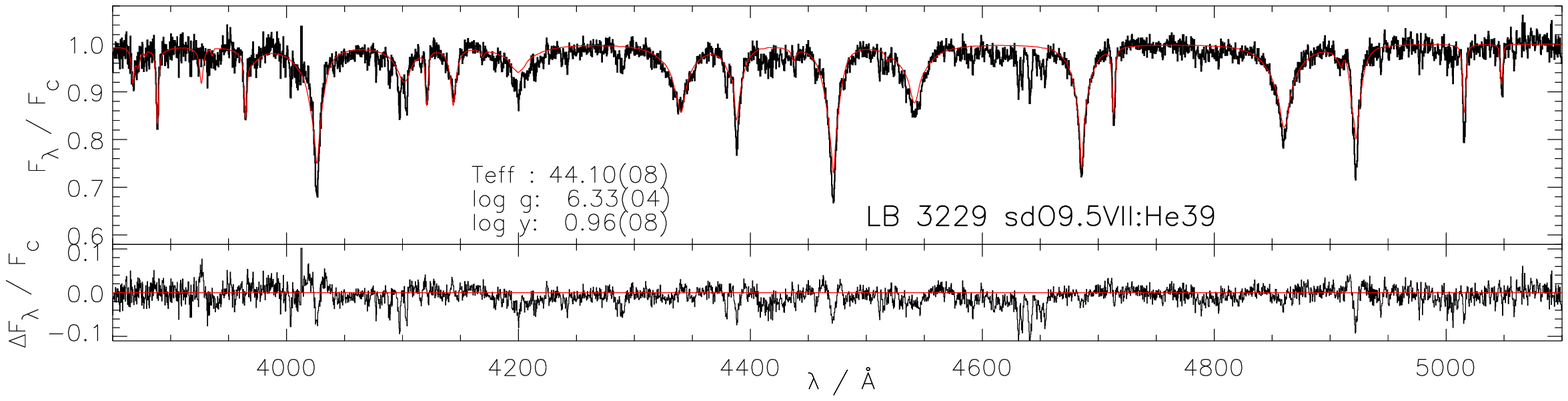}\\
\includegraphics[width=0.85\linewidth]{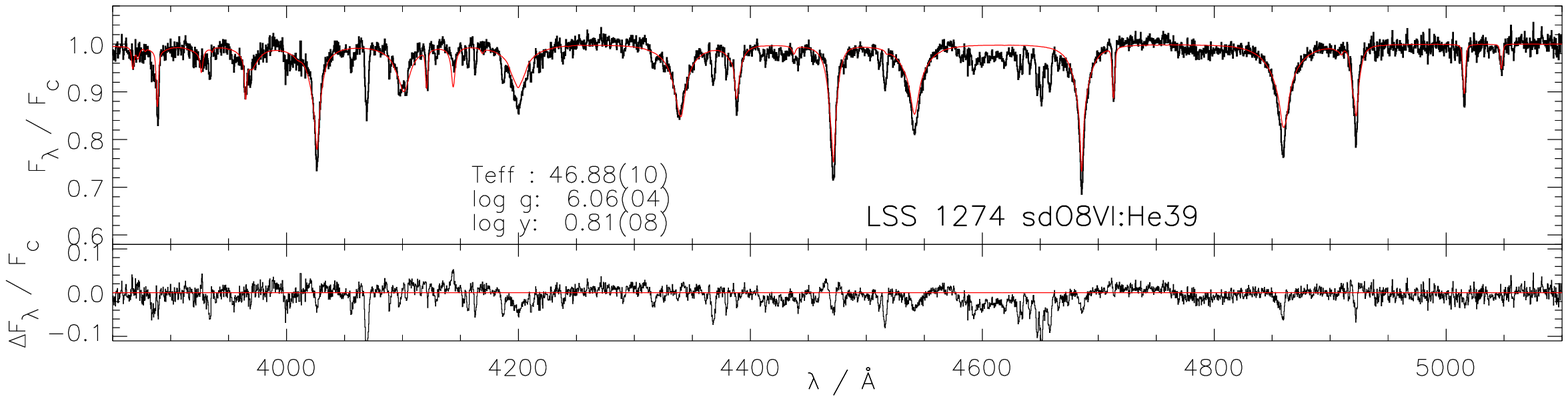}\\
\includegraphics[width=0.85\linewidth]{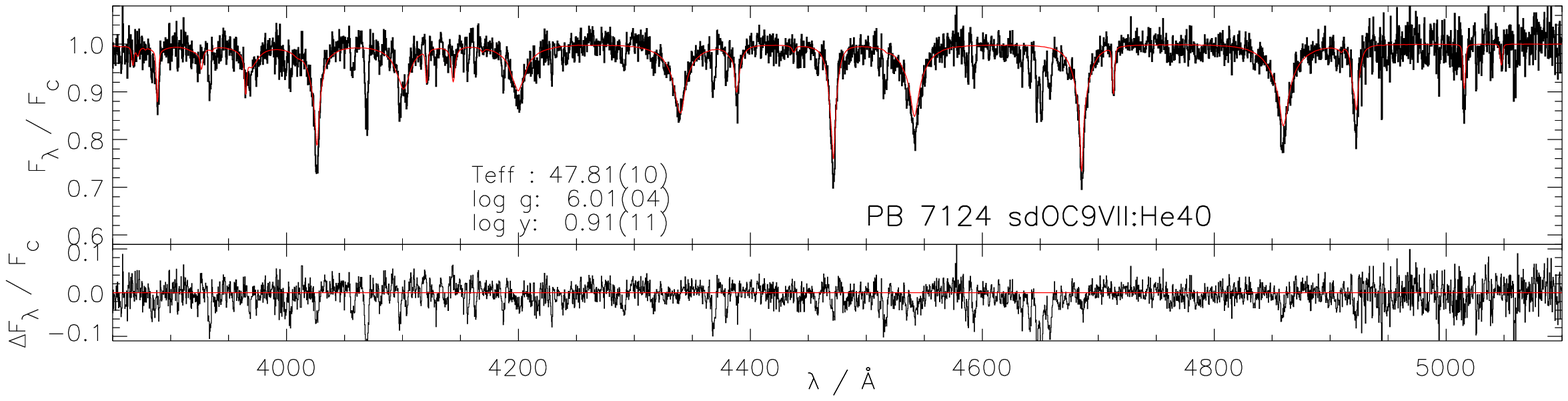}\\
\includegraphics[width=0.85\linewidth]{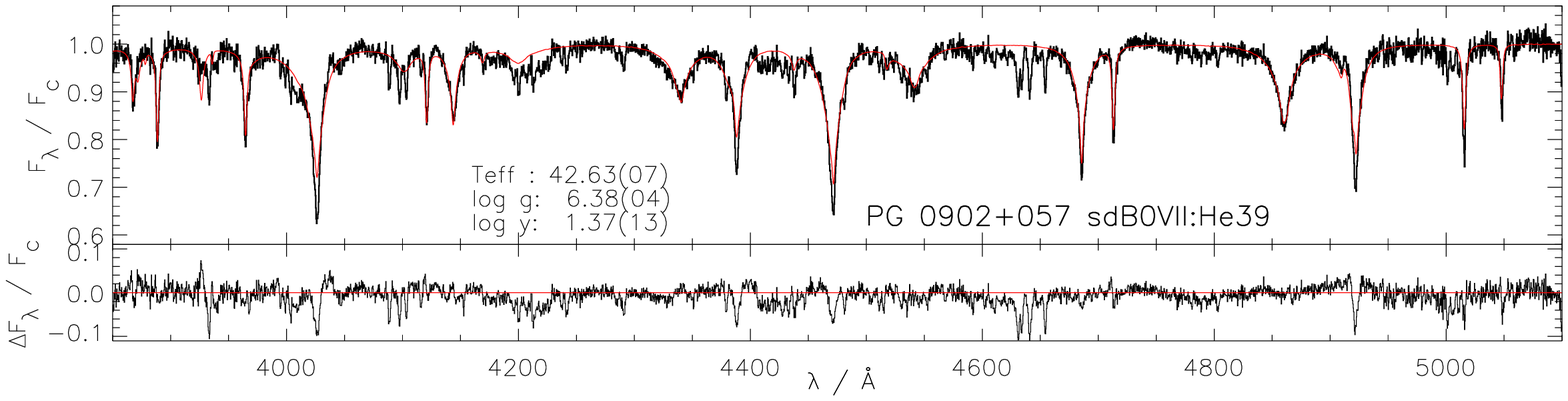}\\
\includegraphics[width=0.85\linewidth]{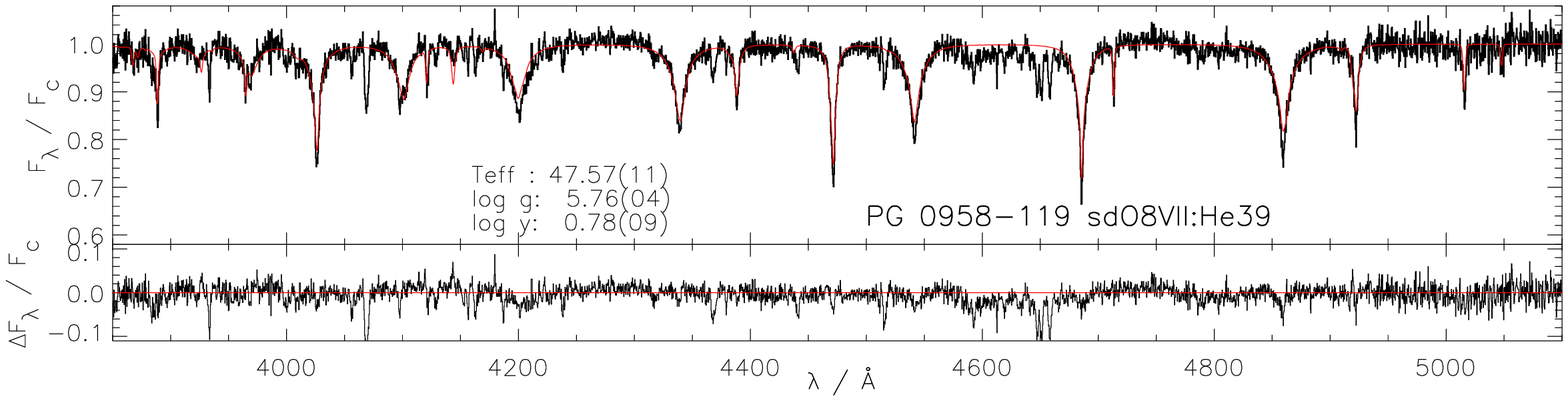}\\
\contcaption{}
\label{f:fit17}
\end{figure*}

\begin{figure*}

\includegraphics[width=0.85\linewidth]{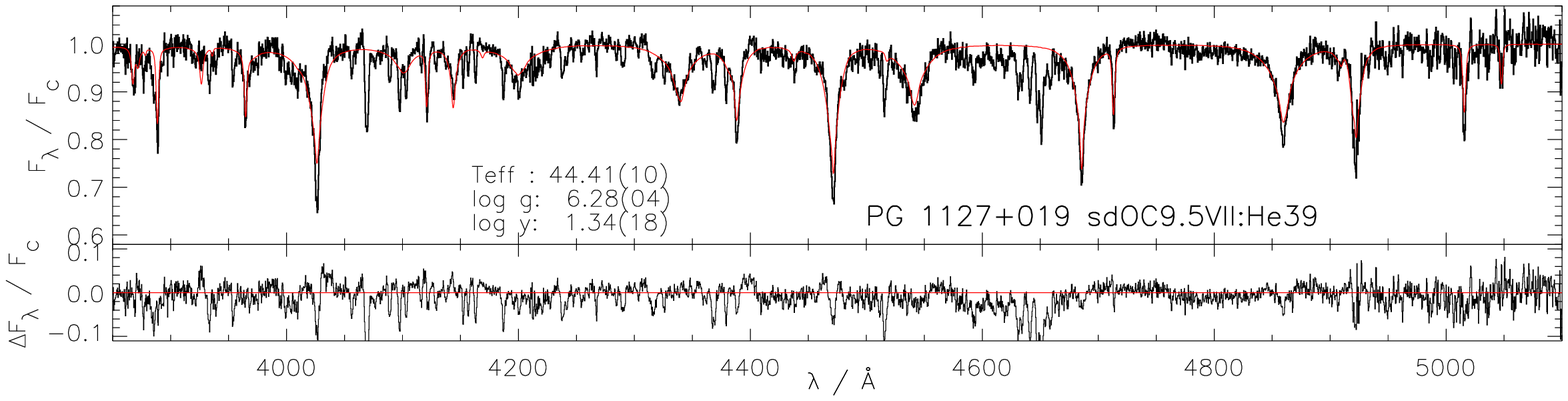}\\   
\includegraphics[width=0.85\linewidth]{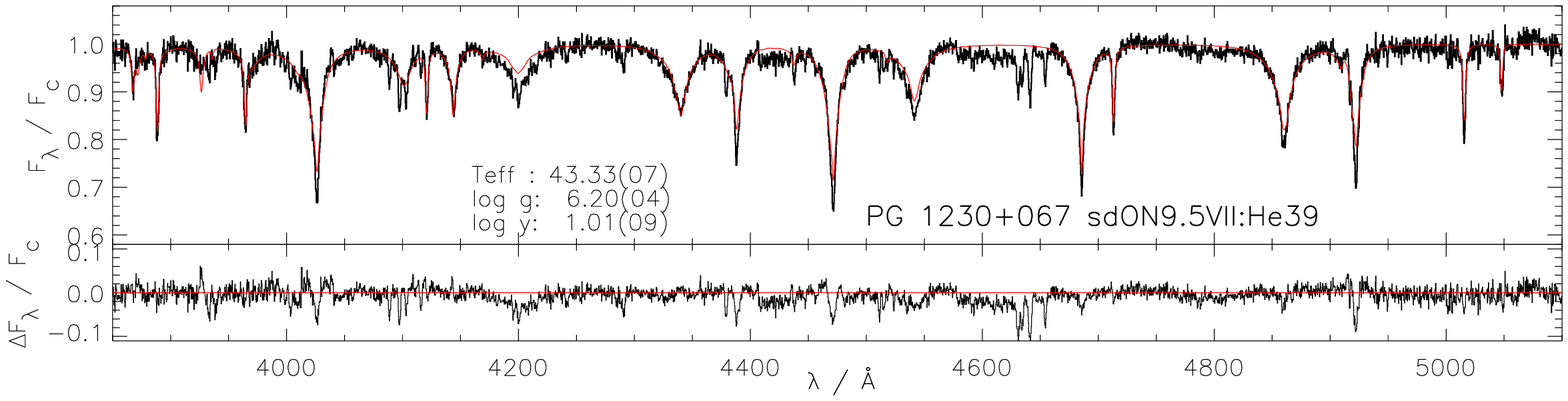}\\
\includegraphics[width=0.85\linewidth]{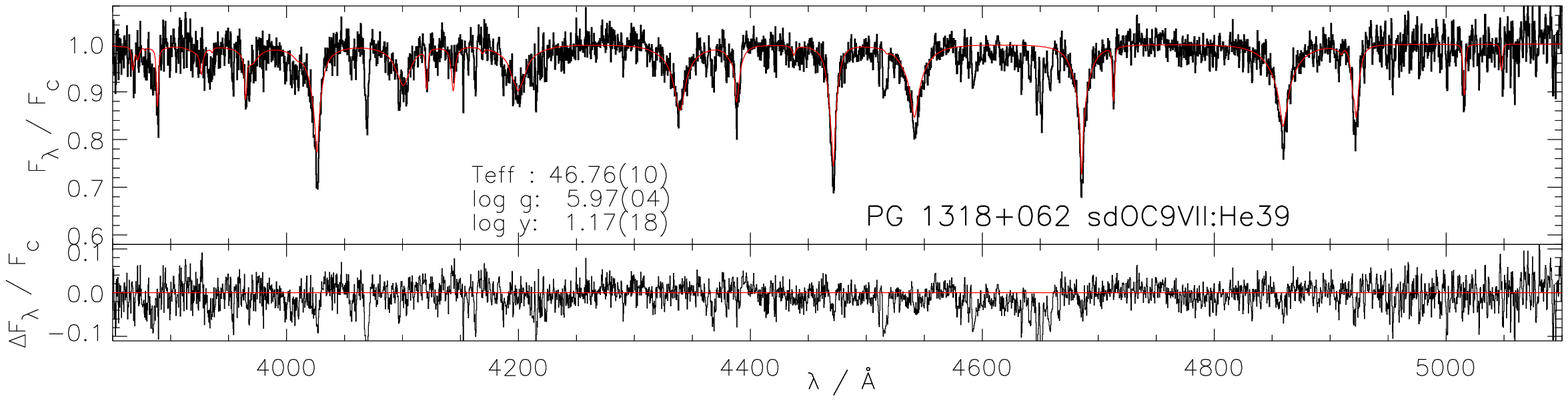}\\
\includegraphics[width=0.85\linewidth]{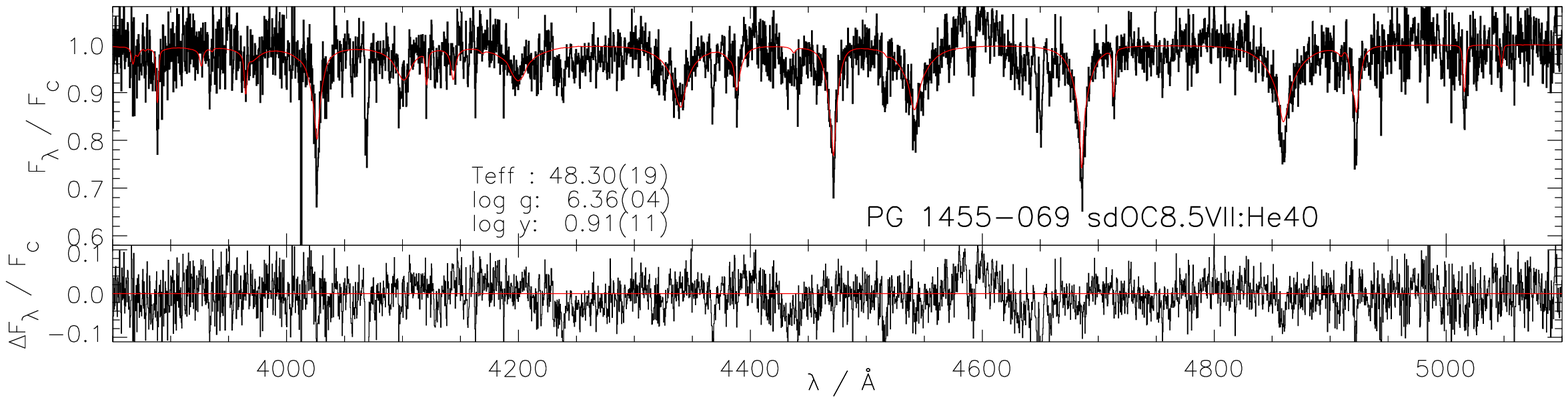}\\
\includegraphics[width=0.85\linewidth]{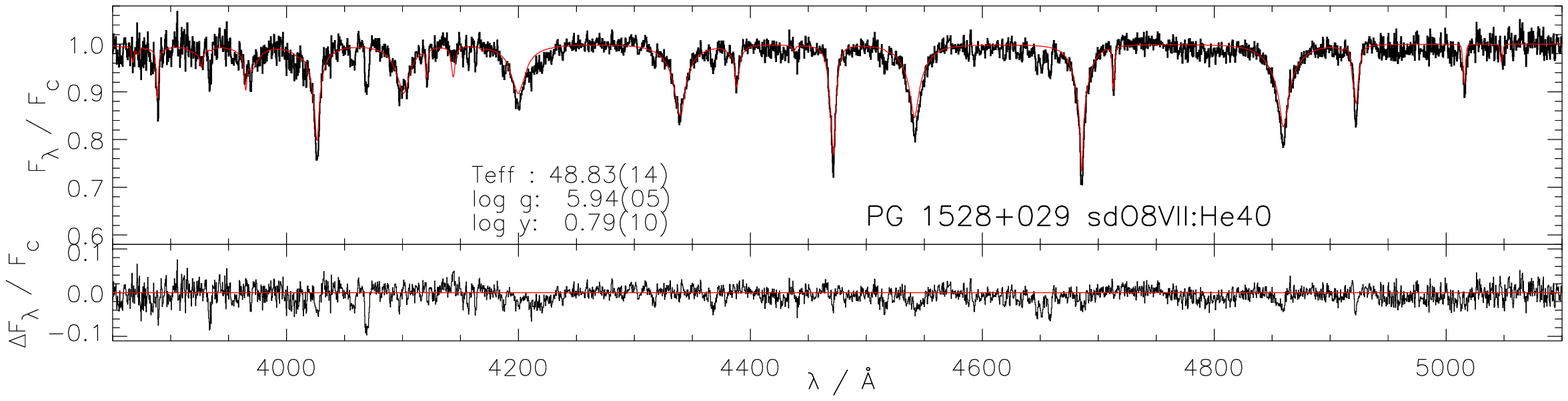}\\
\includegraphics[width=0.85\linewidth]{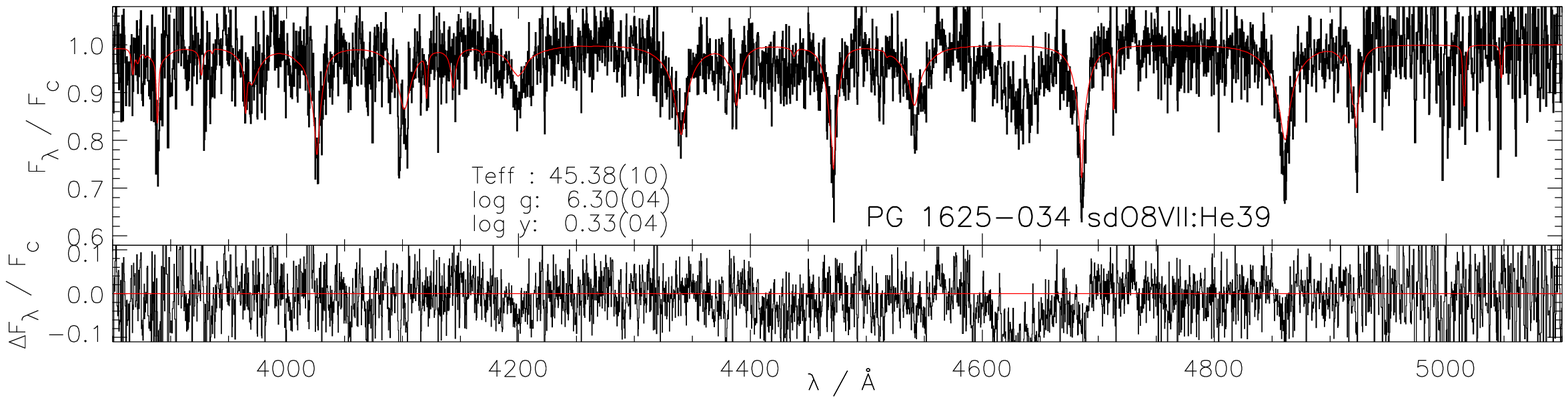}\\
\contcaption{}
\label{f:fit18}
\end{figure*}

\begin{figure*}
\includegraphics[width=0.85\linewidth]{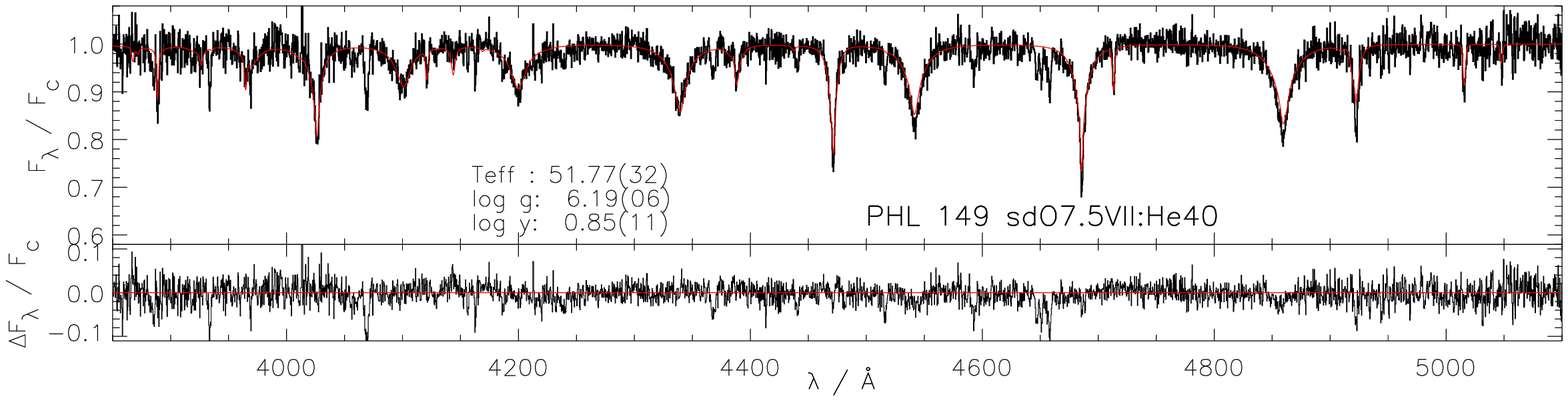}\\
\includegraphics[width=0.85\linewidth]{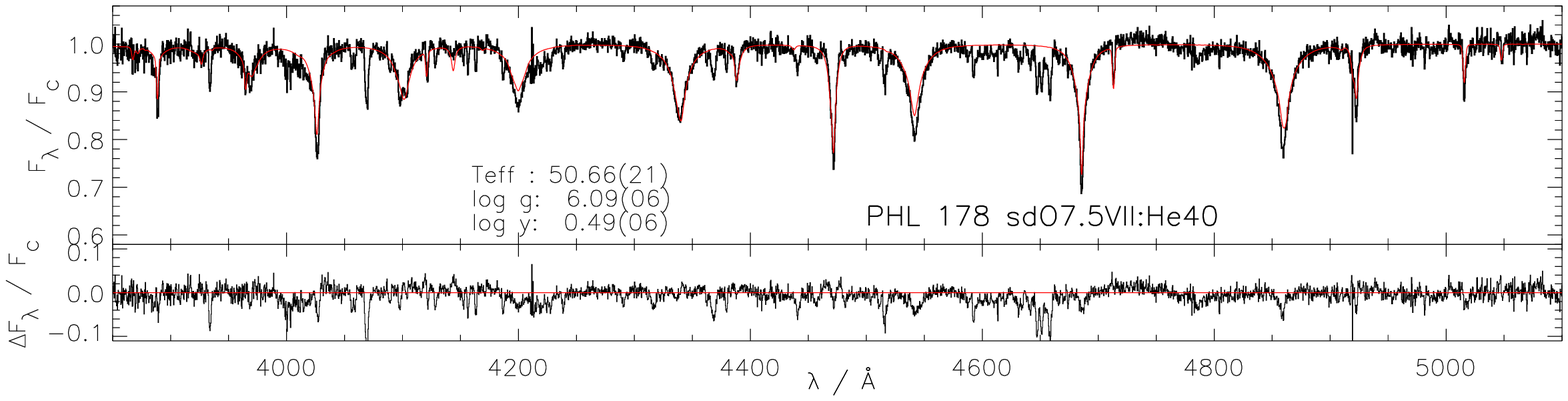}\\
\includegraphics[width=0.85\linewidth]{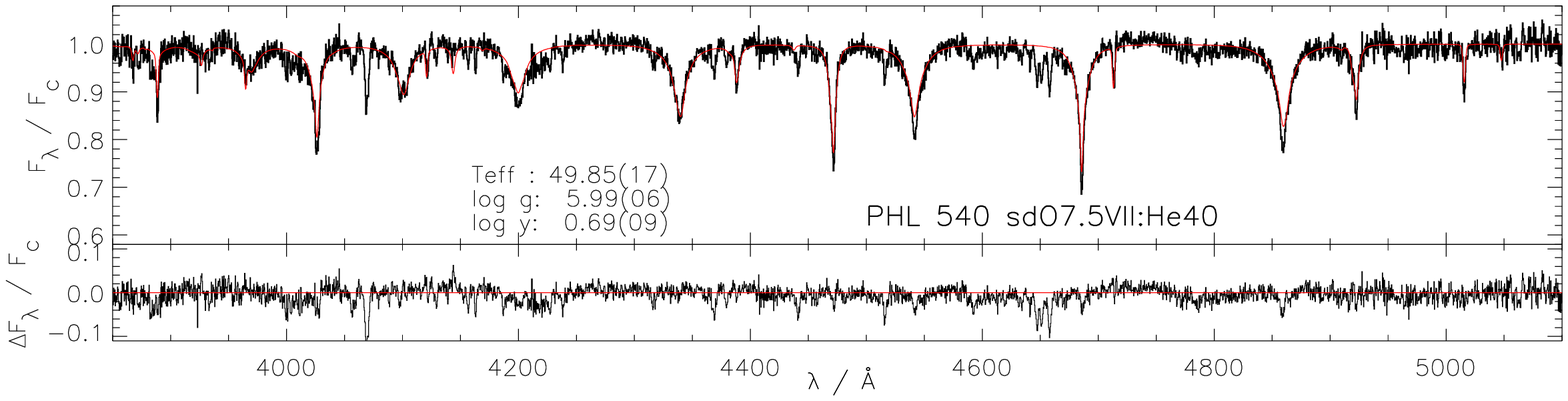}\\
\includegraphics[width=0.85\linewidth]{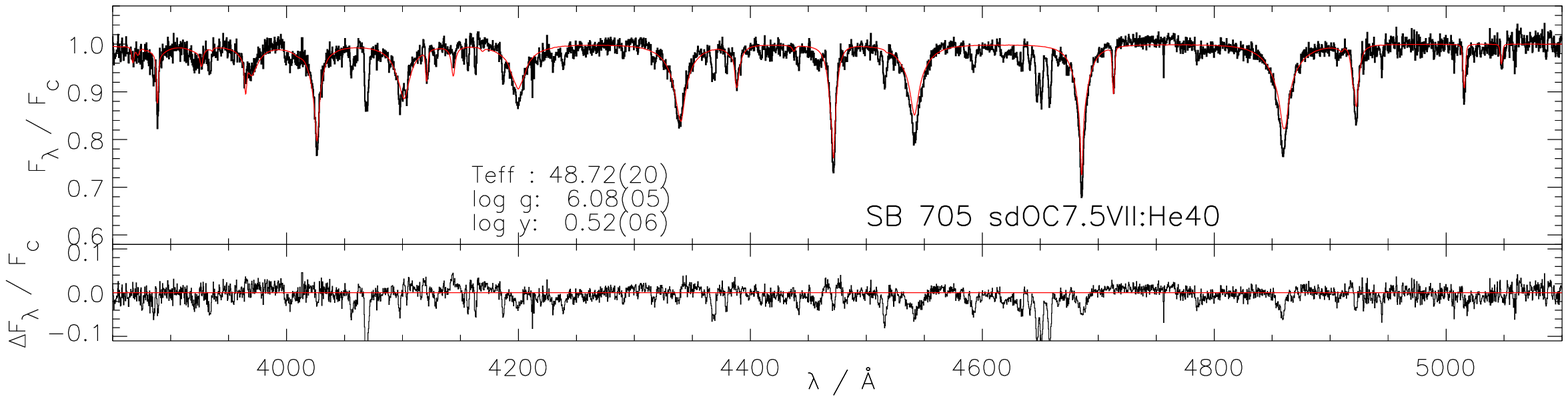}\\
\includegraphics[width=0.85\linewidth]{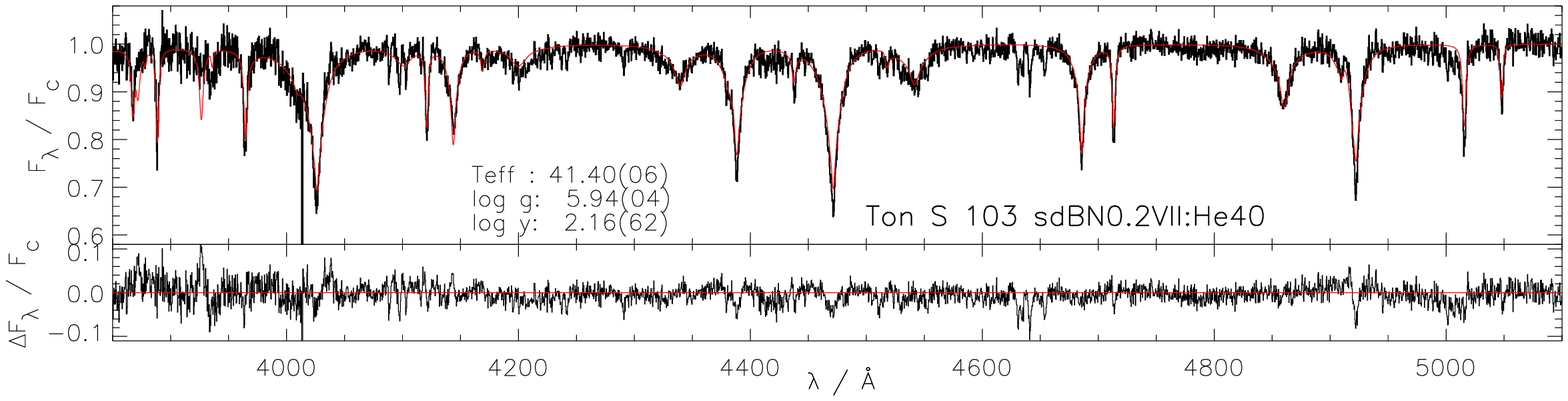}\\
\includegraphics[width=0.85\linewidth]{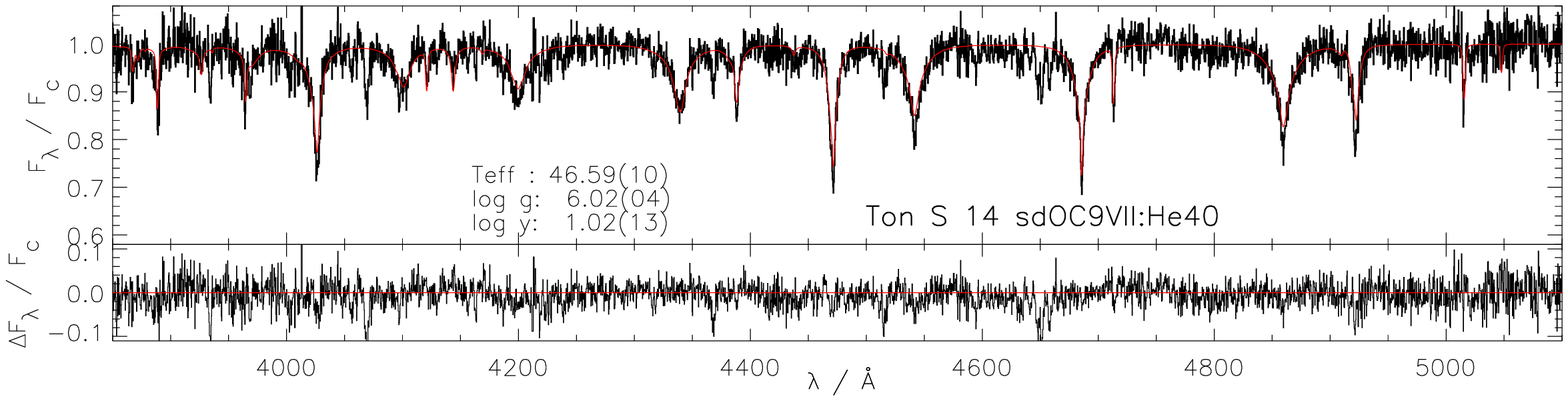}\\
\contcaption{}
\label{f:fit19}
\end{figure*}

\begin{figure*}
\includegraphics[width=0.85\linewidth]{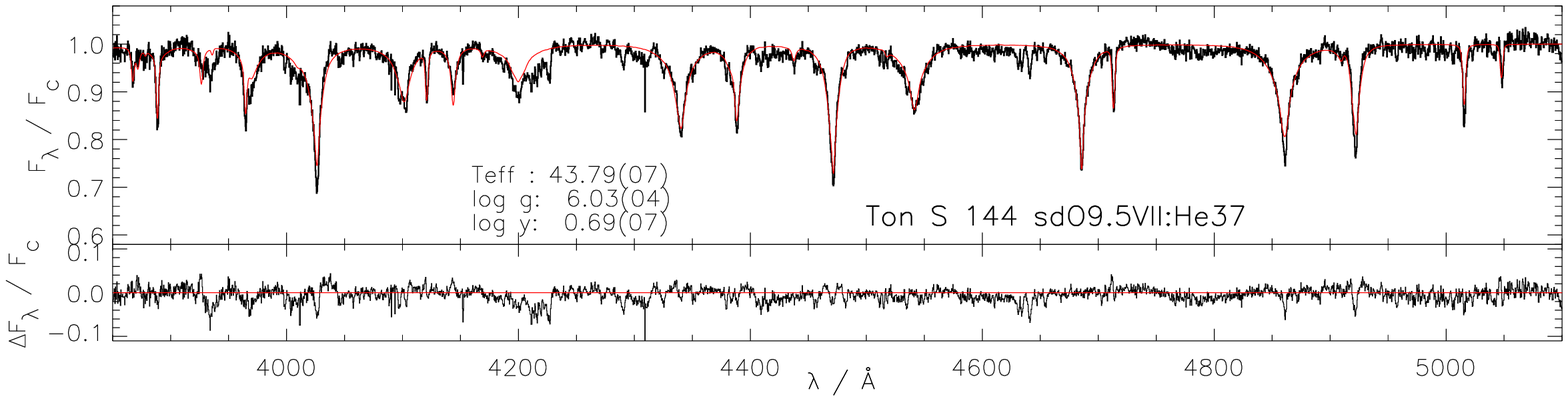}\\
\includegraphics[width=0.85\linewidth]{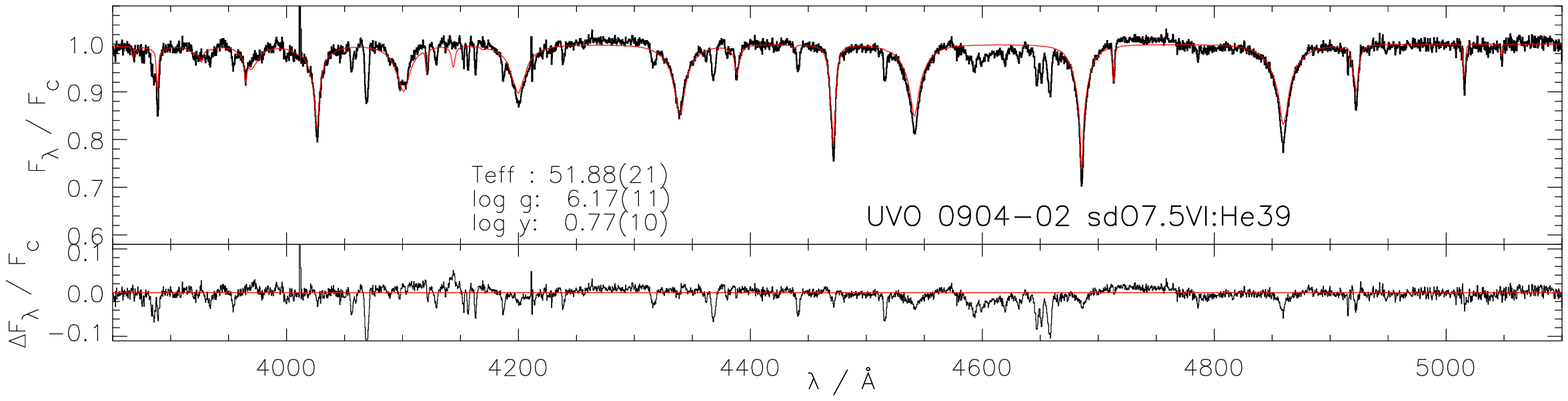}\\
\contcaption{}
\label{f:fit20}
\end{figure*}

\label{lastpage}